\documentclass[12pt]{iopart}
\usepackage{latexsym}
\usepackage{epsfig}
\usepackage{graphics}
\usepackage{amssymb}
\usepackage{color}

\newcommand{\be}{\begin{equation}}
\newcommand{\ee}{\end{equation}}
\newcommand{\bea}{\begin{eqnarray}}
\newcommand{\eea}{\end{eqnarray}}
\newcommand{\nr}{\nonumber\\}
\newcommand{\<}{\langle}
\renewcommand{\>}{\rangle}

\begin{document}
\title{Full-Potential Multiple Scattering Theory with Space-Filling Cells for
bound and continuum states}

\author{Keisuke Hatada$^{1,2}$, Kuniko Hayakawa$^{2,3}$, Maurizio Benfatto$^2$,
Calogero R. Natoli$^{1,2}$}

\address{ $^1$Instituto de Ciencia de Materiales de Arag\'on,
CSIC-Universidad de Zaragoza, 50009 Zaragoza, Spain.,\\
$^2$INFN Laboratori Nazionali di Frascati, Via E. Fermi 40, c.p. 13, I-00044
Frascati, Italy,\\
$^3$Centro Fermi, Compendio Viminale, Roma I-00184, Italy }

\ead{ hatada@unizar.es}

\date{\today}

\begin{abstract}
We present a rigorous derivation of a real space Full-Potential
Multiple-Scattering-Theory (FP-MST) that is free from the drawbacks that up
to now have impaired its development (in particular the need to expand cell shape
functions in spherical harmonics and rectangular matrices), valid both for
continuum and bound states, under conditions for space-partitioning that are not
excessively restrictive and easily implemented.
In this connection we give a new scheme to generate
local basis functions for the truncated potential cells that is simple, fast,
efficient, valid for any shape of the cell and reduces to the minimum the
number of spherical harmonics in the expansion of the scattering wave
function. The method also avoids the need for saturating
'internal sums' due to the re-expansion of the spherical Hankel functions around
another point in space (usually another cell center). Thus this approach,
provides a straightforward extension of MST in the Muffin-Tin (MT)
approximation, with only one truncation parameter given by the classical relation
$l_{\rm max} = kR_b$, where $k$ is the electron wave vector (either in the
excited or ground state of the system under consideration) and
$R_b$ the radius of the bounding sphere of the scattering cell.
Moreover, the scattering path operator of the theory can be found in terms
of an absolutely convergent procedure in the
$l_{\rm max} \rightarrow \infty$ limit. Consequently, this feature
provides a firm ground to the use of FP-MST as a
viable method for electronic structure calculations and makes possible the
computation of x-ray spectroscopies, notably photo-electron diffraction,
absorption and anomalous scattering among others, with the ease and
versatility of the corresponding MT theory. Some numerical applications
of the theory are presented, both for continuum and bound states.

\end{abstract}


%\maketitle

%
\section{ Introduction }

At its most basic, Multiple scattering Theory (MST) is a technique for
solving a linear partial differential equation over a region of space
with certain boundary conditions. It is implemented by dividing the space
into non-overlapping domains (cells), solving the differential equation
separately in each of the cells and then assembling together the partial
solutions into a global solution that is continuous and smooth across the
whole region and satisfies the given boundary conditions.

As such MST has been applied to the solution of many problems drawn both from
classic as well as quantum physics, ranging from the study of membranes and
electromagnetism to the quantum-mechanical wave equation.
In quantum mechanics it has been widely used to solve the Schr\"odinger
equation (SE) ( or the associated Lippmann-Schwinger equation (LSE))
both for scattering and bound states.
It was proposed originally by Korringa and by Kohn and Rostoker (KKR) as a
convenient method for calculating the electronic structure of solids
~\cite{korringa47,kohn54} and was later extended to polyatomic molecules
by Slater and Johnson ~\cite{slater72}.
A characteristic feature of the method is the complete separation between
the potential aspect of the material under study, embodied in the
cell scattering power, from the structural aspect of the problem,
reflecting the geometrical position of the atoms in space.

Applications of the KKR method were first made within the so-called
muffin-tin (MT) approximation for the potential. In this approximation
the potential is confined within non-overlapping spheres, where it is
spherically symmetrized, and takes a constant value in the interstitial
region. Moreover, although spherical symmetry is not formally necessary,
the condition that the bounding spheres do not overlap was thought to be
necessary for the validity of the theory. Despite this approximation the
method is complicated and demanding from a numerical point of view and as
a band-structure method it was therefore superseded by more efficient
linearized methods, such as the linearized muffin-tin-orbital method (LMTO)
~\cite{andersen75} and the linearized augmented-plane-wave method (LAPW).
~\cite{koelling75}

Full-potential versions of these band methods have also been introduced in
recent years. However, none of these methods can match the power and
versatility of
a full-potential method based on the formalism of MST, either in terms of
providing a complete solution of the SE or in the range of problems that
could be treated. In particular, none of these methods leads easily to the
construction of the Green's Function (GF) which is indispensable in the study
of a number of properties of many physical systems.

Due to these reasons, in the last two decades, the KKR method has experienced
a revival in the framework of the Green's function method (KKR-GF). Indeed,
due to the introduction of the complex energy integration, it was found that
the method is well suited for ground-state calculations, with an efficiency
comparable to typical diagonalization methods.
An host of problems became in this way tractable, ranging from solids with
reduced symmetry (like {\it e.g.} isolated impurities in ordered crystal,
surfaces, interfaces, layered systems, etc..) to randomly disordered alloys
in the coherent potential approximation (CPA).

At the same time it soon became clear that the MT approximation was not
adequate to the treatment of systems with reduced symmetry or for the
calculation of lattice forces and relaxation. In order to deal with these
problems a number of groups developed a full potential (FP) KKR-GF method,
obtaining very good results, comparable with full-potential LAPW
method (FLAPW), for what concerns total energy calculations, lattice forces,
relaxation around an impurity,
(~\cite{beiderkellen96,huhne98,asato99,papanik02,ogura05} and Refs.
therein).
Due to their method of generating the single site solutions and the cell
{\it t}-matrix, the additional numerical effort required for the
implementation of the FP-MS scheme scales only linearly with the number
of non-equivalent atoms and is not significantly greater than in the MT case.

In this development the authors took an empirical attitude toward some
fundamental problems related to the extension of MST to the full-potential
case, like the strongly debated question of the {\it l}-convergence of the
theory or the need to converge ``internal'' sums arising from the
re-expansion of the free Green's function around two sites, which entails
the unwanted feature of the introduction of rectangular matrices into the
theory.~\cite{nesbet92}
Without getting involved into {\it ab initio} questions, they just use square
matrices for the structural Green's Function $G^{nn'}_{LL'}(E)$ needed to
calculate the Green's Function of the system (see {\it e.g.} Eqs. (6) and (9)
in Ref.~\cite{papanik02}) and truncate the {\it l}-expansion to
$l_{\rm max} = $ 3 or 4, obtaining in this way the same accuracy as the
FLAPW method.

Some observations are in order at this point. First, the FP method
in the framework of MST has been initially developed only for periodic
systems in two or three dimensions and for states below the Fermi level.
To our knowledge, its extension to treat bound and continuum states of
polyatomic molecules and in general real space applications of the method
have progressed very slowly and have been scarce.
Secondly, the generation of the local solutions of the SE with
truncated cells in the FP extension of the MST has up to now involved the
expansion of the cell shape function in spherical harmonics, which might
create convergence problems, as discussed below.
Thirdly, the FP extension of MST has generated a lot of controversies that
have gone on for more than thirty years ~\cite{butler92}. Some of the
problems have found a solution and we refer the reader to the book of Gonis
and Butler ~\cite{gonis00} for a comprehensive review of the state of the
art in this field (in particular see their chapter 6).
However, questions like the {\it l}-convergence of the theory or the
use of square matrices are still matter of debate and some
rigorous answer should be given to them.

As mentioned above, applications to states well above the Fermi energy,
as required in the simulations of x-ray spectroscopies, like absorption,
photo-emission, anomalous scattering, etc..., have been scarce.
In the words of Ref. ~\cite{gonis00}, ``the feeling that one should
calculate
the ``near-field-corrections'' (NFC), coupled with the need to solve a fairly
complicated system of coupled differential equation to determine the local
(cell) solutions (based on the phase function method) has contributed greatly
to the slow development of a FP method based on MST''.
It was only after it was realized that NFC are not necessary and a new
method to generate local solutions was found that progress became faster,
at least in the calculation of the electronic structure of solids.
(~\cite{beiderkellen96,huhne98,asato99,papanik02,ogura05} and
Refs. therein) The only remaining drawback was and
is the truncation of the potential at the cell boundary which is still
performed via a shape function expanded in spherical harmonics. Added to this
there is the feeling that one should still converge the ``internal'' sums
leading to the use of rectangular matrices in the angular momentum (AM) indexes,
although this last step is sometimes ignored without justifications. Last,
but not least, the question of the {\it l}-convergence of the theory
remains unsettled.

For all these reasons FP codes based on MST for the calculation of x-ray
spectroscopies are not very numerous. We mention here the work by Huhne
and Ebert ~\cite{huhne99} on the calculation of x-ray absorption spectra
using the FP spin-polarized relativistic MST and that of Ankudinov and Rehr
~\cite{Ankudinov05} in the scalar relativistic approximation.
These authors use the potential shape function to generate the local basis
functions which are at the heart of MST. The expansion of the shape function
and the cell potential in spherical harmonics leads to a high number of
spherical components in the coupled radial equations that becomes
progressively cumbersome to handle and time consuming with increasing
energy and in absence of symmetry. This feature might also be at the
origin of another problem related with the saturation of "internal" sums
in the MSE ~\cite{gonis00}, as discussed later in this paper. Moreover
no critical discussion is devoted in their work to the {\it l}-convergence
problems of MST or the use of square matrices in the theory.

Another code based on a version of the MST that uses non overlapping
spherical cells and
treats the interstitial potential in the Born approximation is that of
Foulis {\it et al}. ~\cite{natoli90,foulis90} This method however treats
in an approximate way the potential in the interstitial region and moreover
looses one of the major advantages of the MST, namely the separation between
dynamics and geometry in the solution of the scattering problem. Foulis
~\cite{foulis07} is now developing an exact FP-MS scheme based on distorted
waves in the interstitial region that seem to be promising, but its
numerical implementation is still to come.

There are other codes that simulates x-ray spectroscopies and are not based
on MST: that of Joly ~\cite{joly01} is based on the discretization of the
Laplacian in three dimensions (finite-difference method (FDM)), where
the SE is solved in a discretized form on a three-dimensional grid, the
values of the scattering wave-function being
the unknowns. This method is however limited to cluster sizes
of the order of 20 atoms (without symmetry), due to the high memory
requirement when the number of mesh points increases with the dimensions
of the cluster. Finally a method based on the pseudo-potential theory to
calculate x-ray absorption is worth mentioning.~\cite{cabaret02} It can
easily cope with clusters of many atoms (300 and more) with a computational
effort that scales linearly with the number of atoms. One of its drawback is
its little physical transparency and the fact that it has been applied only
to calculate x-ray absorption spectra. Also, relaxation around the core hole
must be taken into account by super-cell calculations and there is little
flexibility to deal with energy-dependent complex potentials.

The purpose of the present paper is the rigorous derivation of a real space
FP-MST, valid both for continuum and bound states, that is free from the
drawbacks hinted to above, in particular the need to use cell shape
functions and rectangular matrices, under conditions for space-partitioning
that are not excessively restrictive and easily implemented (see beginning
of Section~\ref{msesbs}).
In connection with this we shall present a new scheme to generate local basis
functions for the truncated potential cells that is simple, fast, efficient,
valid for any shape of the cell and reduces to the minimum the number of
spherical harmonics in the expansion of the scattering wave function.
Finally we shall also address the problem of the {\it l}-convergence of the
theory, giving a positive answer to this debated question.

Even though this work is primarily motivated by applications in spectroscopy,
it will be clear from the context that bound states can be treated as well.
Actually the method can also work for complex energy values, so that one can
take advantage of the fact that the solution of the Schr\"odinger equation
is analytical in the energy plane, as is the associated Green's function,
except for cuts and poles on the real axis. Therefore spectroscopy is only
one regime of applications.

Section~\ref{lbfsc} of this paper presents the new scheme to generate local basis
functions and tests it against known solutions for potentials cells
with and without shape truncation.
Section~\ref{msesbs} provides a new derivation of the FP-MST that allows us to work
with square matrices for the phase functions $S_{ LL' }$ and $E_{ LL' }$
and for the cell $T_{ LL' }$ matrix with only one truncation parameter,
contrary to the present accepted view.~\cite{gonis00} Due to their importance
in the theory, various equivalent forms for the Green's function are presented
in this scheme. This latter is extended to the calculation of bound states
of polyatomic molecules and tested against the known eigenvalues of the
hydrogen molecular ion. Section~\ref{cmst} discusses the strongly debated problem of
the {\it l}-convergence of the theory and provides a truncation procedure
that converges absolutely in the $l_{\rm max} \rightarrow \infty$ limit.

Section~\ref{apps} reports one additional application of the present
FP-MS theory besides those already presented in
Ref.s ~\cite{hatada07} and ~\cite{hatada09},
namely the calculation of the absorption cross
section in the case of linear molecules ($Br_2$ diatomic molecule), where
the improvement of over the MT approximation is quite dramatic. Moreover,
with an eye to using the theory to study the performance of model optical
potentials, Section~\ref{apps} also presents a preliminary application of the
non-MT (NMT) approach to the study of the relative performance of the
Hedin-Lundqvist (HL) and the Dirac-Hara (DH) potentials in the case of
a transition metal.
Finally Section~\ref{conc} presents the conclusions of the present work.
A preliminary and partial account of this latter has been presented in
Ref.s ~\cite{hatada07} and ~\cite{hatada09}.

\section{Local Basis functions for single truncated potential cells}
\label{lbfsc}

A characteristic feature of MST is that it does not rely on a finite
basis set for the expansion of the global wave function inside each cell as
all other methods of electronic structure calculations do. Instead it relies
on expanding the global solution in terms of local solutions of the
Schr\"odinger equation at the energy of interest, which can be regarded as
an optimally small, energy adapted basis set~\cite{papanik02}.
Therefore it is essential
for the practical implementation of the theory to devise an efficient numerical
method to generate them. We shall consider Williams and Morgan (WM) basis
functions $\Phi_L ({ \bf r })$ ~\cite{williams74} which inside each cell
are local solutions of the SE and behave at the origin as
$J_{ L }({ \bf r })$ for $r \rightarrow 0$. Throughout the paper we shall
use real spherical harmonics and shall put for short
$J_{ L }({\bf r}; k) \equiv j_l ( kr ) Y_{L} ( \hat {\bf r} )$,
$N_{ L }({\bf r}; k) \equiv n_l ( kr ) Y_{L} ( \hat {\bf r} )$ and
$\tilde{H}_{ L }^+ ({\bf r}; k) \equiv -i k h_l^+ ( kr ) Y_{L}
 ( \hat {\bf r} )$, where $j_l, n_l, h_l$ denote respectively spherical Bessel,
Neumann and Hankel functions of order $l$.
The truncated cell potential $V(r,\hat{\bf r})$ is defined to coincide with
the global system potential inside the cell and to be equal to zero (or to
a constant) outside. As mentioned in the introduction we want to avoid the
expansion of the truncated cell shape function (or equivalently of the
truncated potential) in spherical harmonics due to convergence problems.
However we observe that, even if the potential has a step, the wave function
and its first derivative are continuous, so that its angular momentum
expansion is well behaved and even converges uniformly
in $\hat{\bf r}$.~\cite{kellog54} Therefore we can safely write
$\Phi_L ({ \bf r }) = \sum_{L'} R_{L' L} ( r ) Y_{L'} ( \hat {\bf r} )$
and this expression can be integrated term by term under integral sign.

\subsection{Three-dimensional Numerov method}
\label{3dnum}

In order to generate the basis functions we write the SE in polar coordinates
for the function $P_L ({ \bf r }) = r \Phi_L ({ \bf r })$
\be
\left[ \frac {d^2}{dr^2} + E - V(r,\hat{\bf r}) \right ]P_L (r,\hat{\bf r}) =
\frac {1}{r^2} \tilde{L}^2 P_L (r,\hat{\bf r})
\label{se}
\ee
\noindent where $\tilde{L}^2$ is the angular momentum operator, whose action
on $P_L (r,\hat{\bf r})$ can be calculated as:
\be
  \tilde{L}^2 P_{ L} (r,\hat{\bf r}) \hspace{-1.5mm } = \hspace{-1.5mm }
    \sum_{L'} l' ( l' + 1) r R_{L' L} ( r ) Y_{L'} ( \hat {\bf r} )
     \label{pe}
\ee

Equation (\ref{se}) in the variable $r$ looks like a second order equation
with an inhomogeneous term. Accordingly we use Numerov's method to solve it.
As is well known, putting $f_{ i, j }^L = P_{ L} (r_i,\hat{\bf r}_j)$ and
dropping for simplicity the index $L$, the associated three point recursion
relation is
\be
  A_{ i+1, j } f_{ i+1, j }
  - B_{ i, j } f_{ i, j }
  + A_{ i-1, j } f_{ i-1, j }= g_{ i, j }
  - \frac{ h^6 }{ 240 } f_{ i, j }^{\rm vi}
\label{num3p}
\ee
where,
\bea
  A_{ i, j } &=& 1 - \frac{ h^2 }{ 12 }v_{ i, j } \nr
  B_{ i, j } &=& 2 + \frac{ 5 h^2 }{ 6 }v_{ i, j } = 12 - 10 A_{ i, j } \nr
  v_{ i, j } &=& V ( r_i, \hat {\bf r}_j ) - E \nr
  g_{ i, j } &=& \frac{ h^2 }{ 12 } [ q_{ i+1, j } +
                 10 q_{ i, j } + q_{ i-1, j } ] \nr
  q_{ i, j } &=& \frac{ 1 }{ r_i^2 } \sum_{L'} l' ( l' + 1)
     r_i R_{L' L} ( r_i )Y_{L'} ( \hat {\bf r}_{ j } )
    \label{num3pcf}
\eea
Here $i$ is an index of radial mesh and $j$ an index of angular points on a
Lebedev surface grid. ~\cite{lebedev75} Obviously
$r_i R_{L' L} ( r_i ) = \sum_{j} w_{ j } P_{ L} (r_i,\hat{\bf r}_j) Y_{L'} (
\hat {\bf r}_{ j } )$, where $ w_{ j }$ is the weight function for angular
integration associated with the chosen grid. The number of surface points
$N_{Leb}$ is given by $N_{Leb} \approx (2l_{\rm max} + 1 )^2/3$
as a function of the maximum angular momentum used~\cite{wang03}, taking
into account that one integrates the product of two spherical harmonics.
As it is, we cannot use Eq. (\ref{num3p}) to find  $f_{ i+1,j}$ by
iteration, from the knowledge of $ f_{ i,j} $ and $ f_{ i-1,j} $ at all the
angular points, since the "inhomogeneous" term $ q_{ i+1,j } $ is not
expressible in terms of $ f_{ i+1,j } $ due to the last line of
Eq. (\ref{num3pcf}) and is calculated at the radial mesh point $i+1$.

We first eliminate this point from the expression of $g_{ i,j }$,
observing that
\bea
  g_{ i,j }
  &=& \frac{ h^2 }{ 12 }
    \left[ q_{ i+1,j } + 10  q_i + q_{ i-1,j } \right] \nr
  &=& \frac{ h^2 }{ 12 }
    \left[ \frac{ q_{ i+1,j } - 2 q_i + q_{ i-1,j } }{ h^2 } h^2
    + 12 q_{i,j} \right] \nr
\eea
The second order central difference is given by ~\cite{abramowitz72}
\bea
   q_{ i+1 } - 2 q_i + q_{ i-1 } &=& h^2 q_{ i }''
  + \frac{ h^4 }{ 12 } q_i^{iv} + \frac{ h^6 }{ 360 } q_i^{vi}
  + \frac{ h^8 }{ 20160 } q_i^{viii}+ \cdots
  \label {eqn:h2fmod}
\eea
so that
\bea
   g_{ i,j }&\sim& \frac{ h^2 }{ 12 }
    \left[ \left( q_{i,j}'' + \frac{ h^2 }{ 12 }q_{i,j}^{iv} \right) h^2
    + 12 q_{i,j} \right]
  \label {eqn:gij}
\eea
omitting errors of order $h^6$ and higher.

Now for the second derivative $q_{i,j}''$ we use the backward formula ~\cite{abramowitz72}
\bea
  q_{i,j}'' = \frac{ q_{ i,j } - 2 q_{ i-1,j } + q_{ i - 2,j } }{ h^2 }
    + h q_{i,j}''' - \frac{ 7 h^2 }{ 12 } q_{i,j}^{iv}.
  \label {eqn:qijback}
\eea
to avoid the contribution of the point $i+1$.
Inserting Eq. (\ref{eqn:qijback}) into Eq. (\ref{eqn:gij})
\bea
    g_{ i,j }&\sim& \frac{ h^2 }{ 12 }
    \left[ 13 q_{ i,j } - 2 q_{ i-1,j } + q_{ i - 2,j } \right]
    + \frac{ h^5 }{ 12 } q_{i,j}''' - \frac{ h^6 }{ 24 } q_{i,j}^{\rm iv}
  \label{eqn:inhomo}
\eea
which is the formula we wanted to arrive at.
Therefore our modified Numerov procedure becomes:
\be
  A_{ i+1,j } f_{ i+1,j }
  - B_{ i,j } f_{i,j}
  + A_{ i-1 } f_{ i-1,j }= g_{i,j}
  + \frac{ h^5 }{ 12 } q_{i,j}'''
\label{modnum}
\ee
where,
\bea
  A_{ i,j } &=& 1 - \frac{ h^2 }{ 12 }p_{i,j} \nr
  B_{ i,j } &=& 2 + \frac{ 5 h^2 }{ 6 }p_{i,j} = 12 - 10 A_{ i,j } \nr
  g_{ i,j } &=& \frac{ h^2 }{ 12 }
              \left[ 13 q_{ i,j } - 2 q_{ i-1,j } + q_{ i - 2,j } \right]
\label{modnum1}
\eea
which now needs three backward points to start.

\noindent  The appearance of the third $r$ derivative of $q_i'''$ in Eq. (\ref{modnum}),
which is strictly infinite at the step point, does not cause practical problems.
Although not necessary, one can always assume a smoothing of the potential at the
cell boundary {\it \`{a} la Becke}, ~\cite{becke88} reducing at the same time the
 mesh $h$, so that the error at that particular step point is negligible.

In this way, at the cost of a bigger error $ O ( h^5 )$ compared to the
original Numerov formula and the
introduction of a further backward point (three points $ i $,
$ i-1 $ and $ i-2 $ are now involved in (\ref{modnum1})), the three-dimensional
discretized equation can be
solved along the radial direction for all angles in an onion-like way, provided
the expansion (\ref{pe}) is performed at each new radial mesh point to calculate
$q_{ i,j }$.
We use a log-linear mesh $\rho = \alpha \, r + \beta \, { \ln } \, r $, to reduce
numerical errors around the origin and the bounding sphere.~\cite{brastev66}

%This modified Numerov method is much faster and requires a smaller memory
%space than the corresponding cell methods based on the shape function and/or
%the phase functions to generate the local solutions $P_L ({ \bf r })$.

\subsection{Matrix Numerov method}
\label{mnm}

It is well known that errors in the Numerov difference equation originating
from the unidimensional differential equation
\be
\left[ \frac {d^2}{dr^2} + E - \frac{l(l+1)}{r^2} -V(r) \right ] P_l (r) = 0
\nonumber
\ee
grows exponentially when $E - l(l+1)/r^2 -V(r) \le 0$. Therefore near the
origin and in general for large $r$ meshes  and/or high $l$ values the
method is not suitable.
This is also true for Eq. (\ref{se}). To avoid this problem we use the so
called Gaussian elimination for the difference equation
~\cite{fischer77,mitchell90,amusia97}.
We notice that in the MT sphere lying inside the cell the AM expansion of the
potential is regular and in general only few multipoles are appreciable.
Therefore, by projecting onto $Y_L ( {\hat {\bf r} } )$ we can rewrite
Eq. (\ref{se}) as ~\cite{natoli86}
\be
    \left( - \frac{d^2}{dr^2}
              + \frac{ l ( l + 1 ) }{ r^2 } - E \right)
       X_{ L \, L' } ( r )
       + \int d { \hat {\bf r} } \, Y_L ( {\hat {\bf r} } ) \, V ( {\bf r} )
       P_{L'} ({\bf r} )   =    0 \nonumber
\ee
{\it i.e.}
\bea
    && \sum_{L''} \left[ \left( - \frac{d^2}{dr^2}
              + \frac{ l ( l + 1 ) }{ r^2 } - E \right) \delta_{ L L'' }
       + V_{ LL'' } ( r ) \right] X_{ L' L'' } ( r )  \nr
    && = {\bf F } ( r ) {\tilde {\bf X } } ( r ) = 0
  \label{matnmv}
\eea
where $X_{ L \, L' } ( r ) = \, r \, R_{ L\, L' } ( r )$, $\tilde { X }$
is its transposed,
\be
  V_{ LL' } ( r ) = V_{ L'L } ( r )
    = \int d { \hat {\bf r} } \, Y_L ( {\hat {\bf r} } ) \,
    V ( {\bf r} ) \, Y_{L'} ( {\hat {\bf r} } )
%  = \sum_{L''} \, C \, ( \, L L' \, ; L'' \, ) \, V_{ L'' } ( r )
\ee
and
\bea
  ( {\bf F } ( r ) )_{LL'} &&= ( {\bf F } ( r ) )_{L'L} \nr
  && = \left[ \left( - \frac{d^2}{dr^2}
        + \frac{ l ( l + 1 ) }{ r^2 } - E \right) \delta_{ L L' }
        + V_{ LL' } ( r ) \right]
\label{fop}
\eea

Eq.~( \ref{matnmv} ) is a system of coupled radial Schr\"odinger equations
in matrix form that can be solved simultaneously for all $L, L'$ components
with appropriate initial conditions.

The Numerov recursion relation for the matrix SE~\cite{natoli86} is
(notice the change of sign of the coefficient $B$ for sake of later
convenience)
\bea
  &&{\bf A}_{ i+1 } {\tilde {\bf X} }_{ i+1 }
  + {\bf B}_{ i } {\tilde {\bf X}}_i
  + {\bf A}_{ i-1 } {\tilde {\bf X}}_{ i-1 }= 0 \\
  &&{\bf A}_{ i } = 1 - \frac{ h^2 }{ 12 } {\bf P}_i \nr
  &&-{\bf B}_{ i } = 2 + \frac{ 5 h^2 }{ 6 } {\bf P}_i = 12 - 10 {\bf A}_{ i }
\nr
  &&( {\bf P}_i )_{L \, L'} = V_{ L \, L' } ( r_i )
+ \left[ \frac{ l ( l + 1 ) }{ r_i^2 } - E \right] \delta_{ L \, L' }
\label{gauss1}
\eea
where $i$ is the generic point of the radial mesh. Its explicit matrix form is,
\be
  \left(
    \begin{array}{cccccc}
      {\bf A}_{ 0 } & {\bf B}_{ 1 } & {\bf A}_{ 2 } &&& \mbox{\huge{$O$}}\\
      & {\bf A}_{ 1 } & {\bf B}_{ 2 } & {\bf A}_{ 3 } && \\
      && \ddots & \ddots & \ddots & \\
      \mbox{\huge{$O$}}&&& {\bf A}_{ M - 1 } &  {\bf B}_{ M } & {\bf A}_{ M + 1
}
    \end{array}
  \right)
  \left(
    \begin{array}{l}
     {\tilde {\bf X}}_0 \\
     {\tilde {\bf X}}_1 \\
     \hspace{2mm}\vdots \\
     {\tilde {\bf X}}_{ M + 1 }
    \end{array}
  \right) =
  \left(
    \begin{array}{c}
     {\bf 0} \\
     {\bf 0} \\
     \vdots  \\
     {\bf 0}
    \end{array}
  \right)
\ee
Since the regular solution has the boundary condition, ${\bf X}_0 = {\bf 0}$
we can rewrite this latter equation as
\be
  \left(
    \begin{array}{ccccc}
      {\bf B}_{ 1 } & {\bf A}_{ 2 } && \mbox{\huge{$O$}}\\
      {\bf A}_{ 1 } & {\bf B}_{ 2 } & {\bf A}_{ 3 } && \\
      & \ddots & \ddots & \ddots &  \\
      \mbox{\huge{$O$}}&& {\bf A}_{ M - 1 } &  {\bf B}_{ M }
    \end{array}
  \right)
  \left(
    \begin{array}{l}
     {\tilde {\bf X}}_1 \\
     {\tilde {\bf X}}_2 \\
     \hspace{2mm}\vdots \\
     {\tilde {\bf X}}_{ M }
    \end{array}
  \right) =
  \left(
    \begin{array}{c}
     {\bf 0} \\
     {\bf 0} \\
     \vdots  \\
     - {\bf A}_{ M + 1 } {\tilde {\bf X}}_{ M + 1 }
    \end{array}
  \right)
\ee
This set of equations can be solved by performing forward Gaussian
elimination near the origin,~\cite{fischer77,mitchell90,amusia97}
\be
  \left(
    \begin{array}{ccccc}
      {\bf D}_{ 1 } & {\bf A}_{ 2 } &&& \mbox{\huge{$O$}}\\
      & {\bf D}_{ 2 } & {\bf A}_{ 3 } & \\
      && \ddots & \ddots  \\
      &&& {\bf D}_{ M - 1 } &  {\bf A}_{ M } \\
      \mbox{\huge{$O$}}&&&& {\bf D}_{ M}
    \end{array}
  \right)
  \left(
    \begin{array}{l}
     {\tilde {\bf X}}_1 \\
     {\tilde {\bf X}}_2 \\
     \hspace{2mm}\vdots \\
     {\tilde {\bf X}}_{ M - 1 } \\
     {\tilde {\bf X}}_{ M }
    \end{array}
  \right) =
  \left(
    \begin{array}{c}
     {\bf 0} \\
     {\bf 0} \\
     \vdots  \\
     {\bf 0} \\
     - {\bf A}_{ M + 1 } {\tilde {\bf X}}_{ M + 1 }
    \end{array}
  \right)
\label{gauss2}
\ee
with
\bea
  &&{\bf D}_{ 1 } = {\bf B}_{ 1 }, \nr
  &&{\bf D}_{ 2 } = {\bf B}_{ 2 } -
    {\bf A}_{ 1 } {\bf D}_{ 1 }^{-1}{\bf A}_{ 2 }, \nr
  && ..., \nr
  &&{\bf D}_{ i } = {\bf B}_{ i } -
    {\bf A}_{ i-1 } {\bf D}_{ i-1 }^{-1} {\bf A}_{ i }, \; ( i = 1, ..., M )
\label{gauss3}
\eea
constituting a set of forward recurrence relations for the quantities
${\bf D}_{ i }$. In terms of these latter we finally obtain the following
recurrence relations:
\be
  {\tilde {\bf X}}_{ i } = - {\bf D}_{ i }^{-1} {\bf A}_{ i+1 }
{\tilde {\bf X}}_{ i+1 }, ( i = 1, ..., M)
\label{gauss4}
\ee
the solution of which can be calculated backward starting from
${\tilde {\bf X}}_{ M+1 } = {\bf I}$, modulo a constant normalization
matrix.
As will be clear from the following, this initial matrix in practice will
not be needed.
%
%\bea
%  &&{\tilde {\bf X}}_{ i } {\tilde {\bf X}}_{ M+1 }^{-1}
%    {\tilde {\bf X}}_{ M  +1 }  =
%    {\tilde {\bf X}}_{ i }' {\tilde {\bf X}}_{ M+1 }  =
%   - {\bf D}_{ i }^{-1} {\bf A}_{ i+1 } {\tilde {\bf X}}_{ i+1 }'
%      {\tilde {\bf %X}}_{ M+1 } \nr
%  &&{\tilde {\bf X}}_{ i }' =
%    - {\bf D}_{ i }^{-1} {\bf A}_{ i+1 } {\tilde {\bf X}}_{ i+1 }', \nr
%  &&{\tilde {\bf X}}_{ M+1 }'= {\tilde {\bf X}}_{ M+1 }^{-1}
%    {\tilde {\bf X}}_%{ M+1 } = {\bf I}
%\eea
%
Summarizing, our strategy to generate the cell basis functions
$P_L ({ \bf r }) = r \Phi_L ({ \bf r })$ is the following. In a spherical domain
around the origin, inside which there are no discontinuities of the potential,
we use the matrix Numerov method with Gaussian elimination (GE), since
we can expand the potential in a well behaved series of spherical harmonics.
We use the GE method to avoid the well know instability of the Numerov recursion
relation near the origin when the angular momentum l is high, as mentioned above.
As boundary conditions we use ${\tilde {\bf X}}_0$ = 0 at the origin and
${\tilde {\bf X}}_{ M+1 }$ = I at the radius of the sphere, which is usually
taken to  coincide with the MT sphere inscribed in the cell.
We then take the last three points of the solution so obtained
to start the 3-d modified Numerov procedure outward across the potential
discontinuity up to the cell bounding sphere. Since the local SE we are dealing
with is an homogeneous equation, its solution is determined up to an arbitrary
normalization constant (reflected in the second arbitrary condition
${\tilde {\bf X}}_{ M+1 }$ = I ). For the basis
functions $P_L ({ \bf r })$ we never need such a constant, since only ratios
of these functions appear in MS Theory, as clear in the following.
Instead, when we compare with a definite solution, like in
Fig.s \ref{fig_3dwell} and \ref{fig_mathieu} below, we need to provide the
value of this solution at another point, usually the radius of the sphere.
This means taking a value for ${\tilde {\bf X}}_{ M+1 }$  appropriate for this
solution.

It is also clear that the method can also be applied to generate by inward
integration the irregular solutions needed to calculate the Green's function.

This procedure is quite efficient and was tested against analytically solvable,
separable model potentials, with and without shape truncation, obtaining
very good results. In Ref.~\cite{hatada09} we have shown the
comparison between the analytical solution and the numerical one for certain
directions in the special case of the truncated potential
$V(x,y,z) = a \, \theta ( |x| - R_c ) + b \, \theta ( |y| - R_c ) + c \,
\theta ( |z| - R_c )$,
where $\theta$ is the step function, $R_c$ = 3.78 au = 2.0 \AA $\,$ and
$a = -0.05, \, b= -0.1, \, c = -0.15$ Ryd, for an energy E = 0.3 Ryd.
For this comparison we used an $l_{\rm max} = 7$ and a number of surface
points on the Lebedev grid equal to 266.

Fig.~\ref{fig_3dwell} shows the same comparison in the more stringent case of
a discontinuity of the order of one Ryd. We took indeed
$a = -0.5, \, b= -1.0, \, c = -1.5$ Ryd for E = 0.3 Ryd.
In this case, along the $z$ direction, one can even observe
a kink in the curvature of the solution, which is well reproduced and related
to the discontinuity of the second derivative at the truncation value of
$R_c$ = 2.0 \AA. As expected, for good agreement we had to increase $l_{\rm max}$
up to 11 and take a number of Lebedev points equal to 1454.
Notice here that the numerical method to generate the solution is really
three-dimensional and does not take advantage of the separability of the
analytical one.
For this comparison we used the Matrix Numerov method with GE up to
$R_c = 2.0 \, \AA = 3.78 \, {\rm au}$, then switched to 3-d Numerov. Due
to the high potential step in this case, we took a number N of radial mesh
points given by N = 834.
\begin{figure}
  \begin{center}
    \begin{tabular}{cc}
      {\includegraphics{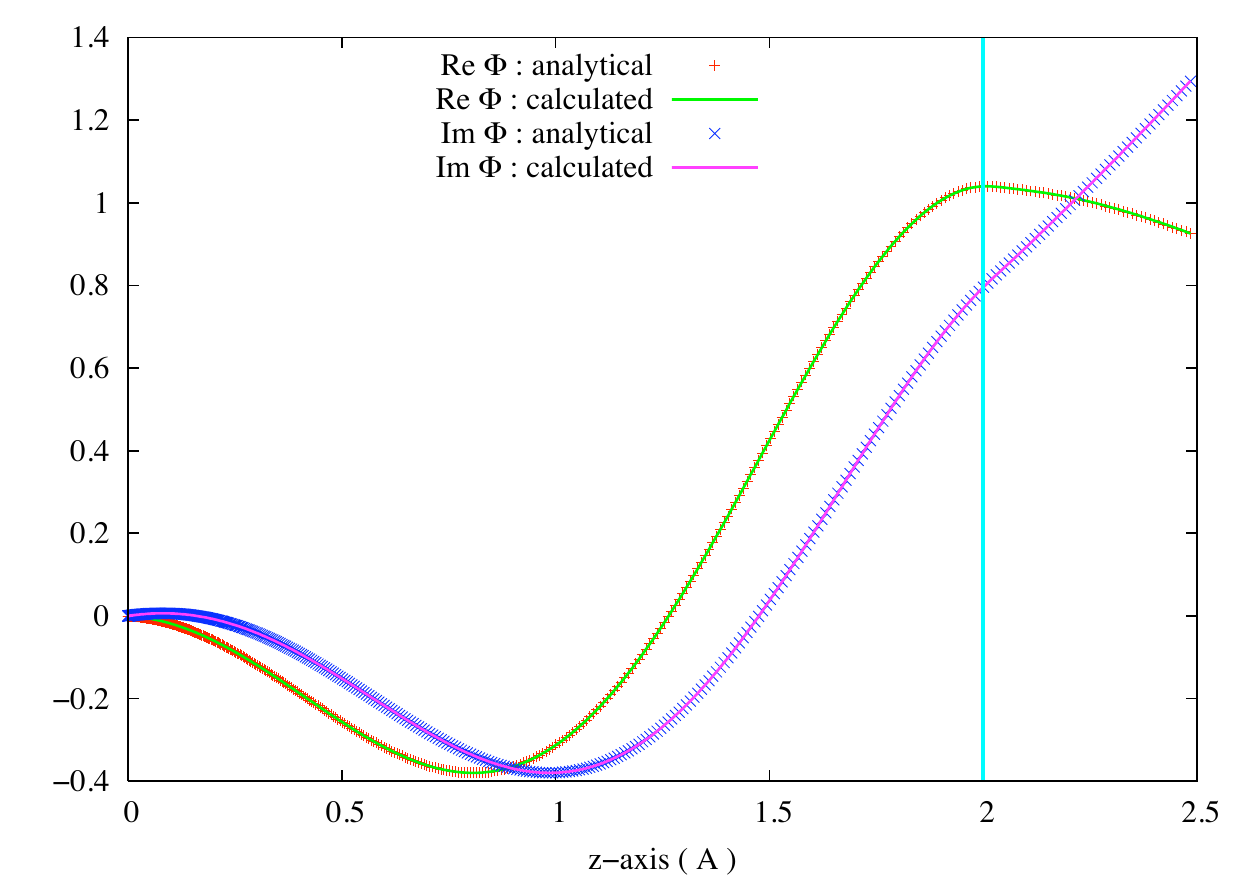}} &
%      \resizebox{70mm}{!}{\includegraphics{wf_3dwell2.pdf}} &
%      \resizebox{70mm}{!}{\includegraphics{../SE_RESULT/3dwell_2.pdf}} \\
%      \resizebox{70mm}{!}{\includegraphics{../SE_RESULT/3dwell_3.pdf}} &
%      \resizebox{70mm}{!}{\includegraphics{../SE_RESULT/3dwell_4.pdf}} \\
    \end{tabular}
  \end{center}
  \caption{Real and imaginary part of the numerical solution of the SE
along $z$ direction for the separable truncated potential given in the
text, compared to the analytical one. (Color online)}
  \label{fig_3dwell}
\end{figure}

In order to test the reliability of the method also in the case of a potential
which is not truncated but varies substantially in sign and magnitude inside
the defining region, we show in Fig. \ref{fig_mathieu} the same comparison for
the Mathieu functions, solution of the separable SE with periodic boundary
conditions
\bea
  && \left[ \frac{ d^2 }{ d x^2 } + \frac{ d^2 }{ d y^2 } + \frac{ d^2 }{ d z^2 }
  \right] \psi ( x, y, z ) \nr
  &&= ( - a_x - a_y - a_z  + 2 q_x \cos 2 x + 2 q_y \cos 2 y + 2 q_z \cos 2 z )
\psi ( x, y, z )
\label{3dmathieu}
\eea
for the case where
\bea
%V = 2 * ( q_x cos ( 2 x ) + q_y cos ( 2 y ) + q_z cos ( 2 z ) ) \nr
%E = a_x + a_y + a_z  \nr
q_x = 1.0,\quad a_x = -0.455139, \qquad {\rm parity = even,\; period = \pi} \nr
q_y = 0.3, \quad a_y = -0.044566,\qquad {\rm parity = even,\; period = \pi} \nr
q_z = 1.0, \quad a_z = +1.859110,\qquad {\rm parity = even,\; period = 2 \pi}
\nonumber
\eea
The energy eigenvalue is E = 1.359405, $l_{max} = 20$ and the number
of surface points is given by 1730. For the convenience of the
reader the Mathieu functions  are described in ~\ref{a1}. The number of
radial mesh points was equal to 250. Also in this case we employed both
methods of integration with a switch radius of $2.45 \,$ au.

In general the minimum number of surface points is chosen according
to the rule that to integrate exactly $\int d\Omega Y_L(\Omega)$ we
need $\approx (l+1)^2/3$ points.~\cite{lebedev75} If we want to
integrate the product of two Spherical Harmonics, $l$ should be the
sum of the individual $l's$; the same for a product three, etc... So
for the 3-d Numerov we need to integrate only a product of two
functions whose expansions are both truncated to a certain $l_{\rm
max}$, therefore the number of points is $(2l_{\rm max}+1)^2/3$,
whereas for the truncated separable potential and the Mathieu
functions we have the product of three functions (one for each space
coordinate), so we need $(3l_{\rm max}+1)^2/3$ points. Moreover for
the matrix Numerov method we again have $(3l_{\rm max}+1)^2/3$ due
to the calculation of $V_{LL'}$.
\begin{figure}
  \begin{center}
    \begin{tabular}{cc}
      \resizebox{70mm}{!}{\includegraphics{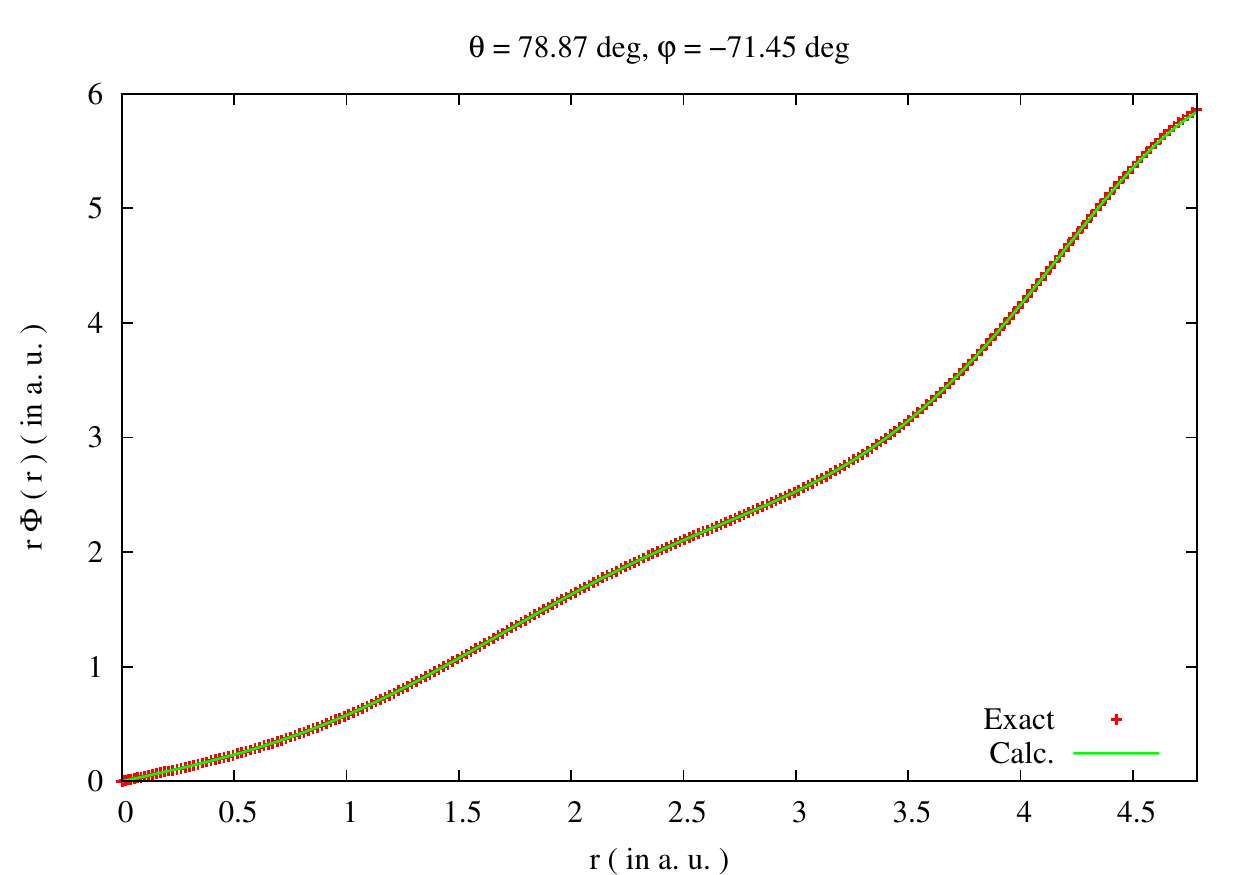}}
      \resizebox{70mm}{!}{\includegraphics{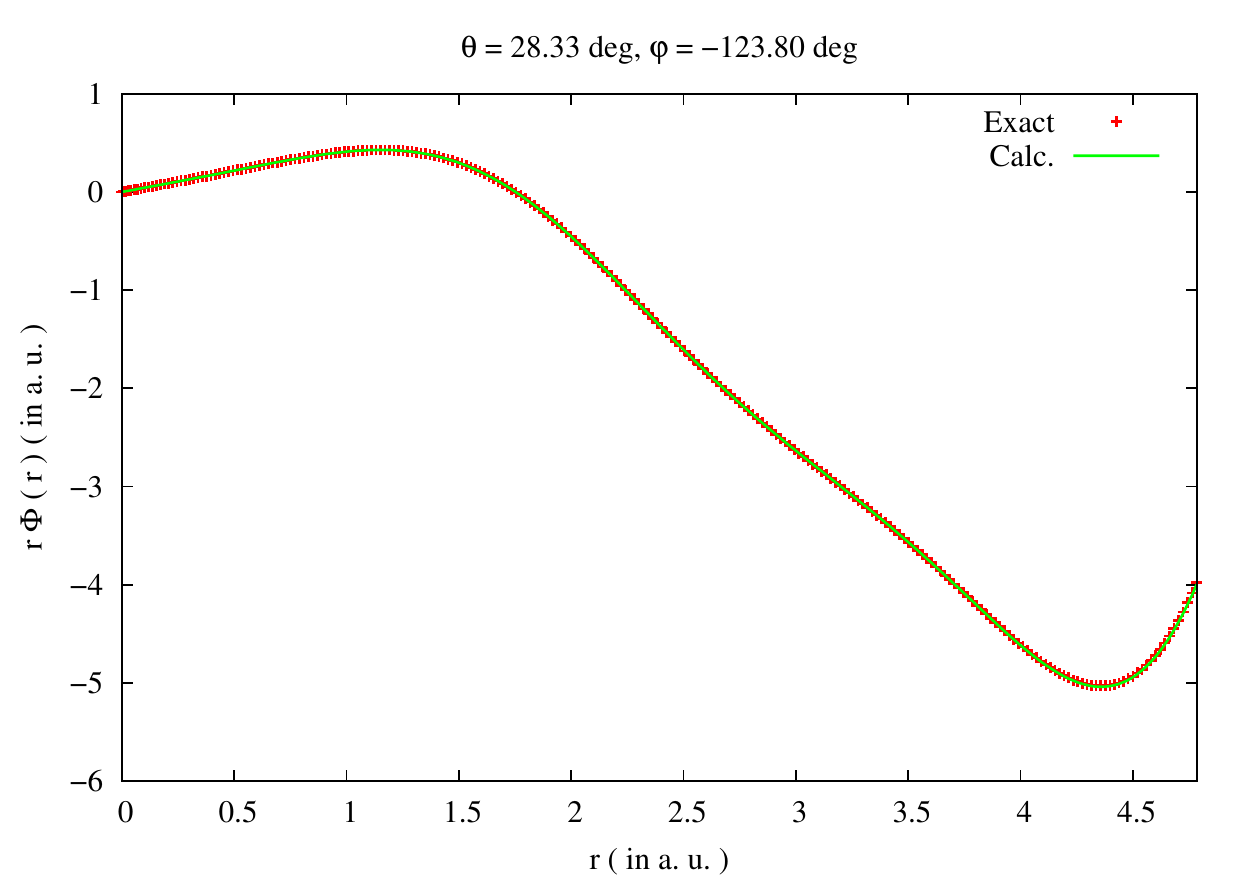}}
    \end{tabular}
    \begin{tabular}{cc}
      \resizebox{70mm}{!}{\includegraphics{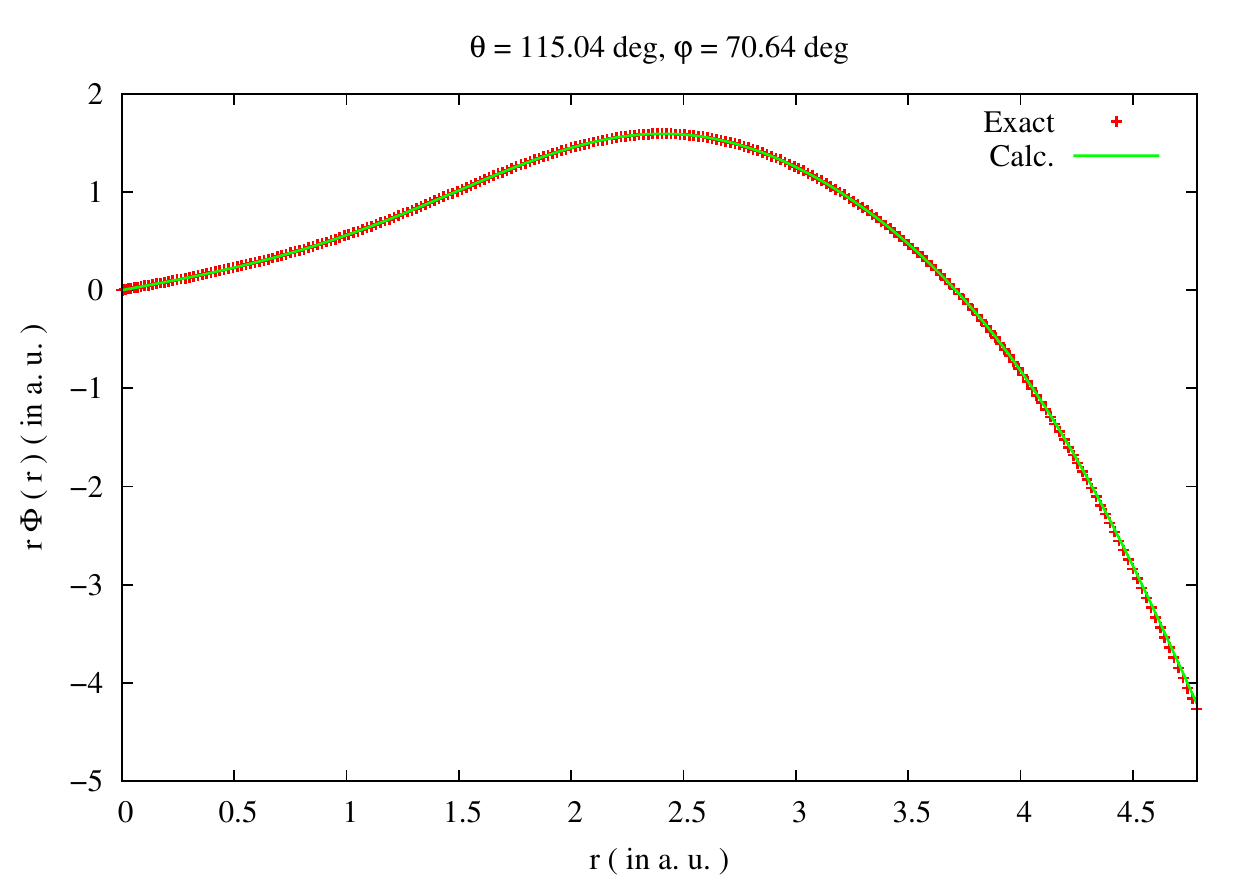}}
      \resizebox{70mm}{!}{\includegraphics{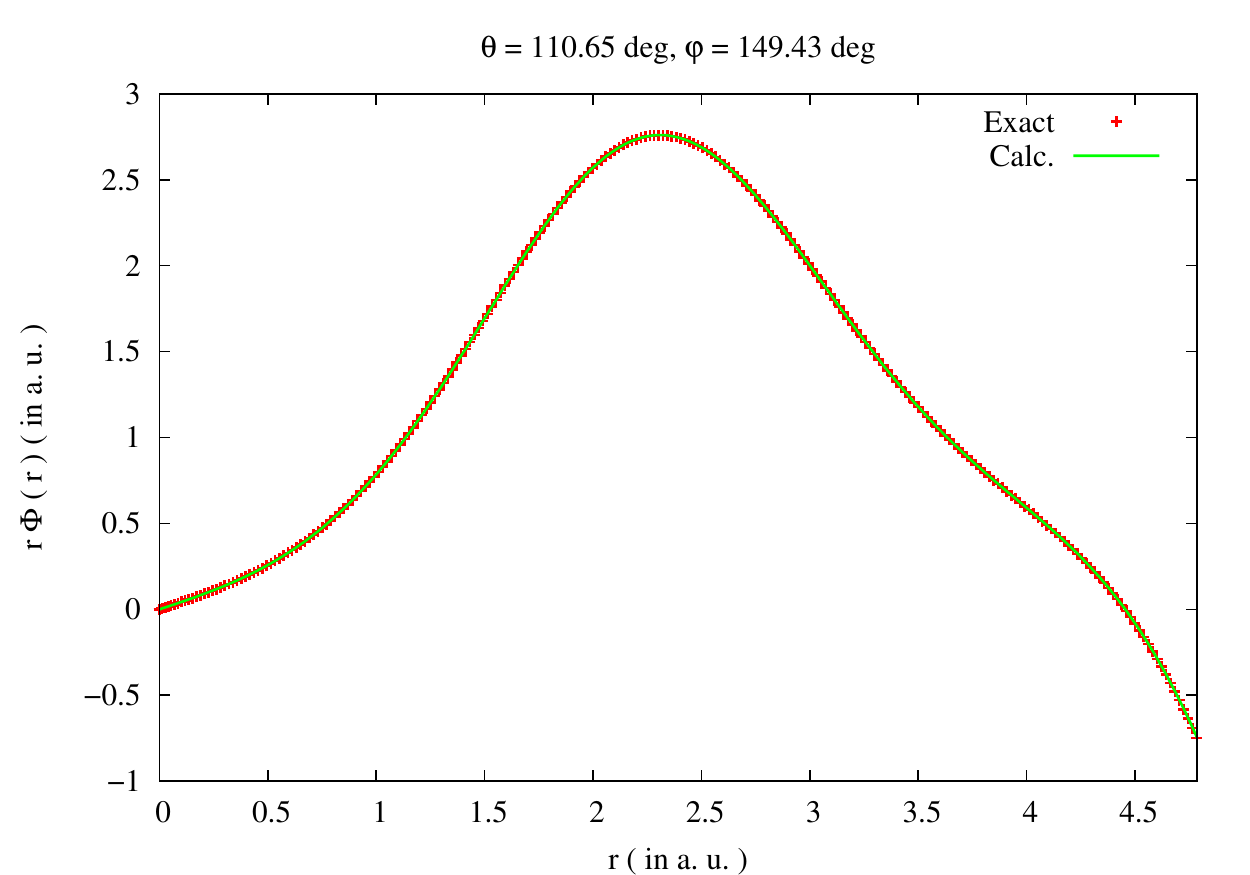}}
    \end{tabular}
  \end{center}
  \caption{Comparison of the particular Mathieu function described in the
text with the one generated by matrix and 3-d Numerov method at
four different angles. The switch radius was at $2.45 \,$ au. (Color online)}
  \label{fig_mathieu}
\end{figure}

\subsection{A linear-logarithmic mesh }

To solve Eq. (\ref{se}) by Numerov procedure, there are several choices
for the radial mesh. Due to the singularity of the potential near the
origin we found that the best strategy in our case was to take a mixed
logarithmic and linear mesh, as usual in atomic physics.
~\cite{brastev66,fischer77,amusia97}
For non-MT calculation, especially with truncated potential, this mesh is
the appropriate choice.
In this case the new radial variable is
\bea
  \rho \, ( \, r \, ) = \alpha \, r + \beta \, { \ln } \, r
  \label{llog}
\eea
with $\alpha$ and $\beta$ constant. A constant mesh size of $\rho$ can be
taken in the interval $\rho_0 \le \rho \le \rho_N$. The initial value of
$\rho_0$ is chosen according to the empirical formula
\be
  \rho_0 = - \beta \, ( \, 10 + \, { \ln } \, Z \, ) \nonumber
\ee
whereas the final $\rho$ is defined as
\be
  \rho_N = \rho_0 + Nh
\ee
so that $\alpha$ is given by
\be
\alpha = (\rho_N - \beta \, {\ln}\, r_N)/r_N
\ee
taking $\beta$, the mesh size $h$ and the number of points $N$ as input
values. In the calculation of local basis functions we choose $r_N$ equal to
the radius of the cell bounding sphere $R_b$, $\beta=1.0$ and put
$N \approx 100R_b$. Instead for Fig.s \ref{fig_3dwell} and \ref{fig_mathieu}
we took respectively $\beta=0.67$ and $\beta=0.05$.
The value of $r = r(\rho)$ corresponding to a given value of $\rho$
can be readily found by application of the Newton technique.~\cite{amusia97}

Following the change of variable in Eq. (\ref{llog}) the SE in Eq. (\ref{matnmv})
becomes ${\bf F } ( \rho ) {\bf Y } ( \rho ) = 0$ where

\bea
  &&( {\bf F } ( \rho ) )_{LL'} \nr
  && = \Bigg[
           \left\{ - \frac{d^2}{d\rho^2}
        + \left( \, \alpha + \frac{\beta }{ r } \right)^{ -2 } \right. \nr
  && \left. \times \left( \frac{ l ( l + 1 ) +
                        \beta \, ( \, \alpha \, r + \beta / 4 \, )
                       ( \, \alpha \, r + \beta \, )^{-2} }{ r^2 }
                  - E \right)
          \right\} \delta_{ L L' }  \nr
  &&     + \left( \, \alpha + \frac{\beta }{ r } \right)^{ -2 }
                V_{ LL' } ( r ) \Bigg] \nonumber\\
  && {\bf Y } ( \rho ) =
    \sqrt{ \alpha + \frac{\beta}{ r} } \, {\bf X } \, ( \, r \, )
\label{ylogm}
\eea
where $r = r(\rho)$. The same, {\it mutatis mutandis}, applies to Eq.
(\ref{se}) for the three-dimensional Numerov method.

A comment is in order at this point. Strictly speaking by changing to the
log-linear mesh the Eq.s (\ref{gauss1}-\ref{gauss4}) are not valid anymore,
since $r = 0$ cannot be realized (it would correspond to $\rho = - \infty)$
and therefore not explicitly implemented as boundary condition.
By working out again the Gaussian elimination process when $X_0$ is not zero,
one arrives at the same equations (\ref{gauss3}), except that in the rhs term
the zero of the i-th row is replaced by $A_{i-1} \, D_{i-1}^{-1} \, A_0 \, Y_0$,
where $Y_0$ is the value of (\ref{ylogm}) calculated at the first point $\rho_0$.
Now
\bea
Y_{LL'}(\rho)=\sqrt{\alpha+{\beta \over r(\rho)}}X_{LL'}(r(\rho)) =
\sqrt{\alpha r(\rho)+\beta} \sqrt{r(\rho)} R_{LL'}(r(\rho))
\eea
Since at the origin $R_{LL'}$ is diagonal in l and behaves like a spherical
Bessel function, $\sqrt{r(\rho)} R_{LL}(r(\rho))$ is of the  order of $10^{-3}$
for l=0 at $\rho_0 \approx 10^{-5}$, $10^{-8}$ for l=1 etc.. We can therefore
take $Y_0 = 0$ and use the simplified Gaussian elimination formulas
Eq.s (\ref{gauss1}-\ref{gauss4}).
We checked that this is a good approximation also for l=0.

\section{ Multiple scattering method for scattering and bound states }
\label{msesbs}

\subsection{Scattering states}
\label{scatstat}

We begin by presenting the derivation of MSE for scattering states. In
this case we seek a solution of the SE continuous in the whole
space with its first derivatives, satisfying the asymptotic boundary condition
\be
\psi({\bf r};{\bf k}) \, \simeq \,
\left( \frac{k}{16\pi^3} \right)^{\frac{1}{2}} \, \left[ {\rm e}^{{\rm i}
{\bf k} \cdot {\bf r}} + f(\hat{\bf r};{\bf k})
 \frac{{\rm e}^{{\rm i} {k r}}}{r} \right]   %\nonumber
\label{asexp0}
\ee
where ${\bf k}$ is the photo-electron wave-vector and $f(\hat{\bf r};{\bf k})$
is the scattering amplitude.
The factor $\left( k / (16\pi^3) \right)^{\frac{1}{2}}$ takes into account
the normalization of the scattering states to one state per Ryd.
In the spirit of MST we partition the space in terms of non overlapping
space-filling cells $\Omega_j$ with surfaces $S_j$ and centers at ${\bf R}_j$.
Accordingly we partition the overall space potential $V({\bf r})$ into cell
potentials, such that $V({\bf r}) = \sum_j v_j({\bf r}_j)$, where
$v_j({\bf r}_j)$ takes the value of $V({\bf r})$ for ${\bf r}$ inside cell
$j$ and vanishes elsewhere. As clear from the following the zero value of
the potential outside the cell is not necessary and can be replaced by any
constant. The results will not depend on this particular value.
Here and in the following ${\bf r}_j = {\bf r} - {\bf R}_j$. The partition
is assumed to satisfy the requirement that the shortest inter-cell vector
${\bf R}_{ij} = {\bf R}_i - {\bf R}_j$ joining the origins of the nearest
neighbors cells $i$ and $j$, is larger than any intra-cell vector ${\bf r}_i$
or ${\bf r}_j$, when ${\bf r}$ is inside cell $i$ or cell $j$. If
necessary, empty cells can be introduced to satisfy this requirement.
We also assume that there exists a finite neighborhood around the origin of
each cell lying in the domain of the cell.~\cite{butler92} We then start
from the following identity involving surface integrals in
${d\hat {\bf r}} \equiv d\sigma$
\bea
\displaystyle{
\sum_{j = 1}^{N} \, \int_{S_{j}} \, \left[
G_{0}^{+}({\bf r}' - {\bf r};{\bf \kappa})
{\bf \nabla} \psi({\bf r};{\bf k}) -
\psi({\bf r};{\bf k}) {\bf \nabla} G_{0}^{+}({\bf r}' - {\bf r};{\bf \kappa})
\right] \cdot {\bf n}_{j} \, {\rm d} \sigma_{j} } && \nr
\displaystyle{= \int_{S_{o}} \, \left[
G_{0}^{+}({\bf r}' - {\bf r};{\bf \kappa})
{\bf \nabla} \psi({\bf r};{\bf k}) -
\psi({\bf r};{\bf k}) {\bf \nabla} G_{0}^{+}({\bf r}' - {\bf r};{\bf \kappa})
\right] \cdot {\bf n}_{o} \, {\rm d} \sigma_{o} \, }. && %\nonumber
\label{one}
\eea
\noindent  Here $\Omega_o = \sum_j \Omega_j$, with surface $S_o$,
centered at the origin $o$ and
$G_{0}^{+}({\bf r}' - {\bf r};{\mathbf \kappa})$ is the free
Green's function with outgoing wave boundary conditions satisfying the
equation
$(\nabla^2 + {\bf \kappa}^2) \, G_{0}^{+}({\bf r}' - {\bf r};{\bf \kappa}) =
\delta({\bf r}' - {\bf r})$, where ${\bf \kappa}^2 = E-V_0$ and
$V_0$ is an arbitrary constant equal to the assumed value of the cell
potential outside the cell domain. The identity (\ref{one})is valid for
all ${\bf r}'$ lying in the neighborhood of the origin of each cell, since in
this case the integrands are continuous with their first derivatives.
In this context we shall use two distinct $k$-vectors, defined respectively
as ${k}= \sqrt{E}$ and ${\kappa}= \sqrt{E-V_0}$. This letter will
appear in the expansion of the Green's function
$G_{0}^{+}({\bf r}' - {\bf r};{\bf \kappa})$ by spherical functions.
~\cite{natoli90} Obviously ${ k}= { \kappa}$ for $V_0 = 0$.

Equation (\ref{one}), with the choice $V_0=0$, can also be derived
from the Lippmann-Schwinger equation
\be
\psi({\bf r}';{\bf k}) = {\rm e}^{{\rm i}{\bf k} \cdot {\bf r}'} +
\int \, G_{0}^{+}({\bf r}' - {\bf r};{ k}) \, V({\bf r}) \,
\psi({\bf r};{\bf k})
\, {\rm d^3r}  %\nonumber
\label{two}
\ee
satisfied by the scattering state
(see  \ref{a3} ). %(see Ref.~[\onlinecite{gonis00}] page 128).
However we prefer to start from the identity Eq. (\ref{one}) to take advantage
of the arbitrariness of the constant $V_0$. For convenience of the reader
we recall the expansions ~\cite{gonis00}
%for $G_{0}^{+}({\bf r}' - {\bf r}; {\bf \kappa}) = - 1/(4\pi) e^{i{\bf \kappa}
%|{\bf r}' - {\bf r}|} / |{\bf r}' - {\bf r}| $
\bea
{\rm e}^{{\rm i}{\bf k} \cdot {\bf r}} &=& 4\pi \sum_{L} i^l
Y_L (\hat {\bf k})  J_L({ \bf r }; k) \label{plwexp}  \\
G_{0}^{+}({\bf r}' - {\bf r}; {\bf \kappa}) &=& - \frac{1}{4\pi} \,
\frac{e^{i{\bf \kappa}
|{\bf r}' - {\bf r}|}} {|{\bf r}' - {\bf r}|} \nr
&=&  G_{0}^{+}({\bf r}_i' - {\bf r}_i; {\bf \kappa}) \nr
&=& \sum_L J_L( {\bf r}_{i}'; {\bf \kappa} )
\tilde{H}_L^{ + } ( { \bf r }_i; {\bf \kappa}) \; (r_i' < r_i)
\label{gfexp1} \\
 &=& \sum_L J_L( {\bf r}_{i}; {\bf \kappa} )
\tilde{H}_L^{ + } ( { \bf r }_i'; {\bf \kappa}) \; (r_i' > r_i)
\label{gfexp2}
\eea

Notice for future reference that in the case ${\bf \kappa} = 0$ the solid
spherical harmonics $J_L( {\bf r}; {\bf \kappa})$ and
$\tilde{H}_L^{ + } ( { \bf r }; {\bf \kappa})$ are to be understood as
$\overline{J}_L ({ \bf r }) = r^l Y_L (\hat {\bf r}) / (2l+1)$ and
$\overline{H}_L ({ \bf r }) = r^{-l-1} Y_L (\hat {\bf r}) $, due to the well
known expansion
\be
\frac {1}{|{\bf r} - {\bf r}'|} = \sum_L \frac{4 \pi}{2l+1} \,
\frac {r_<^l}{r_>^{l+1}} \, Y_L (\hat {\bf r}) \, Y_L (\hat {\bf r}')
\label{gfexp0}
\ee
which is the $\kappa \rightarrow 0$ limit of Eq.s (\ref{gfexp1}) and (\ref{gfexp2}).

The heart of MST is the introduction of the functions
$\Phi_{ L } ({ \bf r }_j; k)$ which inside cell $j$ are local solutions
of the SE with potential $v_j({\bf r}_j)$ behaving as $J_{ L }({\bf r}_j; k)$
for $r_j \rightarrow 0$. They form a complete set of basis functions such
that the global scattering wave function can be locally expanded as
~\cite{butler92}
\be
\psi ( { \bf r }_j;{\bf k} ) =
\sum_{ L } A_{ L }^j ({\bf k}) \Phi_{ L } ( { \bf r }_j; k )
\label{wfexpj}
\ee
where we have underlined the $k$ dependence of $\Phi_{ L } ({ \bf r }_j; k)$
through its behavior at the origin.

In order to find the asymptotic behavior in the outer region
$\mathcal{C} \Omega_o$ we introduce the scattering functions in
response to an exciting wave of angular momentum $L$:
\be
\psi_L ({\bf r}_o;{ k}) = J_L ({\bf r}_o; k) +
\int \, G_{0}^{+}({\bf r}_o - {\bf r}_o';{ k}) \, V({\bf r}_o') \,
\psi_L ({\bf r}_o';{ k}) \,{\rm d^3r_o'}
\label{twol}
\ee
Then, under the assumption of short range potentials ({\it i.e.} potentials
that behave like $1/r^{1+\epsilon}$ with positive $\epsilon$ at great
distances), letting ${\bf r}_{o} \rightarrow \infty $ and using expansion
Eq. (\ref{gfexp2}) in Eq. (\ref{twol}) we find
\bea
\psi ( { \bf r }_o;{\bf k} ) & = &
\sum_{ L } \tilde{A}_{ L }^o ({\bf k}) \Bigg[ J_L({ \bf r }_o; k)  \nr
 &&  + \sum_{L'}
 \tilde{H}_{L'}^{ + } ( { \bf r }_o; k) \int J_{L'} ({ \bf r }_o'; k)
\, V({\bf r}_o') \, \psi_L ({\bf r}_o';{ k}) \,{\rm d^3r_o'} \Bigg] \\ \nr
 & = & \sum_{ L } \tilde{A}_{ L }^o ({\bf k}) \Bigg[ J_L({ \bf r }_o; k)
+ \sum_{L'} \tilde{H}_{L'}^{ + } ( { \bf r }_o; k) T_{L'L}^o \Bigg]
\label{wfexpo}
\eea
where, in order to impose the asymptotic behavior in Eq. (\ref{asexp0}),
$ \tilde{A}_{ L }^o = i^l Y_L(\hat {\bf k})\, ({k/\pi})^{1/2}$ and
$T_{LL'}^o$ is the $T$-matrix for the whole cluster, equal to
\be
T_{L'L}^o = \int J_{L'} ({ \bf r }_o; k)
\, V({\bf r}_o) \, \psi_L ({\bf r}_o;{ k}) \,{\rm d^3r_o}
\label{tomdef}
\ee
In general for short range potentials decaying slowly,
the asymptotic behavior in Eq. (\ref{wfexpo}) is reached only at great distance
from the origin of the coordinates (usually at the center of the atomic
cluster under study). In order to limit the number of cells, so that the
surface $S_o$ just surrounds the cluster, we introduce the local solution
\be
\Phi_L ({ \bf r }_o; k) = \sum_{ L' } \, R_{ L'L}^o  ( r_o ) \,
Y_{ L' } ( \hat {\bf r}_o )
\label{phio}
\ee
in the outer region $\mathcal{C} \Omega_o$, which can be obtained by inward
integration of the SE starting from the appropriate asymptotic value
$\tilde{H}_{L'}^{ + } ( { \bf r }_o; k)$.
Therefore we take here
\be
\psi ( { \bf r }_o;{\bf k} ) = \sum_{ L } \left[ \tilde{A}_{ L }^o ({\bf k})
 J_L({ \bf r }_o; k) + \Phi_{L} ( { \bf r }_o; k) A_{L}^o ({\bf k}) \right]
\label{wfexpo1}
\ee
Notice that the function $\Phi_L ({ \bf r }_o; k)$ in Eq. (\ref{phio})
(and consequently $R_{L'L}^o ( r_o )$ ) is complex, unlike
the functions $\Phi_L ({ \bf r }_i; k )$ that can be taken real, if the
potential is real. If the potential has a Coulomb tail,
the spherical Bessel and Hankel functions should be replaced by the
corresponding regular and irregular solutions $F_L({ \bf r }_o; k)$ and
$G_L({ \bf r }_o; k)$ of the radial SE with a Coulomb potential.
Due to the possibility that the optical potential used for calculating
the spectroscopic response functions be complex, it should be clear from
the context that the formalism works also for complex energies and/or
potentials. The extension to complex energies will come very handy when
exploiting the analytic properties of the Green function.

Insertion of the expressions Eq.s (\ref{wfexpj}) and (\ref{wfexpo1})
into the identity Eq. (\ref{one}) provides a set of algebraic equations
(known as MSE) that determine the expansion coefficients
$A^{j}_{ L } ({\bf k})$ and the $A_{L}^o ({\bf k})$ in such a
way that the local representations are smoothly continuous
across the common boundary of contiguous cells. Indeed, taking
${\bf r}'$ in the neighborhood of the origin of cell $i\neq o$, using
the expansion Eq. (\ref{gfexp1})
(since ${\bf r}$ is confined to lie on the cell surfaces), and putting to
zero the coefficients of $J_L( {\bf r}_{i}'; {\bf \kappa})$ due to their
linear independence, we readily arrive at the MST compatibility equations
for the amplitudes $A^{j}_{ L } ({\bf k})$ and $A_{L'}^o ({\bf k})$
\be
\sum_{jL'} H_{L L'}^{i j} A^{j}_{L'}({\bf k}) =
\sum_{L'} \left[ M_{L L'}^{io} \tilde{A}_{L'}^o ({\bf k})  + N_{L L'}^{i o}
A_{L'}^o ({\bf k}) \right]
\label{si}
\ee
\noindent where
$$
H_{ LL' }^{i j}  =  \int_{ S_j } [ \, \tilde{H}_L^{ + } ( { \bf r }_i;\kappa )
     \nabla \Phi_{ L' } ( { \bf r }_j; k ) - \Phi_{ L' } ( { \bf r }_j; k )
     \nabla \tilde{H}_L^{ + } ( { \bf r }_i;\kappa ) \, ] \cdot { \bf n }_j \,
     {\rm d}  \sigma_j
$$
$$
M_{ L L'}^{i o}  =  \int_{ S_o } [ \, \tilde{H}_L^+( { \bf r }_i;\kappa )
     \nabla J_{L'} ( { \bf r }_o; k ) - J_{L'} ( { \bf r }_o; k )
     \nabla \tilde{H}_L^+( { \bf r }_i; \kappa ) \, ] \cdot { \bf n }_o \,
     {\rm d}  \sigma_o
$$
$$
N_{L L'}^{i o}  =  \int_{ S_o } [ \, \tilde{H}_L^+( { \bf r }_i;\kappa )
     \nabla \Phi_{L'} ( { \bf r }_o; k ) -
     \Phi_{L'} ( { \bf r }_o; k )
     \nabla \tilde{H}_L^+( { \bf r }_i; \kappa ) \, ] \cdot { \bf n }_o \,
     {\rm d}  \sigma_o
$$
%
%\noindent and the rhs term comes from the outer sphere $\Omega_o$.
%(see Appendix A of Ref. ~\cite{sebilleau06} for details).

A further set equation is obtained by taking ${\bf r}'$ inside the outer
region $\mathcal{C} \Omega_o$, using the expansion Eq. (\ref{gfexp2})
(remembering that ${\bf r}_o < {\bf r}_o'$, since ${\bf r}_o$ lies on
$S_o$). By putting to zero the coefficients of
$\tilde{H}_L^{ + } ( { \bf r }_o'; {\bf \kappa})$ we obtain
\be
\sum_{jL'} K_{L L'}^{o j} A^{j}_{L'}({\bf k}) =
\sum_{L'} \left[  \tilde{M}_{L L'}^{o o} \tilde{A}_{L'}^o ({\bf k}) +
\tilde{N}_{L L'}^{o o} A_{L'}^o ({\bf k}) \right]
\label{so}
\ee
\noindent where
$$
K_{ LL' }^{o j}  =  \int_{ S_j } [ \, J_L ( { \bf r }_o;\kappa )
     \nabla \Phi_{ L' } ( { \bf r }_j; k ) - \Phi_{ L' } ( { \bf r }_j; k )
     \nabla J_L ( { \bf r }_o;\kappa ) \, ] \cdot { \bf n }_j \,
     {\rm d}  \sigma_j
$$
$$
\tilde{M}_{ L L'}^{o o}  =  \delta_{L L'}
    \int_{ S_o } [ \,J_L( { \bf r }_o;\kappa )
     \nabla J_{L'} ( { \bf r }_o; k ) - J_{L'} ( { \bf r }_o; k )
     \nabla J_L( { \bf r }_o; \kappa ) \, ] \cdot { \bf n }_o \,
     {\rm d}  \sigma_o
$$
$$
\tilde{N}_{L L'}^{o o}  =
    \int_{ S_o } [ \, J_L( { \bf r }_o;\kappa )
     \nabla \Phi_{ L' } ( { \bf r }_o; k ) -
    \Phi_{ L' }  ( { \bf r }_o; k )
     \nabla J_L( { \bf r }_o; \kappa ) \, ] \cdot { \bf n }_o \,
     {\rm d}  \sigma_o
$$

From the above derivation it is clear that the set of equations in Eq.s (\ref{si})
and (\ref{so}) determines the amplitudes $A^{j}_{ L } ({\bf k})$
and $A_{L}^o ({\bf k})$
independently of the constant $V_0$, since the identity Eq. (\ref{one}) is valid
whatever $V_0$. In general this will be true only if the $L$-expansion is not
truncated, whereas there will be a more or less pronounced dependence
according to the degree of convergence of the truncated expansion. In
general, the lesser the potential jump at the boundaries of the various cells
the faster the convergence.

Notice that these equations remain valid, with no restriction on the
sums over $L$, even in the case $\kappa = 0$, provided $J_{L}$ and $\tilde{H}_L$
are replaced by $\overline{J}_L$ and $\overline{H}_L$, due to the expansion
Eq. (\ref{gfexp0}) of the zero energy limit of the free Green's Function.

The usual derivation of the MSE now proceeds by re-expanding
$\tilde{H}_L^{ + } ( { \bf r }_i; \kappa  )$ and $J_L ( { \bf r }_o ;\kappa )$
around center $j$ under the geometrical conditions stated at the beginning of
this section, by use of the equations ~\cite{gonis00,natoli90}
\be
\tilde{H}_L^{ + } ( { \bf r }_i; \kappa  ) =
\sum_{L'} G^{i j}_{L L'} J_{L'} ( { \bf r }_j; \kappa  ) \; (R_{ij} > r_j)
\label{reexp1}
\ee
\be
J_L ( { \bf r }_o; \kappa  ) =
\sum_{L'} J^{o j}_{L L'} J_{L'} ( { \bf r }_j; \kappa  ) \; (\rm no \; cond.)
\label{reexp2}
\ee
\be
\tilde{H}_L^{ + } ( { \bf r }_i; \kappa  ) =
\sum_{L'} J^{i o}_{L L'} \tilde{H}_{L'}^{ + } ( { \bf r }_o; \kappa  )
\; (r_o > R_{io})
\label{reexp3}
\ee
where $G^{i j}_{L L'}$ are the free electron propagator in the site and
angular momentum basis ( KKR real space structure factors) given by
\be
G^{ij}_{LL'} = 4\pi \sum_{L''} C(L,L';L'')\, i^{l-l'+l''} \,
\tilde{H}^+_{L''}({\bf R}_{ij}; \kappa )
\label{gdef}
\ee
and $J^{i j}_{L L'}$ is the translation operator
\be
J^{ij}_{LL'} = 4\pi  \sum_{L''} C(L,L';L'')\, i^{l-l'+l''} \,
J_{L''}({\bf R}_{ij}; \kappa )
\label{jdef}
\ee
In these formulas the quantities $C(L,L';L'')$ are the real basis Gaunt
coefficients given by
\be
C(L,L';L'') = \int Y_L(\Omega) Y_{L'}(\Omega) Y_{L''}(\Omega) {\rm d}\Omega
\label{cdef}
\ee
In the following we shall also need the quantity
\be
N^{ij}_{LL'} = 4\pi \sum_{L''} C(L,L';L'')\, i^{l-l'+l''} \,
N_{L''}({\bf R}_{ij};\kappa )
\label{ndef}
\ee

Unfortunately the re-expansions Eq.s (\ref{reexp1}), (\ref{reexp2}) and
(\ref{reexp3})
introduce further expansion parameters into the theory (with related
convergence problems) that are actually unnecessary, as shown below.

We in fact observe that the integrals over the surfaces of the various
cells $j$ can be calculated over the surfaces of the corresponding bounding
spheres (with radius $R_b^j$) by application of the Green's theorem, since
both $\tilde{H}_{ L }^+ ({\bf r; \kappa})$ and $\Phi_{ L } ( { \bf r; k })$
satisfy
the Helmholtz equation $(\nabla^2 + {\bf \kappa}^2) \, F( { \bf r }) = 0 $
outside the domain of the cell. We then use the following relations
\be
\int_{ S_j } \, Y_{L'} ( \hat {\bf r}_j ) \, \tilde{H}_L^{ + }
( { \bf r }_i; \kappa )   \, {\rm d}  \sigma_j =
(R_b^j)^2  \, G^{i j}_{L L'} \,  j_{l'} (\kappa R_b^j )
\label{main1}
\ee
\be
\int_{ S_j } \, Y_{L'} ( \hat {\bf r}_j ) \nabla \, \tilde{H}_L^{ + }
( { \bf r }_i )  \cdot { \bf n }_j \, {\rm d}  \sigma_j =
(R_b^j)^2 \, G^{i j}_{L L'} \, \frac{\rm d} {{\rm d} R_b^j} \,
j_{l'} (\kappa R_b^j)
\label{main2}
\ee
 which are exact for all $L$ provided
$|{\bf r}_i - {\bf r}_j| = R_{ij} > r_j $ for ${\bf r}$ lying on the surface
$S_j$. This is a consequence of the fact that under this condition the series
%$\tilde{H}_L^{ + } ( { \bf r }_i ) = \sum_{L'} G^{i j}_{L L'}
%J_{L'} ( { \bf r }_j )$
in Eq. (\ref{reexp1}) converges absolutely and uniformly in the entire angular
domain, as shown in  \ref{a2}, Eq. (\ref{gjineq}) and can therefore
be integrated term by term. This property is also
true for the series derived with respect to ${\bf r}$. Even though not necessary,
we also checked the numerical equality of both sides of Eq.s (\ref{main1})
and (\ref{main2}) for various values of $L, L'$.

Similarly, since the series in Eq. (\ref{reexp3}) converges uniformly and
absolutely, as shown in  \ref{a2}, Eq. (\ref{jhineq}), we also find
\be
\int_{ S_o } \, Y_{L'} ( \hat {\bf r}_o ) \, \tilde{H}_L^{ + }
( { \bf r }_i; \kappa ) \, {\rm d}  \sigma_j = (R_b^o)^2  \, J^{i o}_{L L'} \,
\tilde{h}_{l'}^+ (\kappa R_b^o )
\label{main3}
\ee
\be
\int_{ S_o } \, Y_{L'} ( \hat {\bf r}_o ) \nabla \, \tilde{H}_L^{ + }
( { \bf r }_i )  \cdot { \bf n }_j \, {\rm d}  \sigma_j =
(R_b^o)^2 \, J^{i o}_{L L'} \, \frac{\rm d} {{\rm d} R_b^o} \,
\tilde{h}_{l'}^+  (\kappa R_b^o )
\label{main4}
\ee
provided $R_{io} < R_b^o$, where $R_b^o$ is the bounding sphere of the
outer region $\mathcal{C} \Omega_o$. Therefore $R_b^o$ should be bigger
than any $R_{io}$.

Finally, due to the absolute and uniform convergence of the series in Eq.
(\ref{reexp2}) without conditions, we find the following relations
\be
\int_{ S_j } \, Y_{L'} ( \hat {\bf r}_j ) \, J_L ( { \bf r }_i; \kappa )
\, {\rm d}  \sigma_j = (R_b^j)^2  \, J^{i j}_{L L'} \,  j_{l'} (\kappa R_b^j )
\label{main5}
\ee
\be
\int_{ S_j } \, Y_{L'} ( \hat {\bf r}_j ) \nabla \, J_L ( { \bf r }_i )
\cdot { \bf n }_j \, {\rm d}  \sigma_j =
(R_b^j)^2 \, J^{i j}_{L L'} \,  \frac{\rm d} {{\rm d} R_b^j} \,
j_{l'} (\kappa R_b^j)
\label{main6}
\ee

By inserting in Eq. (\ref{si}) the expression for the basis functions expanded
in spherical harmonics (we shall suppress the site indices whenever a relation
refers to both sites $i$ and site $o$)
%for the site $j$ and $o$ (see (\ref{phio}))
\be
%\Phi_{ L } ( { \bf r }_j ) = \sum_{ L' } R_{ L'L}^j ( r_j )
%Y_{ L' } ( \hat {\bf r}_j )
\Phi_{ L } ( { \bf r }; k ) = \sum_{ L' } R_{ L'L} ( r )
Y_{ L' } ( \hat {\bf r} )
\label{phiexp}
\ee
remembering that this expansion is uniformly convergent in the angular domain
~\cite{kellog54} and using the relations Eq.s (\ref{main1})-(\ref{main6}) we
finally obtain, under the partitioning conditions specified at the beginning of
Section~\ref{msesbs},
\bea
&&\sum_{L'} E_{L L'}^{i} A^{i}_{L'}({\bf k})\, + \sum_{j, L', L''} ^{j \neq i}
G^{i j}_{L L''} S^{j}_{L'' L'} A^{j}_{L'}({\bf k}) \nr
&&=   \sum_{L'} J^{i o}_{L L'} \left[ M_{L' L'}^{o o} \tilde{A}_{L'}^o ({\bf k}) +
\sum_{L''} E_{L' L''}^{o} A_{L''}^o ({\bf k}) \right]
\label{ceq1}
\eea
\noindent where we have put $E_{L' L''}^{o} \equiv N_{L'
L''}^{o o}$, the quantities $M_{L L}^{o o}$ and
$N_{L' L''}^{o o}$ being the same as those following Eq.(\ref{si}),
calculated with  ${ \bf r }_i$ replaced by ${ \bf r }_o$.

Similarly, putting $S_{L L'}^{o} \equiv \tilde{N}_{L L'}^{o o}$, for Eq.
(\ref{so}) we find
\be
\sum_{j, L', L''} ^{j \neq o}J^{o j}_{L L''} S^{j}_{L'' L'}
A^{j}_{L'}({\bf k}) =
\sum_{L' } \left[ \tilde{M}_{L L'}^{o o} \tilde{A}_{L'}^o ({\bf k})
\delta_{L L'} + S_{L L'}^{o} A_{L'}^o ({\bf k}) \right]
\label{ceq2}
\ee
In the above equations we have defined the quantities
\bea
%E_{LL'}^j &=& ({R_b^j})^2 W[-i\kappa h_l^+,R_{ L L' }^j] \\
%S_{LL'}^j &=& ({R_b^j})^2 W[j_l,R_{ L L'}^j ]
E_{LL'} &=& ({R_b})^2 W[-i\kappa h_l^+,R_{ L L' }] \\
S_{LL'} &=& ({R_b})^2 W[j_l,R_{ L L'} ]
\eea
for the cells $\Omega_j$ and for the outer region $\mathcal{C} \Omega_o$.
%\bea
%E_{LL'}^o &=& ({R_b^o})^2 W[j_l,R_{ L L' }^o] \\
%S_{LL'}^o &=& ({R_b^o})^2 W[-i\kappa h_l^+,R_{ L L'}^o ]
%\eea
%for the outer region $\Omega_o$.
The Wronskians $W[f,g] = fg'-gf'$ are
calculated at $R_b^j$ and $R_b^o$  respectively and reduce
to diagonal matrices for MT potentials.

Equations (\ref{ceq1}) and (\ref{ceq2}) look formally similar to the usual
MSE. However we notice that due to the relations Eq.s (\ref{main1})-(\ref{main6})
there are only two expansion parameters in the theory. They are related to
the AM components of $R_{L'L}$ in the expansion Eq. (\ref{phiexp}) in cell $j$
and in the outer region $\mathcal{C} \Omega_o$. No convergence constraints
related to the re-expansion of the various spherical Bessel and Hankel functions around
a different origin Eq.s (\ref{reexp1})-(\ref{reexp3}) are present.

It is interesting to note that the truncation value for both indices is
the same and corresponds to the classical relation $l_{\rm max} = kR_b^j$,
where $R_b^j$ is the radius of the bounding sphere of the cell at site $j$.
This is true for the index $L$, which reminds that the basis function $\Phi_L$
is normalized like $j_l(kr)Y_L$ near the origin. Due to the properties of the spherical
Bessel functions, when $l \gg k R_b^j$, $\Phi_L$ becomes very small inside
the cell, decreasing like $[(2l+1)!!]^{-1}$. Therefore his weight in the
expansion Eq. (\ref{phiexp}) will be negligible. The other index $L'$, as will
be clear from the following, measures the response of the truncated potential
inside the cell to an incident wave $J_{L'}$ of angular momentum $L'$. Due to
the same argument as above, familiar to scattering theory, the scattering
matrix $T_{L'L}^j$ will decrease like $[(2l+1)!! (2l'+1)!! ]^{-1}$ (see Eq.
(\ref{tmat1}) in  \ref{a2} for $l, l' \gg k R_b^j$). As a consequence
$E^j$ and $S^j$ can be considered square matrices. In the case of the outer
sphere region $\mathcal{C} {\Omega_o}$, the situation is inverted, the index
$L$ being related to the response of the entire cluster to an incident wave
of angular momentum $L$, whereas the index $L'$ corresponds to the number of
AM waves mixed in by the potential not only inside $\Omega_o$ but also in
$\mathcal{C} \Omega_o$. The two indices have the same truncation
$l_{\rm max} = k \tilde{R}_b^o$, provided we take $\tilde{R}_b^o$ as the
radius of the sphere that contains the region of space where the potential
is substantially different from zero. This conclusion is reinforced by the
observation that one can cover this same region by empty cells.

Up to this point we have assumed that $V_0 \neq 0$ and derived consequently
the MSE, having in mind the possibility to check the rate of convergence of
the $L$-expansion. However in the continuum case one usually works under the
assumption that $V_0 = 0$.
In this case the Eq.s (\ref{ceq1}) and (\ref{ceq2}) simplify considerably in
the case of short range potentials. Since now $k = \kappa$, we use the
relation
\be
    \int_{ S_o } [ \,  \tilde{H}_{L'}^+ ( { \bf r }_o; k )
     \nabla J_L( { \bf r }_o; k ) - J_L( { \bf r }_o; k )
     \nabla \tilde{H}_{L'}^+ ( { \bf r }_o; k ) \, ] \cdot { \bf n }_j \,
     {\rm d}  \sigma_o =  - \delta_{L L'}
\ee
so that in Eq. (\ref{ceq1}) $ M_{L L}^{o o} = -1$,
and in Eq. (\ref{ceq2}) $\tilde{M}_{L L}^{o o} = 0$.
Moreover one easily finds that
\be
\sum_{L' } \tilde{A}_{L'}^o ({\bf k})  J^{i o}_{L L'} =
i^l Y_L ({\bf k}) \, e^{i {\bf k} \cdot {\bf R}_{io}} \sqrt{\frac{k}{\pi}}
= I_L^i ({\bf k})
\label{ilk}
\ee
which is obtained from Eq. (\ref{jdef}) by observing that
$$
\sum_{L'}C(L,L';L'') Y_{L'}(\Omega) = Y_{L}(\Omega) Y_{L''}(\Omega)
$$
Then the two sets of equations assume the simpler form
\bea
\sum_{L'} E_{L L'}^{i} A^{i}_{L'}({\bf k})\, && + \sum_{j, L', L''} ^{j \neq i}
G^{i j}_{L L''} S^{j}_{L'' L'} A^{j}_{L'}({\bf k}) \nr
&&- \sum_{L' L''} J^{i o}_{L L'} E_{L' L''}^{o} A_{L''}^o ({\bf k})
=  - I^{i}_{L}({\bf k})
\label{sceq1}
\eea
\be
\sum_{j, L', L''} ^{j \neq o}J^{o j}_{L L''} S^{j}_{L'' L'}
A^{j}_{L'}({\bf k}) -  \sum_{L' } S_{L L'}^o A_{L'}^o ({\bf k}) = 0
\label{sceq2}
\ee
The fact that $E$ and $S$ can be taken to be square
matrices leads to another interesting form of the MSE. Under the assumption
that ${\rm Det} \, S \neq 0$, we can introduce new amplitudes
\be
%B^{j}_{L}({\bf k}) = \sum_{L'} S^{j}_{L L'} A^{j}_{L'}({\bf k})
B_{L}({\bf k}) = \sum_{L'} S_{L L'} A_{L'}({\bf k})
%\label{bdef}
\ee
which is equivalent to using new basis functions $\overline{\Phi}_L$ related
to $\Phi_L$ by the relation
\be
\overline{\Phi}_L = \sum_{L'} (\tilde{S}^{-1})_{LL'} \Phi_{L'}
\label{phsbf}
\ee
where $\tilde{S}$ is the transposed of the matrix $S$.

Defining the quantities
\bea
(T^i)^{-1} = - E^i (S^i)^{-1} \\
\overline{T}^o = - E^o (S^o)^{-1}
\label{tdef}
\eea
(notice the asymmetry between sites $i$ and site $o$)
we can write Eqs. (\ref{sceq1}) and (\ref{sceq2}) as
\bea
\sum_{L'} (T^{i})_{L L'}^{-1} B^{i}_{L'}({\bf k})\, && - \sum_{j, L'}
^{j \neq i} G^{i j}_{L L'} B^{j}_{L'}({\bf k}) \nr
&& - \sum_{L' L''} J^{i o}_{L L'} \overline{T}^{o}_{L' L''}
B_{L''}^o ({\bf k}) = I^{i}_{L}({\bf k})
\label{bei}
\eea
\be
\sum_{j, L'} ^{j \neq o} J^{o j}_{L L'} B^{j}_{L'}({\bf k}) -
B^{o}_{L}({\bf k}) = 0
\label{beo}
\ee
The meaning of the amplitudes $B_{L}({\bf k})$ is immediately found from
these equations if we consider only a single truncated potential at center
$i$. In this case $\overline{T}^{o} \equiv 0$, since now the asymptotic
behavior is given by Eq. (\ref{wfexpo}), and
$B^{o}_{L}({\bf k}) \equiv A_{L}^o ({\bf k}) =
\sum_{L'} T_{L L'}^o \tilde{A}_{L'}^o$
where $T_{L L'}^o$ is the $T$-matrix of the potential. Therefore Eqs.
(\ref{bei}) and (\ref{beo}) tell us that $T^{i}_{L L'} \equiv T_{L L'}^o$.
As a consequence $B^{i}_{L}({\bf k})$ is the scattering amplitude of
angular momentum $L$ in response to an exciting plane wave of wave vector
${\bf k}$. Moreover, we find that $T^i = - S^i (E^i)^{-1}$ is symmetric
in the AM indices (remember that we use a real spherical harmonics basis),
a fact already known from general scattering theory.
This is a consequence of the fact that $S E^{-1}$ is a symmetric
matrix.~\cite{foulis88}
%%%

In the case of many cells, it is expedient to work only in terms of the cell
amplitudes $B^{i}_{L'}({\bf k})$. Inserting into Eq. (\ref{bei}) the expression
for $B^{o}_{L'}({\bf k})$ given by Eq. (\ref{beo}) we obtain
\bea
\sum_{L'} (T^{i})_{L L'}^{-1} B^{i}_{L'}({\bf k})\, && - \sum_{j, L'}
^{j \neq i} G^{i j}_{L L'} B^{j}_{L'}({\bf k}) \nr
&& - \sum_{j L'} \sum_{\Lambda \Lambda'}J^{i o}_{L \Lambda}
\overline{T}^{o}_{\Lambda \Lambda'}
 J^{o j}_{\Lambda' L'} B^{j}_{L'}({\bf k}) = I^{i}_{L}({\bf k})
\label{bei1}
\eea
Introducing $\tau$, the inverse of the multiple scattering matrix
$M \equiv T^{-1} - G - J\overline{T}^{o}J$
\be
\tau = (T^{-1} - G - J\overline{T}^{o}J)^{-1}
\label{spotau}
\ee
known as the scattering path operator~\cite{gonis00}, we derive from
Eq. (\ref{bei1}) that
\be
B^{i}_{L}({\bf k}) = \sum_{j L'} \tau_{L L'}^{i j} I^{j}_{L'}({\bf k})
\label{bsca}
\ee
If we insert this expression in Eq. (\ref{beo}) and remember that by
definition
$B^{o}_{L}({\bf k}) = \sum_{L'} T_{L L'}^o \tilde{A}_{L'}^o$,
we easily find for the cluster $T$-matrix
\be
T_{L L'}^o = \sum_{ij} \sum_{\Lambda \Lambda'} J^{o i}_{L \Lambda}
\tau_{\Lambda \Lambda'}^{ij} J^{j o}_{\Lambda' L'}
\label{tmo}
\ee
Since the matrices $G$ and $J$ are also symmetric (see definitions
Eq.s (\ref{gdef}) and (\ref{jdef})), we find that $\tau$ is likewise symmetric,
implying the symmetry of $T_{L' L}^o$, again in keeping with scattering
theory. This quantity represents indeed for the whole cluster the scattering
amplitude into a spherical wave of angular momentum $L$ in response to
an exciting wave of AM $L'$ and is needed for example in electron
molecular scattering.~\cite{natoli86} Finally Eq. (\ref{bsca}) shows that
the quantities $ B^{i}_{L}({\bf k})$ are scattering amplitudes for the
cluster, for which the generalized optical theorem holds (for real potentials)
~\cite{natoli86,natoli90} (see  \ref{a4})
\be
\int {\rm d} \hat{\bf k}\,  B^{i}_{L}({\bf k})  \, \left[ B^{j}_{L'}({\bf k})
\right]^{\ast} = -\frac{1}{\pi} \, \Im \, \tau_{L L'}^{i j}
\label{got}
\ee
This relation is very important, since it establishes the connection
between the photo-emission and the photo-absorption cross section, as
shown in  \ref{a5}. As it will turn out,
$-\Im \, \tau_{L L}^{i i}$ is proportional to the $L$-projected density
of states onto site $i$.

In the case of one single cell located at site $i$, by construction
the solutions inside and outside the cell are continuously smooth so that,
remembering that by definition $T^{i}_{L L'} \equiv T_{L L'}^o$,
for $r_i = r_o =R_b^i$ we have, neglecting for simplicity from now on the $k$
dependence of the local solutions,
\be
 \sum_L B^{i}_{L}({\bf k}) \overline{\Phi}_L( { \bf r }_i) =
\sum_{ L } \tilde{A}_{ L }^o ({\bf k}) \left[ J_L({ \bf r }_o; k) + \sum_{L'}
  \tilde{H}_{L'}^{ + } ( { \bf r }_o; k) T_{L'L}^i \right]
\label{wfcont}
\ee
Using Eq. (\ref{bsca}) for a single site and equating the coefficients of
$\tilde{A}_{ L }^o ({\bf k})$ we find at the bounding sphere the relation
\bea
\sum_{L'} \overline{\Phi}_{L'}( { \bf r }_i) T_{L'L}^i &=&
 \sum_{L'} (\tilde{E})_{LL'}^{-1} \Phi_{L'}  \nr
      &\equiv& \underline{\Phi}_{L} \nr
      &=& J_L({ \bf r }_o; k) + \sum_{L'}
      \tilde{H}_{L'}^{ + } ( { \bf r }_o; k) T_{L'L}^i
\label{smconteq}
\eea
implying that the basis functions $\underline{\Phi}_{L}$
are scattering functions, obeying the Lippmann-Schwinger equation for
the cell potential. Therefore, introducing new expansion coefficients
$C_{L}({\bf k})$ such that locally
\be
\psi ( { \bf r };{\bf k} ) =
\sum_{ L } C_{ L } ({\bf k}) \underline{\Phi}_{ L } ( { \bf r } )
\label{expswf}
\ee
and repeating the steps leading to the MSE in this new basis, we obtain
\be
 C^{i}_{L}({\bf k})\, - \sum_{j, L' L''} ^{j \neq i} G^{i j}_{L L''}
T_{L''L'}^{j} C^{j}_{L'}({\bf k}) -
\sum_{L'} J^{i o}_{L L'} C_{L'}^o ({\bf k})
= I^{i}_{L}({\bf k})
\label{cei}
\ee
\be
\sum_{j, L' L''} ^{j \neq o}J^{o j}_{L L''} T_{L''L'}^{j}C^{j}_{L'}({\bf k})
%=  \sum_{L' } \tilde{A}_{L'}^o ({\bf k})T_{L L'}^o
=  \sum_{L' }  (\overline{T}^o)_{L L'}^{-1} C_{L'}^o ({\bf k})
\label{ceo}
\ee
Comparing these equations with the previous ones in Eqs. (\ref{bei}),
(\ref{beo}) and (\ref{sceq1}), (\ref{sceq2}) we immediately find the
relations
\bea
B^{j}_{L}({\bf k})& = & \sum_{L'} T_{L L'}^{j} C^{j}_{L'}({\bf k}) \\
B^{o}_{L}({\bf k})& = & \sum_{L'} (\overline{T}^o)_{L L'}^{-1} C_{L'}^o
({\bf k}) \\
%C^{j}_{L}({\bf k})& = & \sum_{L'} E_{L L'}^{j} A^{j}_{L'}({\bf k})
C_{L}({\bf k})& = & \sum_{L'} E_{L L'} A_{L'}({\bf k})
\eea
In the present approach, the three forms of pair of equations
(\ref{sceq1})-(\ref{sceq2}), (\ref{bei})-(\ref{beo}) and (\ref{cei})-
(\ref{ceo}) are equivalent and lead to the same result.

The pair of equations (\ref{cei})-(\ref{ceo}) are quite important, since they
provide the formal justification that in MST one can work with square matrices,
provided that the only indexes appearing in the theory are those of the
radial functions $R_{L L'}(r)$. This is a consequence of the relation
(\ref{tmat}) of  \ref{a2} (second equation) and the fact that
the matrix elements $T_{L L'}$ have a common truncation parameter $l_{\rm max}$.
In fact, since $Tr (T^{\dagger}T) < \infty$ due to the asymptotic behavior of the
$T_{L L'}$ matrix elements given by Eq. (\ref{tmat1}) in the same
Appendix, one can safely define an inverse for the matrix $T_{L L'}(E)$
(except at poles on the negative energy axis) and pass from one representation
to the other. In particular one can pass from the set (\ref{cei})-(\ref{ceo})
to the set (\ref{sceq1})-(\ref{sceq2}). In the traditional derivation of MS
equations, that does not rely on the relations (\ref{main1})-(\ref{main6}) but
hinges on the re-expansion formulas (\ref{reexp1})-(\ref{reexp3}),
this equivalence does not hold. In fact the need to saturate the "internal"
sum over $L''$ coming from the re-expansion introduces a further expansion
parameter and therefore rectangular matrices into the theory. This feature makes
it impossible to define a $T$-matrix and to write a closed form for the GF,
loosing all the advantages of MST over other methods. This drawback has been
avoided in our approach, since in each step of the derivation of the MS
equations we have shown that the introduction of summation indices other than
those present in the radial functions $R_{ L'L} ( r )$ is unnecessary.

Another useful consequence of the fact that the theory can be cast in terms
of square matrices is the possibility to exploit the point symmetry of the
cluster under study. Even though many authors have treated the problem
of how to symmetrize the MSE, this was done in the framework of the MT
theory, where the cell T-matrices are diagonal in the AM indices. New
features appear in the more general case (in particular how to calculate
the symmetrized version of the $T_{L L'}$ matrices) and  \ref{a6}
deals with this situation. Needless to say, we checked in all applications
that the symmetrized and unsymmetrized version of the theory gave the same
results. The application of the symmetrization procedure to Green's Functions
or to periodic systems is rather straightforward.

As already anticipated in the introduction, one of the major advantages of
MST is the direct access to the Green's Function  of the system. Having
explicit expressions for this quantity is of the utmost importance both for
writing down spectroscopic response functions
(see Ref.~\cite{sebilleau06})
and for the calculation of ground state properties through contour integration
in the complex energy plane (see {\it e.g.} Ref.~\cite{papanik02} and
references therein).

The GF is solution of the Schr\"odinger equation with a source term
\bea
  (\nabla^2 \, + E -  V \, ( {\bf r} ) \,) \, G( { \bf r }, {\bf r}'; E )
 = \delta ( {\bf r} - {\bf r}' ).
    \label {eqn:gf}
\eea

In the framework of MST and for general (possibly complex) potentials, the
solution of this equation in the case of a finite cluster can be written as
~\cite{gonis00,Faulkner80}
\bea
  G( { \bf r }_i, {\bf r}_{j}'; E ) &=&
    \, \< \, \overline { \Phi } ( { \bf r }_i ) \, |
    \, ( \, \tau^{ ij } - \, \delta_{ ij } \, T^i \, ) \,
    | \, \overline { \Phi } ( { \bf r }_{j}' ) \, \> \nr
    &+& \, \delta_{ ij } \, \< \, \overline { \Phi } ( { \bf r }_< ) \,|
    \, T^i \, | \, { \Psi } ( { \bf r }_{>}' ) \, \>
\eea
where ${ \bf r }_<$ (${ \bf r }_{>}$) indicates the lesser (the greater)
between $r_i$ and $r_i'$. The function ${ \Psi } ( { \bf r } )$ is
the irregular solution in cell $i$ that matches smoothly to
$\tilde{H}_{L'}^+ ({ \bf r })$ at $R_b^i$.
For short we have saturated the sum over the
angular momentum indices using a bra and ket notation ({\it e. g.})
\be
\< \, \overline { \Phi } ( { \bf r }_i ) \, |
  \, \tau^{ ij } | \, \overline { \Phi } ( { \bf r }_{j}' ) \, \>  =
\sum_{L L'} \overline { \Phi }_L ( { \bf r }_i ) \, \tau^{ ij }_{L L'} \,
\overline { \Phi }_{L'} ( { \bf r }_{j}' )
\ee
Moreover, for simplicity of presentation we have assumed no contribution from
the outer region potential ({\it i.e.} $\overline{T}^o \equiv 0$) allowing
empty cells to cover the volume $\Omega_o$ up to the point at which the
asymptotic behavior in Eq. (\ref{wfexpo}) starts to be valid. The
modifications needed in the case $\overline{T}^o \neq 0$ are obvious. In the
case of a crystal we have to work in Fourier space~\cite{papanik02}.

Now, from Eq. (\ref{smconteq}) written as
\be
\overline{\Phi}_{L}( { \bf r }_i)  =
\sum_{L'} J_{L'}({ \bf r }_o; k) (T^{-1})_{L'L}^i +
\tilde{H}_{L}^{ + } ( { \bf r }_o; k)
\label{smconteq1}
\ee
by continuity we derive inside cell $i$ the relation
\be
\underline{\Phi}_{L}( { \bf r }_i)  =
\sum_{L'} \Lambda_{L'}({ \bf r }_i; k) (T^{-1})_{L'L}^i +
\Psi_{L} ( { \bf r }_i; k)
\label{smconteq2}
\ee
where $\Lambda_{L'}({ \bf r }_i)$ is the irregular function joining smoothly
to $J_{L'}({ \bf r }_o; k)$ at $R_b^i$. Therefore the Green's function takes
the form
\bea
  G( { \bf r }_i, {\bf r}_{j}'; E ) &=&
    \, \< \, \overline { \Phi } ( { \bf r }_i ) \, | \, \tau^{ ij } \,
    | \, \overline { \Phi } ( { \bf r }_{j}' ) \, \> \nr
    &-&  \, \delta_{ ij } \, \< \, \overline { \Phi } ( { \bf r }_< ) \,|
     \, { \Lambda } ( { \bf r }_{>}' ) \, \>
 \label {gfeq3}
\eea
For real potentials, both $\overline{\Phi}_{L}$ and $\Lambda_{L}$ are real,
so that the singular atomic term does not contribute to the imaginary part
of the GF.
%Eq. (\ref{gfabs}) reduces to Eq. (\ref{absgot}).
In this case the quantity
$- \Im \,\int_{\Omega_i} \, G( { \bf r }, {\bf r} ; E ) \, {\rm d^3} r
= - \sum_L \, \Im \, \tau_{L L}^{i i} \, (E) \, \int_{\Omega_i}
\overline{\Phi}_{L}^2 ({\bf r}) \, {\rm d^3} r $ is the projected density of
states on site $i$ at energy $E$, expressed as a sum of the partial
densities of type $L$. This relation (not ${\bf r}$-integrated) constitutes
the basis for calculating the system density by contour integration in the
complex energy plane.

Alternative forms of the GF that are independent of the normalization of
the local solutions $\Phi_L ( { \bf r }_i )$ can be easily obtained in
terms of the $S$ and $E$ matrix. For example we have
\bea
 &&G ( { \bf r }_i, {\bf r }_j'; E ) \nr
 &&= - \, \< \, \Phi ( { \bf r }_i ) \, | \,
 \, \{ ( [ \, \tilde{ S } \, E + \tilde{ S } G \, S \, ]^{ -1 } )^{ ij }
 - \delta_{ ij } \, ([ \, \tilde{ S } \, E \, ]^{ -1 })^{ ii } \,  \}
    \, | \, \Phi ( { \bf r }_j' ) \, \> \nr
 &&- \, \delta_{ ij } \, \< \, \Phi ( { \bf r }_< ) \,|
    \, E^{ -1 } \, | \, { \Psi } ( { \bf r' }_> ) \, \>
  \label {gfeq1}
\eea
which is seen to reduce to the following expression, remembering the
definition of $| \, \underline{\Phi}\,\>$,
\bea
  G ( { \bf r }_i, {\bf r }_j'; E ) &=&
     \, \< \, \underline{\Phi} ( { \bf r }_i ) \, |
\, ( \,  [ \,I - \,  G \, T \, ]^{ -1 } \, G \, )^{ i j } \,
    \, | \, \underline{\Phi} ( { \bf r }_j' ) \, \> \nr
  &-& \, \delta_{ ij } \, \< \, \underline{\Phi} ( { \bf r }_< ) \,
    | \, { \Psi } ( { \bf r' }_> ) \, \>
 \label {gfeq2}
\eea
Indeed from the relation,
\bea
  ( A + B )^{-1} -  A ^{-1}
  &=& ( A + B )^{-1} \, ( A - ( A + B ) ) \,{ A }^{-1} \nr
  &=& - ( A + B )^{ -1 } B A^{ -1 } \nr
  &=& - ( B^{ -1 } A + 1 )^{ -1 } A^{ -1 } \nr
  &=& - ( A B^{ -1 } A + A )^{ -1 }
\eea
we find
\bea
  [ \, \tilde{ S } \, E + \tilde{ S } \, G \, S \,  ]^{ -1 }
    - [ \, \tilde{ S } \, E \, ]^{ -1 }
    &=& - [ \, \tilde{ S } \, E + \, \tilde{ S } \, E \,
      [ \, \tilde{ S } \, G \, S \, ]^{ -1 }
      \, \tilde{ S } \, E \, ]^{ -1 } \nr
    &=& - [ \, \tilde{ S } \, E + \, \tilde{ S } \, E \,
      [ \, G \, S \, ]^{ -1 } \, E \, ]^{ -1 } \nr
    &=& - \, E^{ -1 } \, [ \, \tilde{ S } + \, \tilde{ S } \, E \,
      [ \, G \, S \, ]^{ -1 } \, ]^{ -1 } \nr
    &=& - \, E^{ -1 } \, [ \, \tilde{ S } + \, \tilde{ E } \, S \,
      [ \, G \, S \, ]^{ -1 } \, ]^{ -1 } \,  \nr
    &=&  \, E^{ -1 } \, [ \, T -
      G^{ -1 } \, ]^{ -1 } \, \tilde{ E }^{ -1 } \nr
    &=&  \, -E^{ -1 } \, [ \, I - G \, T
       \, ]^{ -1 } \, G \, \tilde{ E }^{ -1 }
\eea
taking into account that $\tilde{ S } \, E \, = \tilde{ E } \, S \,$ and
$T = - S E^{-1} = - \tilde{E}^{-1} \tilde{S}$.
All these forms are equivalent as long as we can treat the matrices
$S$ and $E$ as square.

\subsection{Bound states}
\label{bound}

Even though the essential of this section has been presented in a conference
proceedings ~\cite{hatada09}, we feel that for the sake of
completeness of presentation and convenience of the reader it should
be repeated here.

The MSE in the case of bound states can be derived from those for
scattering states, by simply eliminating the exciting plane wave in
Eq. (\ref{two}) and taking the analytical continuation to negative
energies in free Green's function $G_{0}^{+}({\bf r}' - {\bf r};k)$,
in order to impose the boundary condition of decaying waves when $r'
\rightarrow \infty$. In this case the Lippmann-Schwinger equation
reduces to the eigenvalue equation \be \psi({\bf r}') = \int \,
G_{0}^{+}({\bf r}' - {\bf r}; k) \, V({\bf r}) \, \psi({\bf r}) \,
{\rm d^3r} \label{bse} \ee where we have dropped the label ${\bf k}$
in the wave function $\psi({\bf r}')$. Since the expansion of
$G_{0}^{+}({\bf r}' - {\bf r}; k)$ in terms of spherical Bessel and
Hankel functions in Eqs. (\ref{gfexp1}) and (\ref{gfexp2}) remain
valid under the analytical continuation to negative energies, so
that $k =\sqrt{E} = i \sqrt{|E|} = i \gamma$, we see that $\psi({\bf
r}')$ behaves like $e^{ikr'}/r' = e^{-\gamma r'}/r'$ for $r'
\rightarrow \infty$. We remind that \bea h_l^+(kr) = -{ i}^{-l} \,
K_l^1 (\gamma r); &     & h_l^-(kr) = -{ i}^{-l} (-1)^l \, K_l^2
(\gamma r)  \nr j_l(kr) = { i}^l \, I_l(\gamma r);    &     &
n_l(kr) = i^{l+1} \frac{(-1)^{l+1} K_l^1 + K_l^2 }{2} \label{acbhf}
\eea where $I_l$ is the modified Bessel and $K_l^1$, $K_l^2$ the
modified Hankel functions of first and second kind, respectively.
Not only the expansions in Eqs. (\ref{gfexp1}) and (\ref{gfexp2}),
but also the re-expansion relations in Eqs. (\ref{reexp1}),
(\ref{reexp2}) and (\ref{reexp3}) remain valid under analytical
continuation with the same convergence properties (see  \ref{a2}).
This fact implies that we can derive the MSE for bound states
following the same patterns as for scattering states, except that
now the behavior of the wave function in the outer region
$\mathcal{C} \Omega_o$ is \bea \psi ( { \bf r }_o ) & = & \sum_{ L }
A_{ L }^o \,  \Phi_{ L }^o ( { \bf r }_o )  \nr & = & \sum_{ L } A_{
L }^o \, \sum_{ L' } \, R_{ L'L}^o  ( r_o ) \, Y_{ L' } ( \hat {\bf
r}_o ) \label{bsexpo} \eea The functions $\Phi_{ L }^o ( { \bf r }_o
)$ are now real and can easily be found by inward integration in the
outer region starting from an asymptotic WKB solution properly
normalized, {\it e.g.} like $[(2l+1)!!]^{-1}$.

Working with the $B_L$ amplitudes we easily arrive at the following
condition for the existence of a bound state
\be
\sum_{j L'} \left\{ (T^{i})_{L L'}^{-1}  \delta_{ij}\,
-  (1 - \delta_{ij}) \, G^{i j}_{L L'}   - \sum_{L' L''}
J^{i o}_{L L'} \overline{T}_{L' L''}^{o}  J^{o j}_{L'' L'}
 \right\} B^{j}_{L'} = 0
\label{beiobs}
\ee
which is the same as Eq. (\ref{bei1}), except that the exciting plane wave
term $I^{i}_{L}({\bf k})$ and the ${\bf k}$ dependence have been dropped.
Notice that we have kept the arbitrariness of $V_0$ in the free Green's
function, in order to check that the eigenvalues do not depend on it. In
the spirit of the analytical continuation, we have a definite rule on how
to calculate the various quantities as a function of $\kappa$.

We now define
\be
C_{LL'} = ({R_b})^2 W[n_l,R_{ L L' }]
%S_{LL'}^o &=& ({R_b^o})^2 W[\kappa n_l,R_{ L L'}^o ]
\label{csdef}
\ee
so that, remembering Eq. (\ref{tdef})
\bea
\kappa^{-1} (T^j)^{-1} = (K^j)^{-1} + i = - C^j (S^j)^{-1} + i \label{kdefj} \\
\kappa^{-1} \overline{T}^o = \overline{K}^o + i = - C^o (S^o)^{-1} + i
\label{kdefo}
\eea
Moreover we observe that
\be
\kappa^{-1} G^{i j}_{L L'} =
N^{i j}_{L L'} - i J^{i j}_{L L'}
\label{gdec}
\ee
where $N^{i j}_{L L'}$ is defined in Eq. (\ref{ndef}) and that
$\sum_{L''} J^{i o}_{L L''}  J^{o j}_{L'' L'} = J^{i j}_{L L'} $, since
$J$ is the translational operator. Substituting these relations into Eq.
(\ref{beiobs}) and eliminating the common factor $\kappa^{-1}$ we finally find
\be
\sum_{j L'} \left\{ (K^{i})_{L L'}^{-1}  \delta_{ij}\,
-  (1 - \delta_{ij}) \, N^{i j}_{L L'}  -  \sum_{L' L''}
J^{i o}_{L L'} \overline{K}_{L' L''}^{o}  J^{o j}_{L L'} \right\} B^{j}_{L'} = 0
\label{beiobsr}
\ee
The generic ($L L'$)-element of this MS matrix is either real for real $\kappa$
($E-V_0 > 0$) or proportional to $i^{l-l'+1}$ for imaginary $\kappa$
($E-V_0 < 0$). Indeed, due to the relations Eq. (\ref{acbhf}), putting for short
$K_l = [(-1)^{l+1} K_l^1 + K_l^2 ]/2$, we easily find that
\bea
N_{L L'} &=& 4\pi i^{l-l'+1} \sum_{L''} C(L,L';L'')\, (-1)^{l''} \,
K_{l''} ( |\kappa| R_{ij} )  Y_{L''} ({\bf R}_{ij}) \nr
(K^{i})_{L L'}^{-1} &=& - i^{l-l'+1} \left[ \underline{C}^i
(\underline{S}^i)^{-1}
\right]_{L L'} \nr
\overline{K}_{L L'}^o &=& - i^{l-l'+1} \left[ \underline{C}^o
(\underline{S}^o)^{-1}
\right]_{L L'}
\nonumber
\eea
where $\underline{C}$ and $\underline{S}$ are defined in terms of the
modified spherical Bessel and Neumann functions as the corresponding quantities.

\noindent Therefore the condition for a bound state becomes
${\rm Det}\, \underline{M} = 0$, where $\underline{M}$ is the MS matrix in
Eq. (\ref{beiobsr}) after a unitary transformation that eliminates the imaginary
factors. In the practical numerical implementation we find the zeros of
the determinant of ${\rm Det}\, (K \underline{M})$, excluding the spurious
solutions coming from the zeros of ${\rm Det}\, \underline{S}$. In this form,
the procedure is equivalent to finding the poles of the GF in the form
Eq. (\ref{gfeq2}) on the real negative axis, as it should be. Still numerical
instabilities might come from the inverse of $\underline{S}^o$ present in
the contribution of the outer sphere region. This unwanted feature could be
eliminated by working with the $A_L$, instead of the $B_L$ amplitudes.

\begin{figure}[htbp]
%  \begin{center}
    \includegraphics[width=15cm]{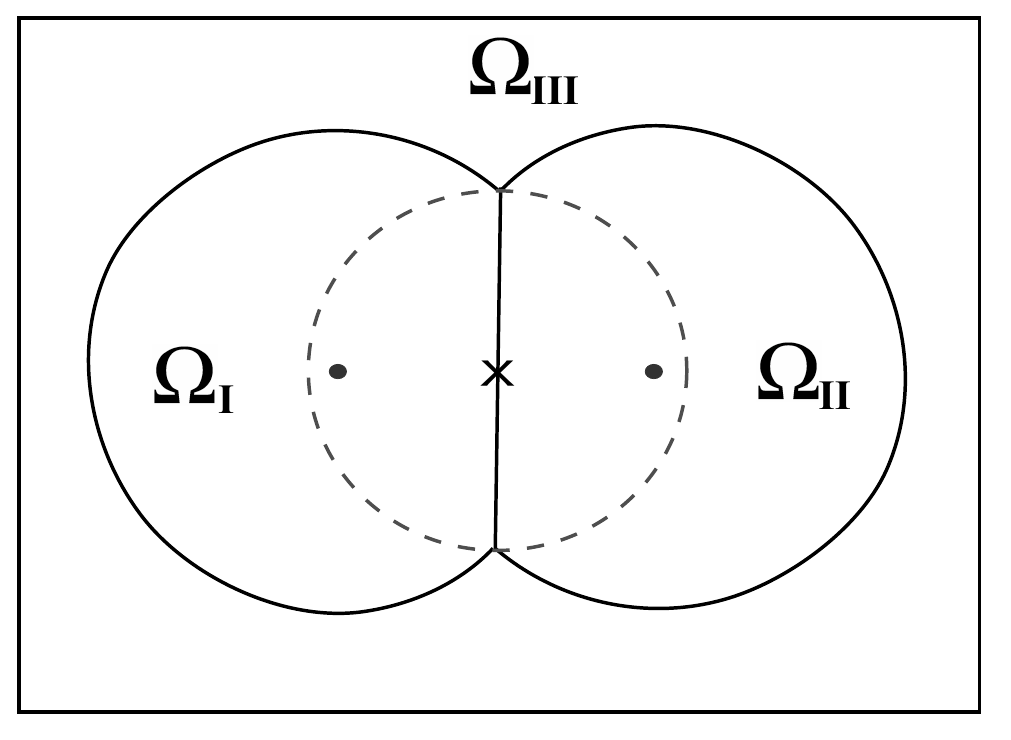}
    \caption{Partitioning of the space for the hydrogen molecular ion
             with no empty cells.}
    \label{fig_3}
%  \end{center}
\end{figure}

                \begin{table}[t]
                \caption{Eigenvalues of the hydrogen molecular ion (in Ryd)}
                \begin{center}
                % \begin{tabular}{|p{20mm}|c|c|c|c|}
\begin{tabular}{|p{8mm}|p{7mm}|p{13mm}|p{14mm}|p{14mm}|p{13mm}|p{13mm}
|p{13mm}|p{13mm}|}
                \hline
 {\scriptsize Mol. orb.} &  {\scriptsize n l m} &  {\scriptsize Exact} & {\scriptsize Smith\,\&  Johnson\cite{smith69}}  &  {\scriptsize Foulis \cite{foulis88} }
&  {\scriptsize 22\,EC V$_0$=-1.90} & {\scriptsize 22\,EC   V$_0$= 0} & {\scriptsize No\,EC V$_0$=-1.90} & {\scriptsize No\,EC V$_0$= 0} \\
                \hline
         {\footnotesize 1a$_{1g}$} &  {\footnotesize 1 0 0} &  {\footnotesize -2.20525} &  {\footnotesize -2.0716} &  {\footnotesize -2.18973} &  {\footnotesize -2.20522} &  {\footnotesize -2.2055}
        & {\footnotesize -2.2050} &  {\footnotesize -2.2048}   \\
        \hline
         {\footnotesize 2a$_{1g}$} &  {\footnotesize 2 0 0} &  {\footnotesize -0.72173} &  {\footnotesize -0.70738} &  {\footnotesize -0.72093} &  {\footnotesize -0.723} &  {\footnotesize -0.724} &
         {\footnotesize -0.731} &  {\footnotesize -0.726} \\
        \hline
         {\footnotesize 3a$_{1g}$} &  {\footnotesize 3 2 0} &  {\footnotesize -0.47155} &  {\footnotesize -0.45574} &  {\footnotesize -0.47102} &  {\footnotesize -0.4727} &  {\footnotesize -0.478} &
         {\footnotesize -0.476} &  {\footnotesize -0.474} \\
        \hline
         {\footnotesize4a$_{1g}$} &  {\footnotesize 3 0 0} &  {\footnotesize -0.35536} &  {\footnotesize -0.34859} &  {\footnotesize -0.35525} &  {\footnotesize -0.356} &  {\footnotesize -0.3550} &
         {\footnotesize -0.357}  &  {\footnotesize -0.356} \\
        \hline
         {\footnotesize 1a$_{2u}$} &  {\footnotesize 2 1 0} &  {\footnotesize -1.33507} &  {\footnotesize -1.2868} &  {\footnotesize -1.33426} &  {\footnotesize -1.3348} &  {\footnotesize -1.3348} &
         {\footnotesize -1.3342}  &  {\footnotesize -1.3343}  \\
        \hline
         {\footnotesize 2a$_{2u}$} &  {\footnotesize 3 1 0} &  {\footnotesize -0.51083} &  {\footnotesize -0.49722} &  {\footnotesize -0.51085} &  {\footnotesize -0.51072} &  {\footnotesize -0.5105}
         &  {\footnotesize -0.5104} &  {\footnotesize -0.5104} \\
        \hline
         {\footnotesize 3a$_{2u}$} &  {\footnotesize 4 1 0} &  {\footnotesize -0.27463} &  {\footnotesize -0.26979} &  {\footnotesize -0.27466} &  {\footnotesize -0.27469} &  {\footnotesize -0.2742} &
         {\footnotesize -0.2745} &  {\footnotesize -0.2745} \\
        \hline
         {\footnotesize 4a$_{2u}$} &  {\footnotesize 4 3 0} &  {\footnotesize -0.25329} &  {\footnotesize -0.24997} &  {\footnotesize -0.25329} &  {\footnotesize -0.254} &  {\footnotesize -0.2536} &
         {\footnotesize -0.2541} &  {\footnotesize -0.25301} \\
        \hline
         {\footnotesize 1e$_{1g}$} &  {\footnotesize 3 2 1} &  {\footnotesize -0.45340} &  {\footnotesize -0.44646} &  {\footnotesize -0.45333} &  {\footnotesize -0.4545} &  {\footnotesize -0.45332} &
         {\footnotesize -0.455} &  {\footnotesize -0.455}   \\
        \hline
         {\footnotesize 1e$_{1u}$} &  {\footnotesize 2 1 1} &  {\footnotesize -0.85755} &  {\footnotesize -0.88866} &  {\footnotesize -0.85585} &  {\footnotesize -0.85754} &  {\footnotesize -0.8561} &
         {\footnotesize -0.870} &  {\footnotesize -0.858}   \\
                \hline
                \end{tabular}
                \end{center}
                \label{tab:egvhmi}
                \end{table}

We applied the theory above to find the exact eigenvalues of the
hydrogen molecular ion, since this test is considered rather stringent
for the validity of the theory due to rapid variation of the potential
in the molecular region and to the awkward geometry of the cells.
 In this case we partition the space in three
regions, as illustrated in Fig. \ref{fig_3}, two truncated spheres around
the protons with a radius of 1.72 a.u. corresponding to cells $\Omega_I$ and
$\Omega_{II}$ and an external region labeled $\Omega_{III}$, corresponding
to the complementary
domain $\mathcal{C} \Omega_o$. The bounding sphere of this latter is
represented by the dashed circle with radius 1.4 a.u., bigger than one
half the distance of the protons, as discussed after Eq. (\ref{main4}).
By calling the region outside this circle $\mathcal{C} \Omega_b$,
the potential is taken to be zero (or constant) into the intersection
of this domain with cells $\Omega_I$ and $\Omega_{II}$, and equal to the
value of the true
potential in the intersection with $\mathcal{C} \Omega_o$. We also did
a calculation with the two atomic cells, 22 empty cells surrounding
them, plus an external region.

It should be noticed that treatment of bound state is done here in analogy
to the X-$\alpha$ MST method~\cite{slater72}, since we intend to put the
theory to a severe test concerning the independence of the eigenvalues from
the value of the interstitial constant $V_0$ and the partitioning of the
space. More modern techniques that avoid finding eigenvalues and eigenstates
of the molecular cluster in the course of a SCF-iteration, exploit the
analyticity of the GF through a contour integration in the complex energy
plane to find directly the molecular density, as mentioned in the
introduction and in Section \ref{scatstat}.

Our findings are listed into Table
(\ref{tab:egvhmi}) and compared with the exact results. The last two
columns show the eigenvalues obtained with two different values of $V_0$,
respectively equal to -1.90 Ryd and 0, showing the 'quasi' independence of
the results from the constant interstitial value $V_0$. The columns with the
label '22 EC' refer to the calculation with two atomic cells, 22 empty cells
and an external region, showing the 'quasi' independence of the result from
the partitioning mode of the space. We attribute the slight dependence of
the eigenvalues on $V_0$ and the partitioning mode to the numerical
instabilities mentioned above and the $L$-truncation of the matrices.

The column labeled 'Smith \& Johnson' refers to the calculation by Smith and
Johnson~\cite{smith69} in the MT approximation, whereas the one labeled
'Foulis' quotes the result by Foulis~\cite{foulis88} obtained within the
distorted wave approximation.

\section{Convergence of Full Potential Multiple Scattering Theory}
\label{cmst}

The inversion of the MS matrix becomes computationally heavy at high
photoelectron energies because of the large number of angular momenta
involved, since $l_{\rm max} \approx k R_b$. A common way to circumvent
this difficulty is to invert the MS matrix by series, whereby
\be
(T^{-1} - G)^{-1} = T \, \sum_n (GT)^n.
\label{msexp}
\ee
While this series is absolutely convergent for nonoverlapping
MT spheres, provided the spectral radius of the matrix $GT$ is less
than one~\cite{garcia86}, it is known to diverge for the case of space-filling
cells. This is easily seen by using the inequality (\ref{gtineq}) in
 \ref{a2}, putting $l = l' \gg l_{\rm max}$, whereby
\be
|G_{ll} T_{ll}| \approx R_b \, \bigg(\frac{2R_b}{R_{ij}}\bigg)^{2l+1}
\label{gtll}
\ee
which signals the divergence of the matrix element $(GT)_{ll}$ (for
space-filling cells $2R_b > R_{ij}$, at least for nearest neighbors).

However due to the behavior shown by Eq. (\ref{gtll}) there is a widespread
belief that the procedure of inverting exactly an $l$ truncated MS matrix and
then letting $l$ go to $\infty$ does not converge in the case of
space-filling cells. We shall show in the following that this is not so,
provided a slight modification of the free propagator $G$ is adopted.

In order to illustrate our point, let us start by solving the Lippmann-Schwinger
equation using the theory of the integral equations, before applying MST.

\be
\psi({\bf r}';{\bf k}) = {\rm e}^{{\rm i}{\bf k} \cdot {\bf r}'} +
\int \, G_{0}^{+}({\bf r}' - {\bf r};{ k}) \, V({\bf r}) \,
\psi({\bf r};{\bf k})
\, {\rm d^3r}
\label{two}
\ee
We cannot use Fredholm theory, since the Kernel for this integral equation
\be
{\bf K}({\bf r}',{\bf r}) = - \frac{1}{4\pi} \,
\frac{{\rm e}^{{\rm i}{\bf k} \cdot |{\bf r}'-{\bf r}|}}{|{\bf r}'-{\bf r}|} \,
V({\bf r})
\label{kernel1}
\ee
is such that
\bea
Tr ({\bf K}^{\dagger} {\bf K}) & = &  \, \int \int  {\rm d}{\bf r} {\rm d}{\bf r}' \,
                     {\bf K}^{\star}({\bf r}',{\bf r}) \, {\bf K}({\bf r}',{\bf r}) \nr
             & = & \left ({\frac{1}{4\pi}}\right)^2 \, \int \int  {\rm d}{\bf r}
                       {\rm d}{\bf r}' \, \frac{V({\bf r})^2}
                       {|{\bf r}'-{\bf r}|^2} \nr
             & \le & \left({\frac{1}{4\pi}}\right)^2 \, \int \int  {\rm d}{\bf r}
                         {\rm d}{\bf r}' \, \frac{|V({\bf r})|^2}
                       {|{\bf r}'-{\bf r}|^2}   \label{tracek}
\eea
and obviously diverges.

However a solution for this problem can be found by following the argument of
section 10.3, page 280 of Ref. 39 in the paper (Newton).
We multiply the Lippmann-Schwinger equation Eq.
(\ref{two}) by $|V({\bf r}')|^{1/2}$ and write
$V({\bf r}) = |V({\bf r})| v({\bf r})$, where $v({\bf r})$ is a sign factor,
equal to $+1$ where the potential is positive and to $-1$ where it is negative.
Then we obtain
\bea
\psi_s({\bf r}';{\bf k}) & \equiv &|V({\bf r}')|^{1/2} \, \psi({\bf r}';{\bf k}) \nr
&=& |V({\bf r}')|^{1/2} \, {\rm e}^{{\rm i}{\bf k} \cdot {\bf r}'} \nr
&+& |V({\bf r}')|^{1/2} \, \int \, G_{0}^{+}({\bf r}' - {\bf r};{ k}) \,
|V({\bf r})|^{1/2} \, v({\bf r}) \,
\psi_s({\bf r};{\bf k})\, {\rm d^3r}
\label{twoprime}
\eea
The Kernel for this integral equation is given by
\be
{\bf K}_s({\bf r}',{\bf r}) = - \frac{1}{4\pi} \, |V({\bf r}')|^{1/2} \,
\frac{{\rm e}^{{\rm i}{\bf k} \cdot |{\bf r}'-{\bf r}|}}{|{\bf r}'-{\bf r}|} \,
|V({\bf r})|^{1/2} \, v({\bf r})
\label{kernel2}
\ee
whereby
\bea
Tr ({\bf K}_s^{\dagger} {\bf K}_s) & = &  \, \int \int  {\rm d}{\bf r} {\rm d}{\bf r}' \,
                     {\bf K}^{\star}({\bf r}',{\bf r}) \, {\bf K}({\bf r}',{\bf r}) \nr
             & = & \left ({\frac{1}{4\pi}}\right)^2 \, \int \int  {\rm d}{\bf r}
                       {\rm d}{\bf r}' \, \frac{V({\bf r}) \, |V({\bf r}')|}
                       {|{\bf r}'-{\bf r}|^2} \nr
             & \le & \left({\frac{1}{4\pi}}\right)^2 \, \int \int  {\rm d}{\bf r}
                         {\rm d}{\bf r}' \, \frac{|V({\bf r})| \, |V({\bf r}')|}
                       {|{\bf r}'-{\bf r}|^2}   \label{traceks}
\eea
which is finite for a large class of potentials (including the molecular ones),
so that the kernel ${\bf K}_s$ is of the Hilbert-Schmidt type and Fredholm
theorem for $L_2$-kernels can be applied. Once the solution
$\psi_s({\bf r}';{\bf k}) $ is found, we can obtain the solution of Eq.
(\ref{two}) simply by dividing it by $|V({\bf r}')|^{1/2}$, except at points
for which $|V({\bf r}')|^{1/2} = 0$, where it can be defined by continuity.

Now, let us apply MST to Eq. (\ref{two}) using the scattering wave functions
$\overline{\Phi}_{ L } ( { \bf r } )$ in Eq. (\ref{smconteq})
as local basis functions. We transform this Lippmann-Schwinger equation into
a set of algebraic equations of infinite dimensions for the coefficients
$C_{ L } ({\bf k})$ in the expansion (\ref{expswf})
\be
 C^{i}_{L}({\bf k})\, - \sum_{j, L' L''} ^{j \neq i} G^{i j}_{L L''}
T_{L''L'}^{j} C^{j}_{L'}({\bf k}) = I^{i}_{L}({\bf k})
\label{cei1}
\ee
where, in comparison with Eq. (\ref{cei}), for simplicity we have
neglected the outer region $\mathcal{C} \Omega_o$, which we can always assume
to be covered by a set of empty cells.
In matricial form we have, putting $K=GT$ and calling $A$ the term in the rhs,
\be
(I - K) \, C = A
\label{mf1}
\ee
The matrix $K$ here is not an operator of the Hilbert-Schmidt type, since
$Tr \, (K^{\dagger} K)$ diverges, due to Eq. (\ref{gtll}) and in keeping with
Eq. (\ref{tracek}). However, following the procedure used above in passing
from Eq. (\ref{two}) to (\ref{twoprime}) and introducing new vector
components $C' = T^{1/2}C$ with a new inhomogeneous term $A' = T^{1/2}A$,
we transform this equation into a new one
\be
(I - K_s) \, C' = A'
\label{mf2}
\ee
where
\be
K_s = T^{1/2} G T^{1/2}
\label{rnspo}
\ee
The square root of the matrix $T$ is defined
in the usual way, by first diagonalizing it with a similarity transformation S,
taking the square root of the diagonal elements and then performing the same
transformation on these letters. In formulas, if $\Lambda = S T S^{-1}$, then
$T^{1/2} = S \Lambda^{1/2} S^{-1}$, so that $T^{1/2} T^{1/2} = T$. There is no
danger in performing these operations with the infinite matrix $T$, since
$Tr \; (T^{\dagger} T) < \infty$, as can be seen from the asymptotic behavior
of its matrix element in  \ref{a2}, Eq. (\ref {tmat1}). Hence the
limiting procedures are well defined.

By virtue of Eq. (\ref {traceks}), \ref{a7} shows that the kernel $K_s$ here is
of the Hilbert-Schmidt type ({\it i.e.} $Tr \, (K^{\dagger}_s K_s)$ is
finite). As is well known~\cite{tricomi85} this letter is the condition for the
existence of the determinant $|I - K_s|$ necessary to define its inverse,
since by Hadamard's inequality, for any finite $L_{\rm max}$, one has
\be
|I - K_s|^2 \le \Pi_L^{L_{\rm max}} \left(1 + \sum_{L'}^{L_{\rm max}} \,
|(K_s)_{LL'}|^2\right)
\ee
and in the limit $L_{\rm max} \rightarrow \infty$ the infinite product will
converge if
$\sum_{LL'} |(K_s)_{LL'}|^2 \equiv Tr \, (K^{\dagger}_s K_s) \le N < \infty$.
~\cite{whittaker65}

%Then the Fredholm type of solution can be constructed (for details see section
%2.5 of Ref. \cite{tricomi85} and footnote on page 66 for $L_2$ Kernels).

This means that the process of truncating the matrix $I - K_s$ to a certain
$l_{\rm max}$ and then taking the inverse, converges absolutely in the limit
$l_{\rm max} \rightarrow \infty$. Once $C'$ is obtained, $C = T^{-1/2} C'$,
thus solving the original problem. Moreover the scattering path operator
$\tau$ (\ref{spotau}) is given by
\be
\tau = T^{1/2} \, (I - K_s)^{-1} \, T^{1/2}
\ee

There is another way to solve Eq. \ref {mf2}, by expanding $(I - K_s)^{-1}$ in
series, i.e. writing
\be
(I - K_s)^{-1} = \sum_n (K_s)^n
\label{mf3}
\ee
However, even if the kernel $K_s$ is of the Hilbert-Schmidt type but
$Tr \, (K_s^{\dagger} K_s) \ge 1$, the series diverges, whereas the process of
truncating and taking the inverse always converges. It goes without saying that
the series $\sum_n K^n$ is always divergent, since $Tr \, (K^{\dagger} K)$ is
infinite. Therefore the series expansion procedure is not always a viable
method to find the inverse of a matrix of the type (I - A).

%%%

In practical numerical applications one does not have to worry about modifying
the structure constants according to Eq. (\ref{t1o2c}) since, for the cell
geometries ordinarily encountered in the applications (see the restrictions
described at the beginning of Section~\ref{msesbs}), $l$-convergence
in the $l$-truncation procedure of the MS matrix
shows up much earlier than what predicted by
the onset of divergence in Eq. (\ref{danger}), written with the unmodified
structure constants ${G}^{ij}_{\Lambda \Lambda'}$. We already found this out
in $GeCl_4$~\cite{hatada07}, where in the first 20 eV an $l_{\rm max} = 3$ was
sufficient to reproduce all spectral features, which did not change by
increasing $l$ up to 10. Similar results were found for other compounds.

\begin{figure}[htbp]
  \begin{center}
\includegraphics[width=14cm]{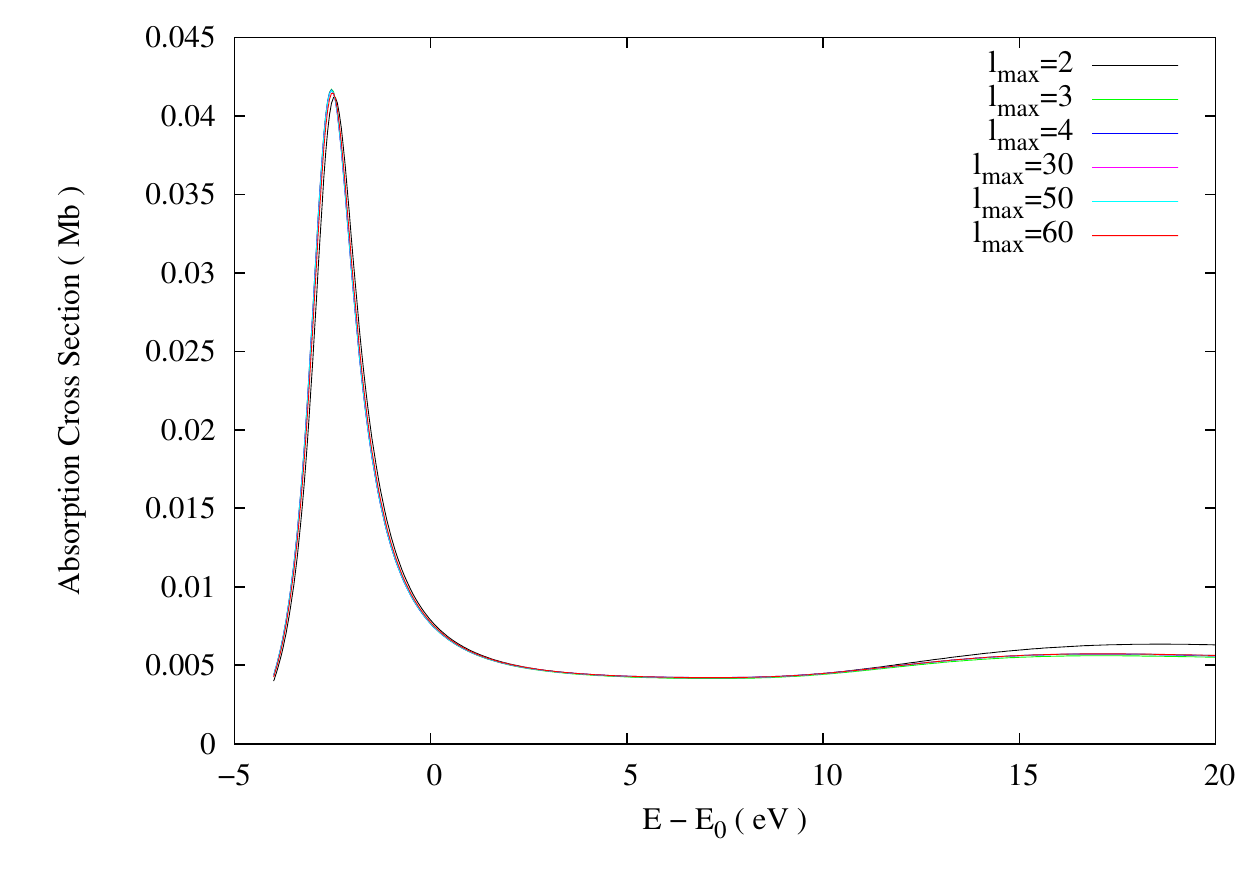}
    \caption{K-edge $z$-polarized absorption cross section for the $\sigma$ state
             of the $Se_2$ molecule, for various $l$-values up to $l_{\rm max}=60$,
             calculated by full inversion of the MS matrix (I-GT).}
    \label{fig_lexp}
  \end{center}
\end{figure}

In this context we did a more stringent test for the $Se_2$ diatomic molecule
formed by two inter-penetrating nonequivalent spheres with 40\% overlap, with
centers on the $z$-axis. We calculated the K-edge $z$-polarized cross section
for the $\sigma$ state ($m=0$) up to $l_{\rm max}=60$ in the energy range
$-4.0 \sim 20.0 \, eV$, using the kernel $K = GT$ and found a convergent
behaviour for the lhs of Eq. (\ref{msexp}) (see Fig. \ref{fig_lexp}).

%, whereas the
%rhs starts diverging for $l_{\rm max}=19$, since there is an eigenvalue of $GT$
%that is greater than one in the whole energy range (see Fig. \ref{fig_ev}).
%If this eigenvalue is only slightly greater than one, the divergence starts
%appearing for high values of the order $n$ (Fig. \ref{fig_se}).
%For values of $l_{\rm max}$ lower than 19, there are eigenvalues bigger than one
%only in the low-energy part of the spectrum and in this range again the rhs
%diverges. Fig. \ref{fig_ev} plots the two or three highest eigenvalues for
%various values of $l_{\rm max}$.

%This finding is not surprising, since as mentioned above, both in the finite
%and infinite dimensional case, if $Tr (K^{\dagger} K) \le N^2$, then the inverse
%of $I - \lambda K$ exists (except in the case when $1/\lambda$ is an eigenvalue of
%$K$), whereas the series $\sum_n (\lambda K)^n$ diverges if $\lambda \ge 1/N$.~
%\cite{tricomi85}

The fact that the full inversion of the MS matrix $(I-GT)$ is stable in this
case up to $l_{\rm max}=60$ is clearly not of general validity, although
indicative of the behavior of the theory. Going to higher values of $l_{\rm max}$
is not easy, because the lack of Lebedev integration formulas for a number of
surface points $\ge 6,000$ prevents us to access such values. Already the slight
discrepancy of the $l_{\rm max}=60$ curve in Fig. \ref{fig_lexp} with the previous
ones (barely visible) is the sign that $\sim$ 6,000 Lebedev points are barely
sufficient in this case. This kind of study for other geometries and bigger
clusters are under way.

%%%%
%%%%

%%%%
%\begin{figure}[htbp]
%  \begin{center}
%\includegraphics[width=14cm]{ev_19.pdf}
%\includegraphics[width=17cm]{se2.eps}
%    \caption{First few eigenvalues of $GT$ for $l_{\rm max}=19$ in the case of the
%             $Se_2$ molecule. }
%    \label{fig_ev}
%  \end{center}
%\end{figure}
%%%%

%%%%
%\begin{figure}[htbp]
%  \begin{center}
%\includegraphics[width=14cm]{abs_tot.pdf}
%\includegraphics[width=17cm]{se2.eps}
%    \caption{K-edge $z$-polarized absorption cross section for the $\sigma$ state
%             of the $Se_2$ molecule, calculated by the series expansion in the
%             rhs of Eq. (\ref{msexp}) for various values of the order $n$ and
%             $l_{\rm max}=19$.}
%    \label{fig_se}
%  \end{center}
%\end{figure}
%%%%

\section{Applications}
\label{apps}

Application of the present FP-MS theory to two cases which, according to our
experience, need significant non-MT corrections for a good reproduction of the
absorption data ({\it i.e.} diatomic-linear molecules and
tetrahedrally coordinated compounds) have already been presented in
Ref. ~\cite{hatada09} for the $K$-edge of $Se_2$
and the $Si$ $L_{2,3}$ edge of crystalline $SiO_2$ ($\alpha$-quartz). There
it was shown that a good description of the anisotropies of the potential
leads to a substantial improvement of the calculated absorption signal  in
comparison with the experimental spectra.

In this section we present another application to the $K$-edge absorption of
$Br_2$ and discuss a preliminary application of the NMT approach to the study
of the performance of two effective optical potentials, the Hedin-Lundqvist (HL)
potential and the Dirac-Hara (DH) in the case of a transition metal.

It should be emphasized that all potentials used here and in
Ref. ~\cite{hatada09} are non-self-consistent,
since the starting charge density is obtained by mere superposition of
atomic densities. Therefore the agreement or disagreement with experiments
might change if a self-consistent charge density were used, although from
our experience the effect of this latter has a minor impact on the spectra
than the elimination of the MT approximation. In any case,
one of the motivations for pursuing  the FP-MS method was exactly the
study of the performance of the various models of optical potential together
with the effect of the self-consistent charge density, once that the
geometrical approximation of the potential had been eliminated. The
application of the present real space theory to the generation of the
self-consistent ground state density using the well known technique of
contour integration in the complex energy plane is under way.

In order to obtain the absorption spectra we start from the well
known expression of the absorption cross
section in terms of the GF, given by
\bea
\sigma_{tot}(\omega)
 & = & - 8 \, \pi \, \alpha \, \hbar \, \omega \,\nr
 &     &\times \sum_{m_c} \, \Im \int \<\phi^c_{L_c}({ \bf r })|{\hat
\varepsilon} \cdot { \bf r } | G({ \bf r },{ \bf r }';E) | {\hat
\varepsilon} \cdot { \bf r }'| \phi^c_{L_c}({ \bf r }')\> {\rm d} {
\bf r } \, {\rm d} { \bf r }' \label{gfabs} \eea \noindent For more
details and other spectroscopies we refer the reader to
Ref.~\cite{sebilleau06}. We used all three forms of GF given by Eqs.
(\ref{gfeq3}), (\ref{gfeq1}) and (\ref{gfeq2}). While the last two
are numerically stable and give almost coincident spectra, the first
one shows occasionally small but noticeable kinks in the calculated
spectrum and sometimes small deviations around maxima and/or minima
of the cross section compared to the other two. This is a known
phenomenon which is now exalted compared to the MT case, where it
was almost unnoticeable. It is due the fact that the singularities
of the $S$-matrix in the definition of the scattering basis
functions $\overline { \Phi } ( { \bf r } )$ in Eq. (\ref{phsbf})
and those of $T^{-1}$ in the inverted MS matrix $\tau = (T^{-1} -
G)$ do not compensate exactly. Therefore, even though the three
forms are formally equivalent, from a computational point of view,
form Eq. (\ref{gfeq3}) is to be avoided.

\begin{figure}[htbp]
  \begin{center}
\includegraphics[width=14cm]{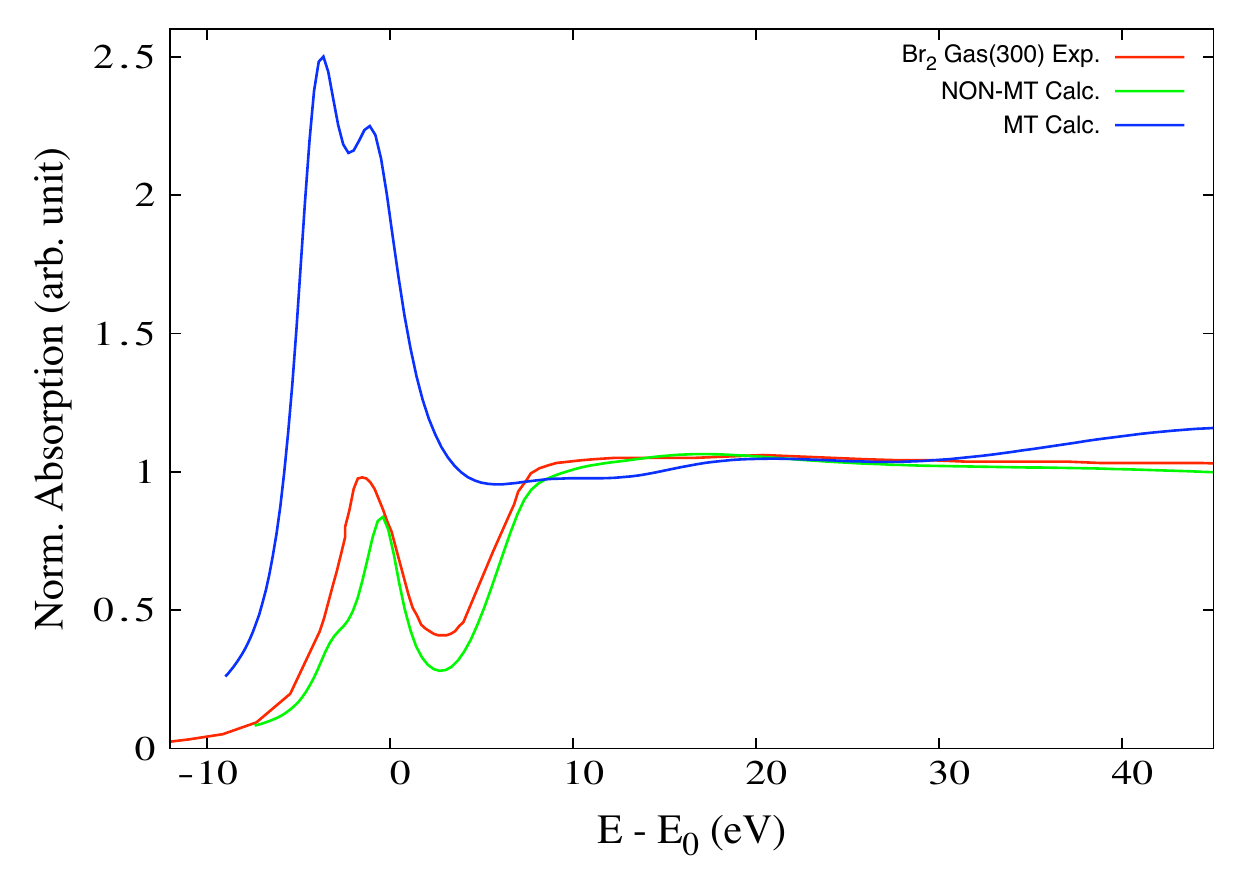}
    \caption{K-edge unpolarized absorption cross section for $Br_2$ molecule,
 showing the comparison between the MT and FP-MS calculations against the
 experimental data}
    \label{fig_cmp}
  \end{center}
\end{figure}

Fig. (\ref{fig_cmp}), shows the
experimental unpolarized K-edge absorption cross section of the diatomic
molecule $Br_2$~\cite{filipponi93} in comparison with a NMT and a MT
calculation as a function of the photo-electron kinetic energy $E$ referred
to $E_0$, the true zero of the non-self-consistent molecular potential
at infinity.
All spectra were normalized at a common energy point between 20 and 30 eV.
For the NMT case we partitioned the space with 24 Voronoi polyhedra arranged
on a BCC lattice: two of them around the physical atoms and 22 empty cells
(EC) to cover the rest of the space where the density (and the potential)
are significantly different from zero. ${\it l}_{\rm max}$ was taken equal to 4
in all polyhedra.
We gave a small finite imaginary part to the energy of the order of
($\sim$ 0.02 eV) in order to be able to use the same
Green's function expression for the cross section Eq. (\ref{gfabs}) both for
bound and continuum states.
To calculate the absorption spectrum, we used the real part of an
Hedin-Lundqvist (HL) potential and then convoluted the result with a
Lorentzian whose width is equal to the that of the core hole (2.52 eV).
We see that the agreement with experiment is rather good.
In contrast, the MT approximation of the potential turns
out to be rather poor.

%%%%
%%%%

\begin{figure}
  \begin{center}
    \begin{tabular}{cc}
      \resizebox{70mm}{!}{\includegraphics{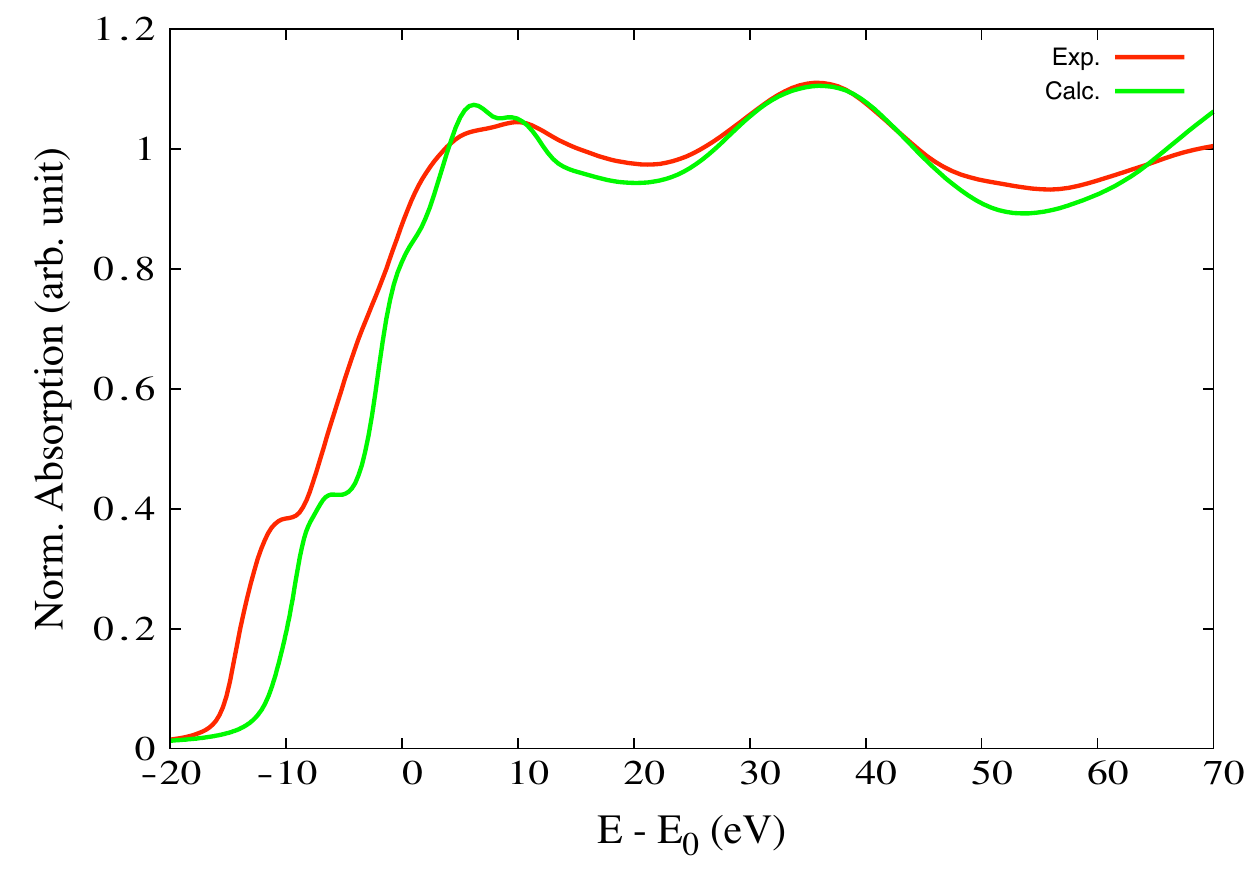}} &
      \resizebox{70mm}{!}{\includegraphics{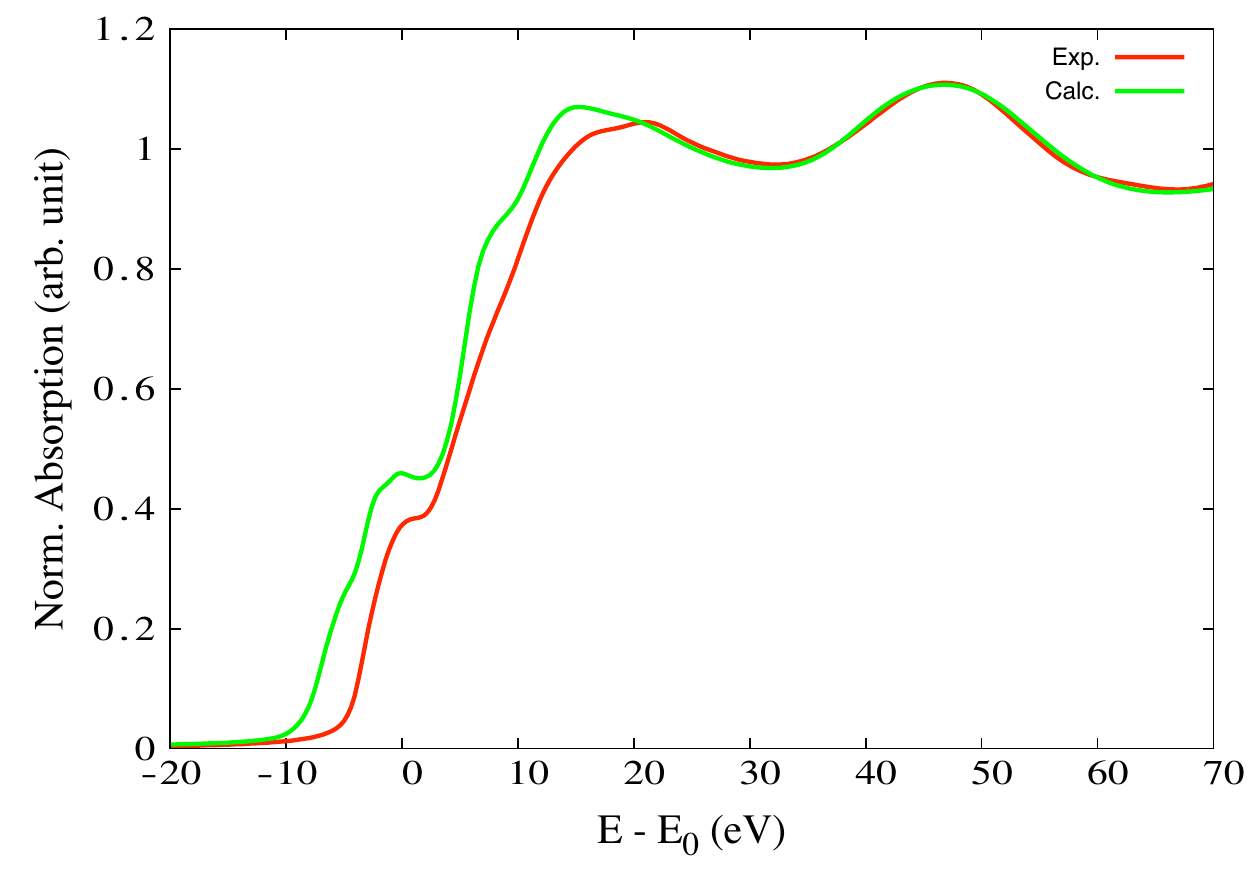}} \\
    \end{tabular}
  \end{center}
  \caption{Comparison between Co K-edge absorption calculated with complex
           HL (left) and DH (right) potentials with experimental results.
           (Color online)}
  \label{comphcp}
\end{figure}

We then present in Fig. (\ref{comphcp}) a preliminary application of
the NMT approach to assess the performance of the HL against a DH
potential, assuming that the losses are sufficiently well described
in both cases by the imaginary part of the HL self-energy, in the
case of HCP Co metal. As is well known, the real part of the HL
potential is composed of two terms: the static Hartree-Fock (HF)
exchange, known also as Dirac-Hara (DH) exchange, coming from the
constant part of the dielectric function  and the dynamically
screened exchange-correlation contribution (HLXC), originating from
the $\omega$-dependent part (see Appendix A of Ref.
~\cite{natoli03}).

This calculation (and other similar along the same line) were performed without
any adjustable parameter. In all cases
the number of atoms forming the cluster is about 140-150, lying inside a sphere of
about 7-8 \AA, enough to obtain spectral convergence in the presence of the complex
part of the potential. The charge density was obtained by superposition of neutral
atom charge densities, from which the Coulomb and the exchange-correlation potential
are calculated. In the case of close-packed structure, this fact should not be
an handicap.

By contour integral of the Green's Function over the energy range of the valence
sates, the Fermi Energy was determined to be around -10 eV with respect to
$E_0 =0$,  {\it i.e.} the zero of the cluster potential at infinity.
It serves to define the local
momentum of the photo-electron in the calculation of the HL (DH) potential, but
the calculated spectra are rather insensitive to small variations of this quantity
by 1-2 eV. No self-consistence loop was attempted to find a self-consistent charge.
The core hole width was taken into account by adding 0.7 eV to the complex part
of the potential.

Surprisingly enough, the comparison shows that the DH potential gives overall
better agreement with the experiments than the HL one. A similar situation is
found for other transition metals and has been reported elsewhere.
~\cite{hatadacs09}
Notice that the same conclusion was drawn in Ref. \cite{hatada07b} for
$Cu_2MnM$, where $M=Al,Sn,In$, although in the MT approximation.

\section{Conclusions}
\label{conc}

We have developed a FP-MS scheme which is a straightforward
generalization of the usual theory with MT potentials and implemented
the code to calculate cross sections for several spectroscopies, like
absorption, photo-electron diffraction and anomalous scattering, as well as
bound states, by a simple analytical continuation.
The key point in this approach is the generation of the cell solutions
$\Phi_{ L }\, (\, { \bf r } \,)$ for a general truncated potential free of the
well known convergence problems of AM expansion together with an alternative
derivation of the MSE which allows us to treat the matrices $S$ and $E$ as
square, with only one truncation parameter, given by the classical relation
$l_{\rm max} \sim k \, R_b$. The fact that the theory can work with square
$S$ and $E$ matrices is of the utmost importance, since this feature allows
the definition of the cell $T$ matrix and its inverse, recuperating in such
way the possibility to define the Green's function and to treat a host of
problems, ranging from solids with reduced symmetry to randomly disordered
alloys in the context of the CPA, as mentioned in the introduction. In
this way one can also show that the wave function and the Green's function
approach provide the same expression for the absorption cross section
for continuum states and real potentials, through the application of the
generalized optical theorem (see  \ref{a5}). For transitions to bound
states the two methods are not equivalent, due to the different normalization
of continuum and bound states, unless one normalizes to one the wave function
for these latter. However this procedure, although feasible, is rather
cumbersome (this was one of the reasons for abandoning the MS method in
favor of the simpler linearized methods in band structure calculations).
Instead, the Green's function
expression for the cross section Eq. (\ref{gfabs}) can always be used, since it gives
the correct normalization in both cases simply by analytical continuation.
We have exploited this fact when calculating the cross section for the
$Se_2$ and $Br_2$ diatomic molecules.

Moreover, in the present paper we have been able to show that the FP-MST
converges absolutely in the $l_{\rm max} \rightarrow \infty$ limit (modulo a
slight modification of the free propagator matrix $G$ which is practically
unnecessary) in the sense that the
scattering path operator of the theory can be found in terms of an absolutely
convergent procedure in this limit. We have thus given a firm ground to its
use as a viable method for electronic structure calculation and at the same
time have provided a straightforward extension of MST in the
Muffin-Tin (MT) approximation for the calculation of x-ray spectroscopies.
Also Quantum Chemistry calculations might benefit from this method in that
it avoids the use of basis functions sets.

Finally it is worth mentioning that in giving a new scheme to generate
local basis functions for truncated potential cells, we have provided
an efficient and fast method for solving numerically a partial differential
equation of the elliptic type in polar coordinates, which can also be used to
solve the Poisson equation in the whole space by the partitioning method.

\section*{Acknowledgements}

%We gratefully acknowledge long and illuminating discussions with Dr. Peter
%Kruger.
We gratefully acknowledge Dr. Peter Kr\"uger for long and
illuminating discussions. We also thank Prof. Isao Tanaka and Dr.
Teruyasu Mizoguchi for drawing our attention to the problem of
$\alpha$-quartz (SiO$_2$). C. R. Natoli acknowledges a financial
support from DGA (Diputaci\'on General de Arag\'on) in the framework
of the promotion action for researcher mobility. This work has been
accomplished in the framework and with the support of the European
Network LightNet.

\appendix

%%%%%%%%%
\section{ The Mathieu functions }
\label{a1}

For the convenience of the reader we give here a brief account of the Mathieu
functions. The solution of the 3-dimensional Mathieu's equation
(\ref{3dmathieu}) of the text
%
%\be
%  \left[ \frac{ d^2 }{ d x^2 } + \frac{ d^2 }{ d y^2 } + \frac{ d^2 }{ d z^2 }
%  \right] \psi ( x, y, z ) = ( - a_x - a_y - a_z  + 2 q_x \cos 2 x +
%  2 q_y \cos 2 y + 2 q_z \cos 2 z ) \psi ( x, y, z )
%\ee
%
is obtained by separation of variables
\be
  \psi ( x, y, z ) = f_x ( x ) f_y ( y ) f_z ( z ).
\ee
in terms of functions $f$ solutions of the one-dimensional Mathieu's equation
\cite{abramowitz72}
\be
  \frac{ d^2 f ( r ) }{ d r^2 } = ( - a + 2 q \cos 2 r ) f ( r )
  \label {eqn:mathieu}
\ee
A solution of Eq. ( \ref{eqn:mathieu} ) having period $ \pi $ or $ 2 \pi $
is  of the form,
\be
  f ( r ) = \sum_{ m = 0 }^{ \infty } ( A_m \cos m r + B_m \sin m r )
\ee
where $ B_0 $ can be taken as zero. If the above expression is substituted
into Eq. ( \ref{eqn:mathieu} ) one obtains
\bea
  &&\sum_{ m = - 2 }^{ \infty } [ ( a - m^2 ) A_m -
    q ( A_{ m - 2 } + A_{ m + 2 } ) ] \cos m r \nr
  &&+ \sum_{ m = - 1 }^{ \infty } [ ( a - m^2 ) B_m -
    q ( B_{ m - 2 } + B_{ m + 2 } ) ] \sin m r = 0  \label {eqn:matrecur}
\eea
with $ A_{ - m } =  B_{ - m } = 0 \quad {\rm if} \quad  m > 0$.
Eq. ( \ref{eqn:matrecur} ) can be reduced to one of four simpler types,
\bea
  && f_0 ( r ) = \sum_{ m = 0 }^{ \infty } A_{ 2 m + p }
    \cos ( 2 m + p ) r, \,\,\,\,\, p = 0 \,\, { \rm or } \,\, 1
\label{eqn:matsol} \\
  && f_1 ( r ) = \sum_{ m = 0 }^{ \infty } B_{ 2 m + p }
    \sin ( 2 m + p ) r, \,\,\,\,\, p = 0 \,\, { \rm or } \,\, 1.
  \label {eqn:matsol1}
\eea
If $ p = 0 $, the solution is of period $ \pi $; if $ p = 1 $,
the solution is of period $ 2 \pi $. $ f_0 $ is an even solution,
and $ f_1 $ is an odd solution.
The recurrence relations among the coefficients of these basic solutions are
easily obtained from the general relations Eq. ( \ref{eqn:matrecur} ).
For even solutions of period $ \pi $ we find
\bea
  && a A_0 - q A_2 = 0 \\
  && ( a - 4 ) A_2 - q ( 2 A_0 + A_4 ) = 0 \\
  && ( a - m^2 ) A_m - q ( A_{ m - 2 } + A_{ m + 2 } ) = 0,
     \,\,\,\,\,\,\,\,  m \ge 3
\eea
and of period $ 2 \pi $,
\bea
  && ( a - 1 ) A_1 - q ( A_1 + A_3 ) = 0 \\
  && ( a - m^2 ) A_m - q ( A_{ m - 2 } + A_{ m + 2 } ) = 0,
     \,\,\,\,\,\,\,\,  m \ge 3.
\eea
For odd solutions of period $ \pi $,
\bea
  && ( a - 4 ) B_2 - q A_4 = 0 \\
  && ( a - m^2 ) B_m - q ( B_{ m - 2 } + B_{ m + 2 } ) = 0,
     \,\,\,\,\,\,\,\,  m \ge 3
\eea
whereas for period $ 2 \pi $,
\bea
  && ( a - 1 ) B_1 + q ( B_1 - B_3 ) = 0 \\
  && ( a - m^2 ) B_m - q ( B_{ m - 2 } + B_{ m + 2 } ) = 0,
     \,\,\,\,\,\,\,\,  m \ge 3
\eea
It is convenient to separate the characteristic values $a$ into two major
subsets:
\bea
a & = & a_r, \;\; {\rm associated \,\, with \,\, even \,\, periodic \,\,
solutions} \nr
a & = & b_r, \;\; {\rm associated \,\, with \,\, odd \,\, periodic \,\,
solutions}
\nonumber
\eea
where $r$ describes the index of the eigenstate.
Table \ref{tab:egv} gives the first three eigenvalues associated to even
periodic solutions and the first two associated to odd periodic solutions
($b_0=0$), for some selected values of $q$. They can serve to generate the
Mathieu functions using the above recurrence relations to determine the
coefficients in the solutions (\ref{eqn:matsol}) and (\ref{eqn:matsol1}).

\begin{table}[t]
  \caption{First few Eigenvalues of Mathieu functions for different $q$ values}
  \begin{center}
%    \begin{tabular}{|p{20mm}|c|c|c|c|}
    \begin{tabular}{|p{9.5mm}|p{29mm}|p{25mm}|p{25mm}|p{25mm}|}
      \hline
        {\footnotesize parity} &\multicolumn{2}{|c|}{{\footnotesize even}}&\multicolumn{2}{|c|}{{\footnotesize odd}}\\
      \hline
        {\footnotesize period} & {\footnotesize $\pi$} & {\footnotesize $2\pi$} & {\footnotesize $\pi$} & {\footnotesize $2\pi$} \\
      \hline
        {\footnotesize q=0.01} & {\footnotesize $a_0$=-4.99995$\times 10^{-6}$} &  {\footnotesize $a_1=1.00999$} &  {\footnotesize $b_2=3.99999$} &  {\footnotesize $b_1=0.989988$} \\
        &  {\footnotesize $a_2=4.00004$ } &  &  &  \\
      \hline
        {\footnotesize q=0.02} &  {\footnotesize $a_0$=-1.99991$\times 10^{-5}$} &  {\footnotesize $a_1=1.01995$} & {\footnotesize $b_2=3.99997$} &  {\footnotesize $b_1=0.97995$} \\
        &  {\footnotesize $a_2=4.00017$} &  &  &  \\
      \hline
        {\footnotesize q=0.03} & {\footnotesize $a_0$=-4.49956$\times 10^{-5}$} &  {\footnotesize $a_1=1.02989$} &  {\footnotesize $b_2=3.99993$} & {\footnotesize $b_1=0.969888$} \\
        &  {\footnotesize $a_2=4.00037$} &  &  &  \\
      \hline
        {\footnotesize q=0.04} & {\footnotesize $a_0$=-7.9986$\times 10^{-4}$} & {\footnotesize $a_1=1.0398$} &  {\footnotesize $b_2=3.99987$} & {\footnotesize $b_1=0.959801$} \\
        &  {\footnotesize $a_2=4.00067$} &  &  &  \\
      \hline
        {\footnotesize q=0.05} & {\footnotesize $a_0$=-1.24966$\times 10^{-3}$} &  {\footnotesize $a_1=1.04969$} &  {\footnotesize $b_2=3.99979$} &  {\footnotesize $b_1=0.949689$ }\\
        &  {\footnotesize $a_2=4.00104$} &  &  &  \\
      \hline
        {\footnotesize q=0.1} & {\footnotesize $a_0$=-4.99454$\times 10^{-3}$} &  {\footnotesize $a_1=1.09873$ } &  {\footnotesize $b_2=3.99917$} &  {\footnotesize $b_1=0.898766$} \\
        &  {\footnotesize $a_2=4.00416$} &  &  &  \\
      \hline
        {\footnotesize q=0.2} & {\footnotesize $a_0$=-1.99133$\times 10^{-2}$} &  {\footnotesize $a_1=1.19487$} &  {\footnotesize$b_2=3.99667$} &  {\footnotesize $b_1=0.795124$} \\
        &  {\footnotesize $a_2=4.01658$} &  &  &  \\
      \hline
        {\footnotesize q=0.3} & {\footnotesize $a_0$=-4.4566$\times 10^{-2}$} &  {\footnotesize $a_1=1.28832$} &  {\footnotesize $b_2=3.9925$} &  {\footnotesize $b_1=0.689166$} \\
        &  {\footnotesize $a_2=4.03706$} &  &  &  \\
      \hline
        {\footnotesize q=1} & {\footnotesize $a_0$=-0.455139} &  {\footnotesize $a_1=1.85911$} &  {\footnotesize $b_2=3.91702$} &  {\footnotesize $b_1=-0.110249$} \\
        & {\footnotesize $a_2=4.3713$} &  &  &  \\
      \hline
        {\footnotesize q=2} & {\footnotesize $a_0$=-1.51396} &  {\footnotesize $a_1=2.3792$} &  {\footnotesize$b_2=3.67223$} &  {\footnotesize $b_1=-1.39068$} \\
        &  {\footnotesize $a_2=5.17267$} &  &  &  \\
      \hline
        {\footnotesize q=5} & {\footnotesize $a_0$=-5.80005} &  {\footnotesize $a_1=1.85819$} &  {\footnotesize $b_2=2.09946$} &  {\footnotesize $b_1=-5.79008$} \\
        &  {\footnotesize $a_2=7.44911$} &  &  &  \\
      \hline
        {\footnotesize q=10} & {\footnotesize $a_0$=-13.937} &  {\footnotesize $a_1=-2.39914$} &  {\footnotesize $b_2=-2.38216$} &  {\footnotesize $b_1=-13.9366$} \\
        &  {\footnotesize $a_2=7.71737$} &  &  &  \\
      \hline
    \end{tabular}
  \end{center}
  \label{tab:egv}
\end{table}
%

%%%%%%%%%

\section{Asymptotic behavior of KKR Structure Factors }
\label{a2}

For $\nu \rightarrow \infty$ through real positive numbers (in practice
for $\nu \gg |z|$), the other variables being fixed, one has
~\cite{abramowitz72}
\bea
J_{\nu} & \approx & \bigg(\frac{1}{2\pi\nu}\bigg)^{1/2} \,
\bigg(\frac{ez}{2\nu}\bigg)^{\nu}  ; \qquad \nr
& & -iH_{\nu}^{\pm}  \approx  \pm \bigg(\frac{2}{\pi\nu}\bigg)^{1/2} \,
\bigg(\frac{ez}{2\nu}\bigg)^{-\nu}
\eea
where $e$ is the Neper number. Remembering that
\bea
j_n (z) & = & \sqrt{\frac{\pi}{2z}} \, J_{n+1/2}(z) ; \qquad \nr
& & h^{\pm}_n(z) = \sqrt{\frac{\pi}{2z}} \, H_{n+1/2}^{\pm}(z)
\eea
we find for the asymptotic behavior of the spherical Bessel and Hankel functions
\bea
j_n (z) & \approx & \frac{z^n}{\sqrt{2}} \, e^{n+1/2}
\,\bigg(\frac{1}{2n+1}\bigg)^{n+1} ; \qquad \nr
& & -ih^{\pm}_n(z) \approx \frac {\sqrt{2}}{z^{n+1}} \frac{1}{e^{n+1/2}}(2n+1)^n
\label{asexp}
\eea

We need to find un upper limit for $G^{ij}_{LL'}$ given by
\be
G^{ij}_{LL'} = -4\pi i k \sum_{L''} \, i^{l - l' + l''} \, C(L,L';L'')\,
h^+_{l''}(\rho)\, Y_{L''}(\hat{\rho})
\ee
where $\rho = kR_{ij}$, $\hat{\rho} = \hat{R_{ij}}$ and $C(L,L';L'')$ are
the Gaunt coefficients. To establish an upper limit for this expression
when $l$ is fixed and $l' \gg \rho$ we replace each $|h^+_{l''}(\rho)|$ in
the sum by its maximum value $|h^+_{l+l'}(\rho)|$, use the asymptotic
value in Eq. (\ref{asexp}) and the relation $\sum_{L''} C(L,L';L'')\,
Y_{L''}(\hat{\rho}) = Y_{L}(\hat{\rho}) Y_{L'}(\hat{\rho})$ to obtain
\bea
|G^{ij}_{LL'}| & \le & 4\pi k |h^+_{l+l'}(\rho)| \, \sum_{L''} |C(L,L';L'')\,
                             Y_{L''}(\hat{\rho})|  \nr
               &\approx& 4\pi k |h^+_{l+l'}(\rho)| \, |\sum_{L''} C(L,L';L'')\,
                             Y_{L''}(\hat{\rho})|  \nr
    &=& 4\pi k |h^+_{l+l'}(\rho)| \, |Y_{L}(\hat{\rho}) Y_{L'}(\hat{\rho})| \nr
               & \le & [(2l+1)(2l'+1)]^{1/2}
 \frac {\sqrt{2}}{\rho^{l+l'+1}} \frac{k}{e^{l+l'+1/2}}[2(l+l')+1]^{l+l'}
\label{gineq}
\eea
since $|Y_{L}(\hat{\rho})| \le \sqrt{(2l+1)/(4\pi)}$. Notice that the
approximation $\sum_{L''} |C(L,L';L'') Y_{L''}(\hat{\rho})| \approx
|\sum_{L''} C(L,L';L'')\, Y_{L''}(\hat{\rho})|$ entails only errors
$O(1)$ in all $l$ variables, as can be verified by explicit calculation, and
therefore completely negligible with respect to the power behavior of the rest
of the factors. In any case, since
$\sum_{L''} |C(L,L';L'') Y_{L''}(\hat{\rho})| \le [(2l+1)(2l'+1)]^{1/2} /(4\pi)
\sum_{L''} (2l'' + 1)$, at the cost of introducing a non influential extra factor
$[2(l+l')+1]^2$ in Eq. (\ref{gineq}) we would get a rigorous inequality.
This expression is obviously also valid for $l \gg \rho$.

Under the same conditions, assuming $l' \gg l$ we derive
\bea
|J^{ij}_{LL'}|
    & \le & 4\pi |j_{l'-l}(\rho)| \, |Y_{L}(\hat{\rho}) Y_{L'}(\hat{\rho})|
\nr
               & \le & [(2l+1)(2l'+1)]^{1/2} \,
 \frac {\rho^{l'-l}}{\sqrt{2}} \, e^{l'-l+1/2} \,
\frac{1}{[2(l'-l)+1]^{l'-l+1}}
\label{jineq}
\eea

The inequalities Eqs. (\ref{gineq}) and (\ref{jineq}) can be used to obtain
other useful inequalities used throughout the paper. For example, for fixed
$l$, using again Eq. (\ref{asexp}), one obtains
\be
|G^{i j}_{L L'} J_{L'} ( { \bf r }_j )| \le
\left(\frac{r_j}{R_{ij}}\right)^{l'} \, \frac{k}{(kR_{ij})^{l+1}} [2(l'+l)+1]^l \,
\sqrt{\frac{2l+1}{4\pi}} \, e^{-l}
\label{gjineq}
\ee
implying that the series $\tilde{H}_L^{ + } ( { \bf r }_i ) =
\sum_{L'} G^{i j}_{L L'} J_{L'} ( { \bf r }_j )$ is absolutely and uniformly
convergent in the angular domain. The uniform convergence comes from the
application of the Weierstrass criterion (see Ref.~\cite{whittaker65},
sect. 3.34, pag. 49).

Similarly one finds
\be
|J^{i o}_{L L'} \tilde{H}_{L'}^+ ( { \bf r }_o )| \le
\left( \frac{R_{io}}{r_o} \right)^{l'+1} \frac{k}{(kR_{io})^{l+1}}\;
[2(l'-l)+1]^l \, \sqrt{\frac{2l+1}{4\pi}} \, e^{-l}
\label{jhineq}
\ee
showing that the series $\tilde{H}_L^{ + } ( { \bf r }_i; \kappa  ) =
\sum_{L'} J^{i o}_{L L'} \tilde{H}_{L'}^{ + } ( { \bf r }_o; \kappa  )$
 is also absolutely and uniformly convergent if $(r_o > R_{io})$.

\noindent Along the same lines we can estimate an upper bound for the atomic
T-matrix for $l,\,l' \gg kR_b$. We find from Eq. (\ref{tomdef}) to first
order
\bea
T_{LL'} & = &  \int_0^{R_b} \, J_{L'}({\bf r}) \, V({\bf r}) \,
              \psi_L({\bf r}) \, {\rm d^3r} \nr
        & = &  \sum_{L''L'''} C(L',L'';L''') \,
              \int_0^{R_b} r^2 \, j_{l'}(kr)
              \, V_{L'''}(r) \, R_{L''L}(r) \, {\rm dr} \nr
        & \approx &  \sum_{L''} C(L',L;L'') \,
                \int_0^{R_b} r^2 \, j_{l'}(kr)
              \, V_{L''}(r) \, j_l(kr) \, {\rm dr}
\label{tmat}
\eea
where the last step follows from the fact that
under the above assumptions $R_{L'L} \approx \, j_{l} \delta_{LL'}$.
Taking into account that
$C(L',L;L'') \approx 1/\sqrt{4\pi}\, O(1)$ for all L-values and using again
Eq. (\ref{asexp}) we obtain
\bea
|T_{LL'}| &\le& 4ll'  \, \int_0^{R_b} r^2 \, |j_{l'}(kr)|
              \, |V_{|l-l'|}(r)| \, |j_l(kr)| \, {\rm dr} \nr
          &\approx& 4ll' k^{l+l'} \, \frac{e^{l+l'+1}}
                    {(2l+1)^{l+1} \, (2l'+1)^{l'+1}} \nr
          & & \times \int_0^{R_b} r^{l+l'+2} \,  \, |V_{|l-l'|}(r)| \,  \, {\rm dr} \nr
          &\le& \frac{Z_{eff}}{k^2} \, \frac{4ll'}{l+l'+2} \,
           \frac{(kR_b)^{l+l'+2} \, e^{l+l'+1}}{(2l+1)^{l+1} \, (2l'+1)^{l'+1}}
\label{tmat1}
\eea
with the understanding that $V_{l} \equiv V_{l0}$, assuming that
$|V_{l}(r)| \le 2Z_{eff}/r$ in atomic units and that $|V_{l}(r)|$ is
decreasing with $l$.

\noindent Based on the above inequalities we easily obtain
\bea
|G^{i j}_{L L'}T_{LL'}| & \le & 8\sqrt{2} e^{1/2} Z_{eff} R_b
\frac{(ll')^{3/2}}{l+l'+2} \, \nr
 &   & \times \left( \frac{R_b}{R_{ij}}\right)^{l+l'+1}
\frac{(2l+2l'+1)^{l+l'}}{(2l+1)^{l+1} \, (2l'+1)^{l'+1}}
\label{gtineq}
\eea
Specializing to the case where $l$ is fixed and $l'$ is running, we also
find
\bea
|G^{i j}_{L L'}T_{L'L'}| & \le & 4\frac{(ll')^{1/2}l'}{l'+1} Z_{eff} \, \nr
 &     & \times \frac{(kR_b)^{2l'+2}}{(kR_{ij})^{l+l'+1}} \, \frac{e^{2l'+1}}{e^{l+l'+1/2}}
\frac{(2l+2l'+1)^{l+l'}}{(2l'+1)^{2l'+1}}
\label{gtineq1}
\eea
which is useful in discussing questions related to the convergence of MST.

Finally we note that all the above inequalities and convergence conditions
remain valid for complex arguments $\rho$, provided it is replaced by its
module $|\rho|$.

\section{Surface identity for scattering states}
\label{a3}

In the case of short range potentials ({\it i.e.} potentials that behave like
$1/r^{1+\epsilon}$ with positive $\epsilon$ as $r \rightarrow \infty$) the
Lippmann-Schwinger equation for scattering states at energy $ E = k^2$
\bea
  \psi ( {\bf r}; {\bf k}  ) = \phi_0 ( {\bf r}; {\bf k} ) +
  \int d{\bf r}' G_0 ( {\bf r} - {\bf r}'; k ) V ( {\bf r}' )
                \psi ( {\bf r}'; {\bf k} )
\label{lsequ}
\eea
is a consequence of the Schr\"odinger equation
\bea
  ( \nabla^2 + E - V ( {\bf r} ) ) \psi ( {\bf r}; {\bf k} ) = 0
  \label{schre}
\eea
together with the relations ($\phi_0 ( {\bf r}; {\bf k} ) \equiv {\rm e}^{{\rm
i}{\bf k} \cdot {\bf r}}$)
\bea
  && ( \nabla^2 + E ) \phi_0 ( {\bf r}; {\bf k} ) = 0  \label{frqu1} \\
  && ( \nabla^2 + E )  G_0 ( {\bf r} - {\bf r}'; k  ) =
     \delta ( {\bf r} - {\bf r}' )
\label{frqu2}
\eea

Starting from Eq. (\ref{lsequ}), we derive the identity
\bea
  \int_{\Omega} d{\bf r}'\left[  G_0 ( {\bf r} - {\bf r}'; k )
                 V ( {\bf r}' )  - \delta ( {\bf r} - {\bf r}' ) \right]
                 \psi ( {\bf r}'; {\bf k} ) = -
                 \phi_0 ( {\bf r}; {\bf k} )
\eea
where $\Omega$ indicates the whole space.
Using  Eq. (\ref{frqu2}) to replace the delta function, and the Schr\"odinger
equation (\ref {schre}) to eliminate $V ( {\bf r}' )$ we obtain
\bea
   \sum_{j=1}^{N+1} &  & \int_{\Omega_j}  \{ G_0 ( {\bf r} - {\bf r}'; k )
                        (\nabla^2 + E) \psi ( {\bf r}'; {\bf k} ) \nr
   & &     - \psi ( {\bf r}'; {\bf k} )
                       (\nabla^2 + E) G_0 ( {\bf r} - {\bf r}'; k )\}
                d{\bf r}_j' = - \phi_0 ( {\bf r}; {\bf k} )
  \nonumber
\eea
where we have decomposed the whole space as $\Omega =
\sum_{j=1}^{N+1} \Omega_j$, such that $\Omega_{N+1} \equiv \Omega_o =
\mathcal{C} \sum_{j=1}^{N} \Omega_j $.

Transforming to surface integrals by application of the Green's theorem
\bea
  \sum_{j=1}^{N+1} & & \int_{S_j}  \left[ G_0 ( {\bf r} - {\bf r}'; k )
   \nabla \psi ( {\bf r}'; {\bf k} ) \right. \nr
   & & \left. - \psi ( {\bf r}'; {\bf k} ) \nabla G_0 ( {\bf r} - {\bf r}'; k ) \right]
     \cdot {\bf n}_j' d \sigma_j' = - \phi_0 ( {\bf r}; {\bf k} )
  \label{surfint}
\eea

We now observe that the surface integral over the surface $S_{N+1} $
of the volume $\Omega_{N+1} \equiv \Omega_o$ has two contributions, one
coming from the surface $S_o$ of $\sum_{j=1}^{N} \Omega_j$, the other one
$S_o^{\infty}$ at infinity, as the limit as $R \rightarrow \infty$ over the
surface of a sphere $S_o^{R}$, of radius $R$. This latter is easily
calculated on the basis of the asymptotic behavior of
$\psi ( {\bf r}; {\bf k})$ in Eq. (\ref {wfexpo}) and the expansion
(\ref {gfexp2}) and gives exactly $- \phi_0 ( {\bf r}; {\bf k} )$,
canceling the rhs term in Eq. (\ref {surfint}).
Therefore we recover the identity (\ref {one}) of Section~\ref{scatstat}
\bea
  & & \sum_{j=1}^N \int_{S_j}  \{ G_0 ({\bf r} - {\bf r}', k )
                        \nabla \psi ( {\bf r}'; {\bf k} )  \nr
  & & {\hspace*{22mm}} - \psi ( {\bf r}'; {\bf k} )
                       \nabla G_0 ( {\bf r} - {\bf r}'; k )\}
                \cdot {\bf n}_j' d \sigma_j'  \nr
  & & = \int_{S_o}  \{ G_0 ({\bf r} - {\bf r}', k )
                        \nabla \psi ( {\bf r}'; {\bf k} ) \nr
  & & {\hspace*{22mm}} - \psi ( {\bf r}'; {\bf k} )
                       \nabla G_0 ( {\bf r} - {\bf r}'; k )\}
                \cdot {\bf n}_O' d \sigma_O'
\eea

\section{The Generalized Optical Theorem}
\label{a4}

For convenience of the reader we give here a proof of Eq. (\ref {got})
in the case where $\overline{T}^o \equiv 0$, {\it i.e.} when
empty cells cover the volume $\Omega_o$ up to the point at which the
asymptotic behavior in Eq. (\ref{wfexpo}) begins to be valid. We start
by observing that
\be
\int {d\hat{\bf k}}  \, I_{L'}^i({\bf k}) \, \left[ I_L^j({\bf k})
\right]^{\ast} = J_{L L'}^{i j} \, \frac{k}{\pi}
\ee
so that, using the relation Eq. (\ref{bsca}), we find
\be
\int {\rm d} \hat{\bf k} \, B^{i}_{L'}({\bf k}) \,
\left[ B^{j}_{L}({\bf k}) \right]^{\ast} = \sum_{L \Lambda} \,
\tau_{L \Lambda'}^{i m} \, J_{\Lambda \Lambda'}^{m n} \,
(\tau_{\Lambda' L'}^{n i})^{\ast} \,
\frac{k}{\pi}
\label{got1}
\ee
where we have used the symmetry of $\tau$. Based on the relations
Eqs. (\ref{csdef}), (\ref{kdefj}) and (\ref{gdec}), valid at any energy,
%\bea
%(T^j)^{-1} = (K^j)^{-1} + i &=& - C^j (S^j)^{-1} + i \nr
%G^{i j}_{L L'} &=& N^{i j}_{L L'} - i J^{i j}_{L L'}
%\eea
and due to the reality of the matrices $K$, $N$ and $J$ for real potential,
we can write
\be
\tau = k^{-1} \, \left[ K - N + iJ \right]^{-1}
\ee
so that the rhs of Eq. (\ref{got1}) becomes
\be
\frac{k}{\pi} \, \left\{ \tau \, J \, \tau \right\}_{L L'}^{i j}=
\frac{1}{\pi} \, \frac{1}{2i}
\left\{ \tau^{\ast} - \tau \right\}_{L L'}^{i j} = - \frac{1}{\pi} \,
\Im \tau_{L L'}^{i j}
\ee
in keeping with Eq. (\ref {got}).

\section{Wave function and GF equivalence for absorption cross section}
\label{a5}

In the independent electron approximation, the core level photoelectron
diffraction (PED) cross-section for the ejection of a photoelectron along
the direction $\hat{\bf k}$ and energy $E = k^2$ from an atom situated at
site $i$ is given by~\cite{sebilleau06}
\be
\frac {d\sigma}{d\hat{\bf k}} = 8 \, \pi^2 \, \alpha \, \hbar \, \omega \,
\sum_{m_c} \left|\<\Theta \psi ( { \bf r }_i;{\bf k})
|{\hat \varepsilon} \cdot { \bf r }_i|
\phi^c_{L_c}({ \bf r }_i)\> \right|^2
\ee
Here $\Theta$ is the time-reversal operator, $\hat {\varepsilon}$ the
polarization
of the incident photon and $\phi^c_{L_c}({ \bf r }_i)$ the initial core
state of angular momentum $L_c$ (we neglect for simplicity the spin-orbit
coupling, which can be easily taken into account). Due to the localization
of the core state, we need only the expression of the continuum scattering
state in the cell of the photoabsorber, given by
\be
\psi ( { \bf r }_i;{\bf k}) = \sum_L B^{i}_{L}({\bf k})
\overline{\Phi}_L ( { \bf r }_i )
\ee
so that
\be
\frac {d\sigma}{d\hat{\bf k}} = 8 \, \pi^2 \, \alpha \, \hbar \, \omega \,
\sum_{m_c} \left | \sum_L M_{L_c L} (E) \, B^i_L ({\bf k}) \right |^2
\ee
where $B^i_L ({\bf k})$ is given by Eq. (\ref{bsca}) and we have defined the
atomic transition matrix element
\be
 M_{ L_c \, L } \, ( E ) \, = \, \int_{{\Omega}_i} d \, { \bf r} \,
 \phi^c_{ L_c } ( { \bf r } ) \, { \hat \varepsilon } \cdot { \bf r } \,
 \overline{\Phi}_L ( { \bf r } )
\ee

The total absorption cross-section, in the case of real potentials, is
obtained by integrating the PED cross-section over all directions of
photoemission
\bea
\int {d\hat{\bf k}} \, \frac {d\sigma}{d\hat{\bf k}} & = &
8 \, \pi^2 \, \alpha \, \hbar \, \omega \, \sum_{m_c}
 \int {d\hat{\bf k}} \, \left | \sum_L M_{L_c L} (E) \, B^i_L ({\bf k})
\right |^2 \nr
& = & - 8 \, \pi \, \alpha \, \hbar \, \omega \, \sum_{m_c} \sum_{L L'} \,
M_{ L_c \, L } \, ( E ) \Im \tau_{L L'}^{i i} M_{ L_c \, L' }
\label{absgot}
\eea
by application of the optical theorem (\ref{got}). This is exactly the form
that one would obtain starting from Eq. (\ref{gfabs}) and using the expression
(\ref{gfeq3}) for the GF.

\section{Exploitation of point symmetry}
\label{a6}

In a cluster (to which we shall also refer as a molecule), point symmetry
can be used to advantage to simplify the problem and reduce the size of the MS
matrix. Specifically we consider the case where the cluster remains invariant
under a finite group of transformations $\mathcal G$ relative to the molecular
center $R_o$. This group has a finite number of finite-dimensional irreducible
representations (irreps) $\Gamma_j$ ($j = 1, 2, ..., g$). Due to the
symmetry, the cluster will consist of $\mathcal P$ groups of equivalent
atoms, transforming into one another under the operations of the group; and
for each group $p = 1, 2, ...., \mathcal P$, there are $N_p$ atoms labeled by
$i_p$. If $N$ is the total number of atoms in the cluster, then
$N = \sum_{p=1}^{\mathcal P} N_p$.

Under these assumptions there exists a unitary transformation $\mathcal C$
that block-diagonalizes the MS matrix according to the irreps. For each
value of the angular momentum index $l$, this transformation is labeled by
the angular projection $m$ and the site $i_p$ on one side, and on the other
the irrep, the row $\rho$ of the irrep and a further index $n$ that
distinguishes independent orthogonal symmetrized basis functions with the same
$l$. The matrix elements of $\mathcal C$ are easily obtained by applying
the projection operator $\sum_R M_{\rho \rho}^{\Gamma_j} (R) \, P_R$ to a
spherical harmonic function
$Y_{ l m } ( \, \hat {\bf r}_{ i_p }) \, \theta (R_b - r_{i_p})$
centered on the site $i_p$ and defined on the
surface of the bounding sphere $R_b$ of the cell $\Omega_{i_p}$
Here $P_R$ is the generic operation belonging to the group
$\mathcal G$ corresponding to the coordinate transformation $R$, and $M_{\rho
\rho}^{\Gamma_j} (R)$ is the matrix element corresponding to $R$ in the matrix
representation of irrep $\Gamma_j$ of the group, $\rho$ labeling the row of
the irrep. As usual in group theory~\cite{tinkham64} the effect of $P_R$ on
the function $f({\bf r})$ is given by $f(R^{-1}\, {\bf r})$. In this way,
if the result is not zero, one generates a symmetrized spherical harmonic
function given by
\bea
K_{l n}^{ \Gamma_j^{\rho},\, p} (\,\hat {\bf r}_p \, ) &\equiv& \sum_R
M_{\rho \rho}^{\Gamma_j} (R) \, P_R \, Y_{ l m } (\,\hat {\bf r}_{ i_p}) \nr
&=& \sum_R \, M_{\rho \rho}^{\Gamma_j} (R) \, \sum_{\mu} D^l_{\mu m} (R) \,
    Y_{ l \mu } (\,\hat {\bf r}_{ i_p'}) \nr
&=& \sum_{m, i_p} C_{l n, \, m }^{ \Gamma_j^{\rho}, \, { i_p } } \,
    Y_{ l m } ( \, \hat {\bf r}_{ i_p } \, )
\eea
where  $D^l_{\mu m} (R)$ is the Wigner rotation matrix corresponding to the
transformation $R$~\cite{tinkham64} and $i_p' = R \, i_p$.
Due to the orthogonality of the basis functions
\bea
\int \, Y_{ l m } (\,\hat {\bf r}_{ i_p}) \,
Y_{ l m' } (\,\hat {\bf r}_{ i_p'}) \,{\rm d}\Omega
&=& \delta_{m m'} \,\delta_{i_p i_p'} \nr
\int  \, K_{l n}^{ \Gamma_j^{\rho},\, p} (\,\hat {\bf r}_p \, )
K_{l' n'}^{ \Gamma_{j'}^{\rho'},\, p} (\,\hat {\bf r}_p \, ) \,{\rm d}\Omega
&=& \delta_{l l'} \, \delta_{n n'} \, \delta_{\Gamma_j \Gamma_{j'}} \,
\delta_{\rho \rho'}
\eea
we obtain ${\mathcal C}\, \tilde{\mathcal C} = \tilde{\mathcal C}\,
{\mathcal C} = I$, {\it i.e.}
\bea
\sum_{m, \, i_p} C_{l n, \, m }^{ \Gamma^{\rho}, \, { i_p } } \,
              C_{l' n', \, m }^{ \Gamma'^{\rho'}, \, { i_p } } &=&
     \delta_{l l'} \, \delta_{n n'} \, \delta_{\Gamma \Gamma'} \,
\delta_{\rho \rho'} \nr
\sum_{\Gamma \rho} \sum_{n} C_{l n, \, m }^{ \Gamma^{\rho}, \, i_p } \,
C_{l n, \, m' }^{ \Gamma^{\rho}, \, i_p' } &=& \delta_{m m'} \,
\delta_{i_p i_p'}
\eea
%$\Lambda = (l, \, n)$,
where for simplicity we have dropped the index $j$ from the symbol $\Gamma$ of
the irreps.

Now, if $M \equiv M_{L L'}^{i j} \equiv (T^{-1} - G)_{L L'}^{i j}$ is the MS
matrix in the non-symmetrized
site and angular momentum indices, its symmetrized version is given by
$M_s = {\mathcal C} \, M \tilde{\mathcal C}$. Therefore for
any representation $\Gamma$ we have, putting for short $\Lambda = (l, \, n)$
and remembering that $L \equiv (l, m)$,
\bea
   T^{ \Gamma, \, p }_{ \Lambda, \, \Lambda' } &=&
    \sum_{ i_p }^{ N_p } \sum_{ m, \, m' } \,
    C_{ \Lambda, \, L }^{ \Gamma, \, { i_p } } \,
    T^{ i_p }_{ L, \, L' } \,
    C_{ \Lambda', \, L' }^{ \Gamma, \, { i_p } } \label{st} \\
  G^{ \Gamma, \, p \, q }_{ \Lambda, \, \Lambda' } &=&
    \sum_{ i_p }^{ N_p } \sum_{ i_q }^{ N_q } \sum_{ m, \, m' } \,
    C_{ \Lambda, \, L }^{ \Gamma, \, { i_p } } \,
    G^{ i_p \, i_q }_{ L, \, L' } \,
    C_{ \Lambda', \, L' }^{ \Gamma, \, { i_q } } \label{sg}
\eea
Here $T^{ \Gamma, \, p }_s$ describes total scattering power of group $p$ and
$G^{ \Gamma, \, p \, q }_s$ is the symmetrized matrix of the KKR structure
factors. The presence of an outer sphere contribution $J\overline{T}^o J$ is
treated on the same footing and contributes
${\mathcal C}\, J\overline{T}^o J\,\tilde{\mathcal C} \equiv
{\mathcal C}\, J \tilde{\mathcal C}\, {\mathcal C} \,\overline{T}^o \,
\tilde{\mathcal C}\, {\mathcal C}\, J\, \tilde{\mathcal C} \equiv
J_s \, \overline{T}^o_s \, J_s$ for each representation
$\Gamma$. Since the outer sphere is centered at the origin of the cluster,
it has no partner spheres equivalent to itself.

All these matrices are labeled only by the groups of
equivalent atoms (prototypical atoms), the angular momentum $l$ and
possibly the index $n$ mentioned above, realizing a sizable reduction in
dimensions. Notice that, since the molecular hamiltonian is invariant
under the operations of $\mathcal G$, the symmetrized matrix elements do
not depend on the row $\rho$ of the representation $\Gamma$.
Moreover, in order to find the symmetrized
$T$-matrix relative to an equivalent group of atoms, we do not
need to calculate $T_{L L'}^{i_p}$ for all sites in the group, since these
are related to one another by the relation
\be
T_{L L'}^{i_p'} = \sum_{\mu \mu'} \, D^l_{m \mu} (R) \,
T_{l\mu \, l'\mu'}^{i_p}
D^{l'}_{m' \mu'} (R) \label{sti}
\ee
where, as before, $i_p' = R \, i_p$. This relation is a consequence of the
invariance of the potential and the $T$-matrix
$T^{i_p}(\hat {\bf r}, \hat {\bf r}')$ under the operations of the group
$\mathcal G$. Specifically
\bea
V_{L L'}^{i_p} &=& \int  Y_{L} (\hat {\bf r})\, V^{i_p}(r, \hat {\bf r}) \,
Y_{L'} (\hat {\bf r})\, {\rm d}\,\hat {\bf r} \nr
           &=& \int  Y_{L} (\hat {\bf r})\, V^{i_p'}(r, R\, \hat {\bf r})\,
Y_{L'} (\hat {\bf r})\, {\rm d}\,\hat {\bf r} \nr
           &=& \int  Y_{L} (R^{-1} \hat {\bf r})\, V^{i_p'}(r, \hat {\bf r})\,
Y_{L'} (R^{-1} \hat {\bf r})\, {\rm d}\,\hat {\bf r} \nr
           &=& D^l_{\mu m} (R) \, V_{l\mu \, l'\mu'}^{i_p'}\,
               D^{l'}_{\mu' m'} (R) \nonumber
\eea
valid also for the matrix elements
$T_{L L'}^{i_p} = \int  Y_{L} (\hat {\bf r})\,
T^{i_p}(\hat {\bf r}, \hat {\bf r}') \,
Y_{L'} (\hat {\bf r}')\, {\rm d}\,\hat {\bf r} \,{\rm d}\,\hat {\bf r}'$.
Notice that the transformation and symmetrization properties are the same for
$T$ and $T^{-1}$, so we can act directly on this latter.

In the MT case the $T$-matrices are angular momentum diagonal and
$m$ and site independent within a set of equivalent atoms, so that
\bea
  T^{ \Gamma, \, p }_{ \Lambda, \, \Lambda' } &=&
    T^{ p }_{ l } \,
    \sum_{ i_p }^{ N_p } \sum_{ m} \,
    C_{ \Lambda, \, L }^{ \Gamma, \, { i_p } } \,
    C_{ \Lambda', \, L}^{ \Gamma, \, { i_p } } \nr
  &=& T^{ p }_{ l } \,
    \delta_{ \Lambda, \, \Lambda' } \,
\eea
The group $T^p$-matrix can also be calculated directly from the
symmetrization of the radial part of the basis function $R_{L L'}(r_{i_p})$,
which transforms as $T$ in Eqs. (\ref{st}) and (\ref{sti}). Even though the
number and type of operations to perform are exactly the same as for obtaining
$T^p$, in this case there is the added advantage of generating
a symmetrized form of the scattering wave function, needed for example to
calculate the absorption or photo-emission cross-section.

The symmetrization of the local wave function in a group of equivalent
atoms for an irrep $\Gamma$ is obtained by observing that the following
function $\psi({\bf r}_p)$ is invariant under any operation of the
group
\bea
\psi({\bf r}_p) &=&  \, \sum_{i_p} \sum_{L L'} \, A_L^{i_p} \,
R_{L' L}^{i_p}(r_{i_p}) \, Y_{L'} ( \, \hat {\bf r}_{ i_p } \, ) \nr
&\equiv& \<A|R|Y\> \nr
&=& \<A|\tilde{\mathcal C}{\mathcal C} R \tilde{\mathcal C} {\mathcal C} |Y\> \nr
&=& \sum_{\Lambda \Lambda'} A_{\Lambda}^{\Gamma,\, p} \,
R_{\Lambda' \Lambda}^{\Gamma,\, p} (r_p) \,
K_{\Lambda'}^{\Gamma,\, p} ( \, \hat {\bf r}_p \, )
\eea
whereby
\bea
A_{\Lambda}^{\Gamma,\, p} &=& \sum_{m \, i_p} C_{\Lambda, m}^{\Gamma,\, i_p}\,
A_{l m}^{i_p} \nr
R_{\Lambda \Lambda' }^{\Gamma,\, p} (r_p) &=& \sum_{ i_p }^{ N_p }
\sum_{ m, \,m' } \, C_{ \Lambda, \, L }^{ \Gamma, \, { i_p } } \,
    R^{ i_p }_{ L, \, L' } \,(r_p) \,
    C_{ \Lambda', \, L' }^{ \Gamma, \, { i_p } } \nonumber
\eea

Inside the MT sphere
$ X_{\Lambda' \Lambda}^{\Gamma,\, p} (r_p) \equiv r\,
R_{\Lambda' \Lambda}^{\Gamma,\, p} (r_p)$
is solution of the symmetrized equation (\ref{matnmv})
\be
\sum_{\Lambda''} \left[ \left( \frac{d^2}{dr^2} + E
- \frac{ l ( l + 1 ) }{ r^2 } \right) \delta_{\Lambda  \Lambda'' }
- V_{\Lambda  \Lambda''}^p ( r ) \right]
X_{ \Lambda'' \Lambda' }^{\Gamma,\, p} ( r ) = 0
\label{matnmvsym}
\ee
where we have written for simplicity $r$ for $r_p$ and
$V_{\Lambda  \Lambda'' }^p ( r )$ is given by
\be
 { V }^{ \Gamma_1, \, p }_{ \Lambda, \, \Lambda' } ( r ) =
    \sum_{ i_p }^{ N_p } \sum_{ m, \, m' } \,
    C_{ \Lambda, \, L }^{ \Gamma_1, \, { i_p } } \,
    V^{ i_p }_{ L, \, L' } ( r ) \,
    C_{ \Lambda', \, L' }^{ \Gamma_1, \, { i_p } }
\ee
$\Gamma_1$ being the identical representation of the group $\mathcal G$.
Near the origin $X_{ \Lambda \Lambda' }^{\Gamma,\, p} , ( r ) \sim
r \, j_l (kr) \, K_{\Lambda}^{\Gamma,\, p} ( \, \hat {\bf r}_p \, ) \,
\delta_{\Lambda \Lambda' }$.

Across the truncated boundary, we instead use the symmetrized version
of Eq. (\ref{se}), so that putting
$P_{\Lambda}^{\Gamma,\, p} ({ \bf r }_p) =
r \Phi_{\Lambda}^{\Gamma,\, p} ({ \bf r }_p)$ and dropping again the index $p$
\be
\left[ \frac {d^2}{dr^2} + E - V(r,\hat{\bf r}) \right ]\,
P_{\Lambda}^{\Gamma} (r,\hat {\bf r })  =
\frac {1}{r^2} \, \tilde{L}^2 \, P_{\Lambda}^{\Gamma} (r,\hat{\bf r})
\label{sesym}
\ee
where
\be
  \tilde{L}^2 P_{\Lambda}^{\Gamma} (r,\hat{\bf r}) =
    \sum_{\Lambda'} l' ( l' + 1) r R_{\Lambda' \Lambda}^{\Gamma} (r)
K_{\Lambda'}^{\Gamma} ( \, \hat {\bf r} \, )
     \label{pesym}
\ee
and we use starting values given by Eq. (\ref{matnmvsym}). Equation
(\ref{sesym}) is obtained from Eq. (\ref{se}) by applying on the left the
projection operator $\sum_R M_{\rho \rho}^{\Gamma_j} (R) \, P_R$, taking into
account that $V({\bf r}_{i_p}) = V(R^{-1} \, {\bf r}_{i_p'})$.

In terms of $R_{\Lambda \Lambda'}^{\Gamma} (r)$ is then possible to
define the symmetrized version of the matrices $E_{L L'}^{i_p}$ and
$S_{L L'}^{i_p}$ as \bea E_{\Lambda \Lambda'}^p &=& ({R_b^p})^2
W[-i\kappa h_l^+,R_{\Lambda \Lambda'}^p] \\
S_{\Lambda \Lambda'}^p &=& ({R_b^p})^2 W[j_l,R_{\Lambda \Lambda'}^p ]
\eea
and derive the symmetrized equivalent of all the quantities introduced toward
the end of section (\ref{scatstat}). In particular the amplitudes
$B^p_{\Lambda} ({\bf k})$ are solutions of the symmetrized MSE
\be
\sum_{q \, \Lambda'} \left[
{( T^{-1} )}^{ \Gamma, \, p }_{ \Lambda, \, \Lambda' } \,
\delta_{p q} + G^{ \Gamma, \, p \, q }_{ \Lambda, \,
\Lambda'} \right]\, B^q_{\Lambda'} ({\bf k}) = I^p_{\Lambda} ({\bf k})
\ee
where $I^p_{\Lambda} ({\bf k}) = \sum_{m \, i_p} C_{\Lambda, m}^{\Gamma,\,
i_p}\, I_{l m}^{i_p} ({\bf k})$.
Assuming that the photo-absorber is located in cell $\Omega_o$ at the origin
of the coordinates $R_o$, the symmetrized PED cross section for the final state
irrep $\Gamma$ with degeneracy $d(\Gamma)$ takes the form
\be
\frac {d\sigma^{\Gamma}}{d\hat{\bf k}} = 8\, \pi^2\, \alpha\, \hbar\, \omega\,
d(\Gamma) \sum_{n_c} \left | \sum M_{\Lambda_c \Lambda}^{\Gamma_c \Gamma} (E) \,
B^o_{\Lambda} ({\bf k}) \right |^2
\ee
where the atomic dipole transition matrix element
\be
 M_{ \Lambda_c \, \Lambda }^{\Gamma_c \Gamma} \, ( E ) \, = \, \int_{{\Omega}_o}
 \, \phi^{\Gamma_c}_{\Lambda_c } ( {\bf r} ) \, D^{\Gamma_d} ({\bf r})
 \,\overline{\Phi}_{\Lambda}^{\Gamma} ( { \bf r } )\, {\rm d} \, { \bf r}
\ee
obeys the selection rules of the Wigner-Eckart theorem~\cite{tinkham64} for the
finite group $\mathcal G$. We have assumed that the dipole operator transforms
according to the irrep $\Gamma_d$.

%%%
\section{Finiteness of $Tr \, (K_s^{\dagger} K_s)$}
\label{a7}

In  this appendix we show that $Tr \, (K_s^{\dagger} K_s)$ is finite.
Starting from Eq. (\ref{traceks}), we partition the space and the
potential in the way described at the beginning of Section \ref{scatstat} and
define a new kernel $\tilde{K}$ that coincides with the kernel ${\bf K}_s$
for ${\bf r}$ and ${\bf r}'$ in different cells and vanishes
identically when ${\bf r}$ and ${\bf r}'$ happen to be in the same cell.

For this new kernel, $Tr (\tilde{K}^{\dagger} \tilde{K})$ is finite
($\le N < \infty$)  for an even larger class of potentials than that defined
by Eq. (\ref{traceks}) so that, rewriting
${\tilde K}({\bf r},{\bf r}')$ in operator notation, we have
\bea
N & \ge & \, \int {\rm d}{\bf r}  \,
\left < {\bf r} |{\tilde K}^{\dagger}{\tilde K} | {\bf r} \right >  \nr
& = & \,  \int \int \int  {\rm d}E {\rm d}E' {\rm d}E'' \, \sum_{LL'L''}
\, \int  {\rm d}{\bf r} \, \left < {\bf r}|J_L(E) \right > \, \nr
&     & \times \left < J_L(E)|{\tilde K}^{\dagger}|J_{L'}(E') \right >
\left < J_{L'}(E')|{\tilde K}|J_{L''}(E'')\right >
\left < J_{L''}(E'')|{\bf r}\right > \nr
& = & \int \int  {\rm d}E {\rm d}E' \, \sum_{LL'} |{\tilde K}_{LL'}(E,E')|^2 \nr
&\ge& \sum_{LL'} |{\tilde K}_{LL'}(E,E)|^2   \label{tracekt}
\eea
taking into account that the functions
$J_L(E)({\bf r}) = (k/\pi)^{1/2} j_l(kr) Y_L(\hat{\bf r})$, with
the normalization to one state per Rydberg, form a complete orthonormal set.

Now it is easy to see that the matrix ${\tilde K}_{LL'}$ is the asymptotic
form of the kernel $K_s$ in Eq. (\ref{rnspo}) for high values of the
indices $LL'$, since
\bea
& & {\tilde K}_{LL'}(E,E) \nr
& & =  \sum_{i \neq j} \int_{\Omega_i} {\rm d}{\bf r}_i  \,
   J_L(E)({\bf r}_i ) |V_i ({\bf r}_i )|^{1/2} v_i({\bf r}_i ) \, \nr
& & \times \int_{\Omega_j} {\rm d}{\bf r}_j' G_{0}^{+}({\bf r}_i - {\bf r}_j' +
{\bf R}_{ij};{ k})
|V_j ({\bf r}_j' )|^{1/2} v_j({\bf r}_j' ) J_L(E)({\bf r}_j' ) \nr
& & =  \sum_{i \neq j} \sum_{\Lambda \Lambda'} \int_{\Omega_i} {\rm d}{\bf r}_i
 \, J_L(E)({\bf r}_i ) |V_i ({\bf r}_i )|^{1/2} v_i({\bf r}_i )
J_{\Lambda}(E)({\bf r}_i ) \; \tilde{G}^{ij}_{\Lambda \Lambda'} \, \nr
& & \times \int_{\Omega_j} {\rm d}{\bf r}_j  \,J_{\Lambda}(E)({\bf r}_j' ) \,
|V_j ({\bf r}_j' )|^{1/2} v_j({\bf r}_j' ) J_L(E)({\bf r}_j' ) \nr
&  & \sim  \sum_{i \neq j} \sum_{\Lambda \Lambda'}
   \left[T^i_{L \Lambda}\right]^{1/2} \, \tilde{G}^{ij}_{\Lambda \Lambda'} \,
   \left[T^i_{\Lambda' L'}\right]^{1/2}
\label{fundineq}
\eea
the last line following by the fact that asymptotically, when
$L \Lambda, L' \Lambda' \gg k R_b^i \; (\forall i)$,
\be
 \left[T^i_{L \Lambda}\right]^{1/2} \sim
\int_{\Omega_i} {\rm d}{\bf r}_i  \,
J_L(E)({\bf r}_i ) |V_i ({\bf r}_i )|^{1/2} v_i({\bf r}_i )
J_{\Lambda}(E)({\bf r}_i )
\label{t1o2}
\ee
Notice also that we have used the two center expansion for the free Green's
Function
\be
G_{0}^{+}({\bf r} - {\bf r}';{ k}) = J_{\Lambda}(E)({\bf r}_i ) \,
\tilde{G}^{ij}_{\Lambda \Lambda'} \, J_{\Lambda}(E)({\bf r}_j' )
\label{danger}
\ee
which might diverge if $r_i + r_j > R_{ij}$, {\it e. g.} for neighboring cells.
A similar problem is encountered when formulating the variational derivation of
MST (see, for example, section 6.5.3, page 140 and following of
Ref. \cite{gonis00}). One way to solve it is to use the
displaced cell approach~\cite{gonis00}, whereby one can write
\bea
&&{\tilde K}_{LL'}(E,E) \nr
&&= \sum_{i \neq j} \sum_{\underline{\Lambda}}
\left\{ \sum_{\Lambda \Lambda'}  \left[T^i_{L \Lambda}\right]^{1/2} \,
J_{\Lambda \underline{\Lambda}}({\bf b}){G}_{\underline{\Lambda} \Lambda'}
({\bf R}_{ij} + {\bf b}) \, \left[T^i_{\Lambda' L'}\right]^{1/2} \right \}
\label{t1o2c}
\eea
provided that $|{\bf R}_{ij} + {\bf b}| > R_b^i + R_b^j$ and the sums inside the
curly brakets be performed first. Here $J_{\Lambda \underline{\Lambda}}({\bf b})$
is the usual translation operator in MST, given by Eq. (\ref{jdef}) with the
vector ${\bf R}_{ij}$ replaced by the vector ${\bf b}$. The tilde over the symbol
${G}^{ij}_{\Lambda \Lambda'}$ in Eq. (\ref{danger}) was meant to be a reminder
to use this procedure. Notice that the vector ${\bf b}$ depends only on the
geometry of the partition of the space in cells and is independent on $l$.

In this way the expression Eq. (\ref{t1o2c}) is always convergent and is such
that $\sum_{LL'} |{\tilde K}_{LL'}(E,E)|^2 = Tr ({\tilde K}^{\dagger} {\tilde K})$
is finite. Consequently also $Tr \, (K_s^{\dagger} K_s)$ is finite.

%%%%

\section*{References}

\bibliographystyle{iopart-num}
\bibliography{myrefs09}

\end{document}